\documentclass[prd,aps,showpacs,nofootinbib,floatfix,superscriptaddress]{revtex4}
\usepackage{graphicx}
\usepackage{axodraw}
\usepackage{epsfig}
\usepackage{rotating}
\usepackage{amssymb}
\usepackage{dsfont}
\usepackage{psfrag}
\usepackage{amsmath,euscript,array,mathrsfs}
\usepackage{epstopdf}

\newcommand{\SP}[1]{\begin{equation}\begin{split} #1
\end{split}\end{equation}}

\newcommand{\beq}{\begin{equation}}
\newcommand{\eeq}{\end{equation}}
\newcommand{\beqs}{\begin{eqnarray}}
\newcommand{\eeqs}{\end{eqnarray}}
\newcommand{\lsim}{\mathrel{\raisebox{-
.6ex}{$\stackrel{\textstyle<}{\sim}$}}}
\newcommand{\gsim}{\mathrel{\raisebox{-
.6ex}{$\stackrel{\textstyle>}{\sim}$}}}
\newcommand{\Tr}{{\rm Tr}}

\def\hbar{\hspace{0pt}\raisebox{1pt}{$-$} \hspace{-7pt} h}

\def\di{\mbox{d}}
\def\r{\rho}

\newcommand{\be}{\begin{equation}}
\newcommand{\ee}{\end{equation}}
\newcommand{\bea}{\begin{eqnarray}}
\newcommand{\eea}{\end{eqnarray}}

\def\lbldef#1#2{\expandafter\gdef\csname #1\endcsname {#2}}

\def\href#1#2{#2}

\newcommand{\ber}{\begin{eqnarray}}
\newcommand{\eer}{\end{eqnarray}}

\newcommand{\beqar}{\begin{eqnarray}}

\newcommand{\eeqar}{\end{eqnarray}}

\newcommand{\dsl}
  {\kern.06em\hbox{\raise.15ex\hbox{$/$}\kern-.56em\hbox{$\partial$}}}

\newcommand{\eeqarr}{\end{eqnarray}}
\newcommand{\ZZ}{{\rm \kern 0.275em Z \kern -0.92em Z}\;}

\def\CC{{\mathchoice
{\rm C\mkern-8mu\vrule height1.45ex depth-.05ex
width.05em\mkern9mu\kern-.05em}
{\rm C\mkern-8mu\vrule height1.45ex depth-.05ex
width.05em\mkern9mu\kern-.05em}
{\rm C\mkern-8mu\vrule height1ex depth-.07ex
width.035em\mkern9mu\kern-.035em}
{\rm C\mkern-8mu\vrule height.65ex depth-.1ex
width.025em\mkern8mu\kern-.025em}}}

\def\RR{{\rm I\kern-1.6pt {\rm R}}}

\def\ZZ{{\rm Z}\kern-3.8pt {\rm Z} \kern2pt}
\def\IB{\relax{\rm I\kern-.18em B}}
\def\ID{\relax{\rm I\kern-.18em D}}
\def\II{\relax{\rm I\kern-.18em I}}
\def\IP{\relax{\rm I\kern-.18em P}}

\newcommand{\bear}{\begin{eqnarray}}
\newcommand{\eear}{\end{eqnarray}}

\def\to{\rightarrow}

\def\to{\rightarrow}

  \def\w{\omega}
\def\r{\rho}                                     
\def\t{\tau}
\def\u{\upsilon}

\def\lab{\label}
\def\6{\partial}

\def\bea{\begin{eqnarray}}
\def\eea{\end{eqnarray}}

\def\beqx{\begin{displaymath}}
\def\eeqx{\end{displaymath}}

\newcommand{\bmat}{\left(\begin{array}}
\newcommand{\emat}{\end{array}\right)}

\def\r{\rho}

\def\t{\tau}

\def\bo{{\raise-.3ex\hbox{\large$\Box$}}}               
\def\face{{\raise.2ex\hbox{$\displaystyle \bigodot$}\mskip-2.2mu \llap {$\ddot
        \smile$}}}                                   
\def\>{\rangle}                                      
\def\<{\langle}                                      

\def\leftrightarrowfill{$\mathsurround=0pt \mathord\leftarrow \mkern-6mu
        \cleaders\hbox{$\mkern-2mu \mathord- \mkern-2mu$}\hfill
        \mkern-6mu \mathord\rightarrow$}        
\def\dvec#1{\vbox{\ialign{##\crcr
        \leftrightarrowfill\crcr\noalign{\kern-1pt\nointerlineskip}
        $\hfil\displaystyle{#1}\hfil$\crcr}}}           
\def\Tr{{\rm Tr \,}}                                    

\def\-{\hphantom{-}}

\begin{document}
\title{Towards multi-scale dynamics on the baryonic branch of Klebanov-Strassler}

\author{Daniel Elander}
\affiliation{Department of Theoretical Physics, Tata Institute of Fundamental Research, Homi Bhabha Road, Mumbai 400 005, India}
\author{J\'er\^ome Gaillard}
\affiliation{Swansea University, School of Physical Sciences,
Singleton Park, Swansea, Wales, UK. 
Departamento de Fisica de Particulas, 
Universidade de Santiago de 
Compostela and Instituto Galego de Fisica de Altas 
Energias (IGFAE), E-15782, Santiago de Compostela, Spain.}
\author{Carlos N\'u\~nez}
\affiliation{Swansea University, School of Physical Sciences,
Singleton Park, Swansea, Wales, UK.
Max-Planck Institut f\"ur Physik (Werner-Heinsenberg-Institut),\\
F\"ohringer Ring, D-80805, M\"unchen, Germany.}
\author{Maurizio Piai}
\affiliation{Swansea University, School of Physical Sciences,
Singleton Park, Swansea, Wales, UK}

\date{\today}


\begin{abstract}
We construct explicitly a new class of backgrounds in type-IIB supergravity which generalize the baryonic branch of Klebanov-Strassler. We apply a solution-generating technique that, starting from a large class of solutions of the wrapped-D5 system, yields the new solutions, and then proceed to study in detail their properties, both in the IR and in the UV. We propose a simple intuitive field theory interpretation of the rotation procedure and of the meaning of our new solutions within the Papadopoulos-Tseytlin ansatz, in particular in relation to the duality cascade in 
the Klebanov-Strassler solution. The presence in the field theory of different VEVs for 
operators of dimensions 2, 3 and 6 suggests that this is an important step towards 
the construction of the string dual of a genuinely multi-scale (strongly coupled) dynamical model.
\end{abstract}

\pacs{11.25.Tq
}

\maketitle

\tableofcontents

\section{Introduction}

The modern formulation of gauge-string dualities offers a new
computational tool allowing to study field theories
in the strong coupling regime, which exhibit very non-trivial dynamical features,
inaccessible to standard (perturbation-theory) methods.
The most celebrated example of such a correspondence~\cite{AdSCFT}
relates a superconformal four-dimensional theory (${\cal N}=4$ super-Yang-Mills with $SU(N_c)$ gauge group) 
to type-IIB superstring theory, on a background with $AdS_5\times S^5$ geometry (AdS/CFT).
In particular, the regime of large 't Hooft coupling and large $N_c$
of the field theory is related to the weakly-coupled, classical supergravity approximation 
of the ten-dimensional dual  (see also~\cite{reviewAdSCFT} for a review).

Since the discovery of this correspondence, a large amount of effort has been put into looking 
for its generalizations to theories that have less supersymmetry, and that are not conformal, 
with the aim of applying some of the technology developed in the AdS/CFT 
context
to situations closely related to phenomenologically relevant field theories.
Among many, three main examples~\cite{KW,KS,MN} exist of regular 
type-IIB backgrounds 
that are dual to ${\cal N}=1$ 
four-dimensional theories (see also the Klebanov-Tseytlin (KT)
~\cite{KT} background and the Baryonic Branch solution in ~\cite{BGMPZ}), 
in which the six-dimensional internal space 
is related to the conifold and its variations~\cite{CO}. 
The Klebanov-Witten (KW) solution is dual to a (super-)conformal 
theory~\cite{KW}, 
having metric of the form $AdS_5\times T^{1,1}$.
The Klebanov-Strassler (KS)~\cite{KS} 
and wrapped-D5 \cite{MN}  ones
are non-singular backgrounds yielding confinement in the IR.
The  confining field theories dual to these two models are quite non-trivial,
 characterized by one  dynamically generated scale, that appears explicitly
in many interesting physical quantities (such as the gaugino condensate, the string tension and the glueball spectrum).

The next order of complexity is to find the gravity dual of strongly-coupled field theories in which 
two or more (distinct and parametrically separated) scales are generated dynamically.
Besides being an interesting field theory problem per se, this line of research
has a possible field of application in  the context
of dynamical electro-weak symmetry breaking, or technicolor~\cite{TC}, in particular 
in what goes under the name of walking technicolor (WTC)~\cite{WTC} and of extended technicolor~\cite{ETC} 
(see~\cite{reviewsWTC} for reviews on the subject).
These theories are strongly coupled, multi-scale theories, in which many operators develop
condensates and large anomalous dimensions, and hence they are peculiarly difficult to study.
Many  phenomenological aspects of these models of electro-weak symmetry breaking
are not well understood. For instance, it is an open problem whether
they predict the existence of a light composite state 
(dilaton) in the spectrum~\cite{dilatonWTC,EP},
which might have couplings very  similar to those of the Higgs particle of the minimal version of the Standard Model~\cite{dilatonpheno},
and hence very similar LHC signatures.
It is hence useful to try to use the techniques of gauge-string dualities in order to study 
the non-perturbative aspects of multi-scale field theories.

A proposal in the direction of studying the dual of a supersymmetric 
field theory yielding the emergence 
of two dynamical scales is contained in~\cite{NPP,ENP,NPR}.
The starting point of this proposal 
is the type-IIB background generated by a stack of D5-branes
wrapping a compact internal two-cycle. The background consists of 
a metric $g_{\mu\nu}$, dilaton $\Phi$ and 
flux for the RR three-form $F_3$. 
In the specific case of a solution in the form of~\cite{MN},
a suitable definition of the gauge coupling,
in terms of the geometry, complemented by a 
specific radius-energy relation~\cite{VLM}, yields a beta-function that is compatible with
the NSVZ beta-function of SYM~\cite{NSVZ}.
Besides the solution in~\cite{MN}, 
there exist several classes of solutions of the same
equations for the wrapped-D5 system~\cite{HNP}, for which the same definition of gauge coupling 
yields a beta-function that exhibits the features expected in a walking theory~\cite{NPP,ENP,NPR}. 
As a function of the value of the radial direction $\r$
at which the coupling is computed, three very different behaviors appear. 
The coupling runs towards small values going above some value $\r_{\ast}$ of
the radial direction, it is approximately constant over a finite range $\r_I<\r<\r_{\ast}$,
and grows indefinitely below $\r_I$, diverging for $\r\rightarrow 0$.
In one specific class of solutions, the behavior of the Wilson loops 
shows that the theory confines in the conventional sense of producing a linear quark-antiquark static 
potential~\cite{NPR}, although a peculiar behavior similar to a phase transition appears.
Interestingly, for the same class of backgrounds 
one finds that the spectrum of scalar excitations (glueballs)
contains a parametrically light state, whose mass is suppressed as a function of $\r_{\ast}$~\cite{ENP}.

Analyzing in detail solutions in the class of~\cite{NPP} is non-trivial. The geometry is very far from 
being  AdS at all values of the radial direction.
Furthermore, the background is singular: while the Ricci scalar and the square of the Ricci tensor are finite, 
the Kretschmann scalar diverges (in spite of this,  the calculation
of the Wilson loops and of the glueball spectrum yield physically sensible results).
It is difficult to understand the dual 
field theory in detail  and what its dynamical
properties are, including the role of the scale $\r_{\ast}$.
In particular, it is not known what the precise nature of the light state found in~\cite{ENP} is.
In this paper, we construct a more general class of type-IIB backgrounds
which share the interesting features of the class in~\cite{NPP}, but that are
easier to analyze. This is to be understood as a further step towards the 
formulation of the (UV-complete and IR-smooth) 
gravity dual of a genuinely multi-scale
field theory.

We will rely on many known results and build upon them, making extensive use of the vast
amount of knowledge cumulated over the years about type-IIB backgrounds that 
are related to the conifold.
We briefly summarize here the main elements that will be needed in the body of the paper.
All the solutions~\cite{KW,KT,KS,MN,BGMPZ,NPP,ENP,NPR} are special cases of the Papadopoulos-Tseytlin (PT) ansatz~\cite{PT}
(see also~\cite{BHM}), and they can all be obtained by lifting to ten dimensions the solutions
of a specific five-dimensional scalar sigma-model coupled to gravity containing eight dynamical scalars. 
The PT ansatz has recently been shown to
yield a consistent truncation, a subsector of the more general consistent 
truncation on $T^{1,1}$ of ten-dimensional type-IIB  supergravity,
down to ${\cal N}=4$ five-dimensional gauged-supergravity with non-compact gauge group $U(1)\times {\rm Heis}_3$~\cite{CF,BGGHO}.

The generic solution of the BPS equations for the wrapped-D5 system can be obtained
by solving a non-linear second-order equation for a generating function $P$~\cite{HNP},
all the other functions in the background being algebraically related to $P$.
Furthermore, elaborating on~\cite{MM}, it was recently shown that given a solution
of the wrapped-D5 system, subject to some restriction on its 
UV behavior, it is possible to 
algorithmically generate a whole class of more general solutions,
still satisfying the PT ansatz, but in which 
a non-trivial flux  for the RR five-form  $F_5$  
and  the NS two-form $B_2$ are present~\cite{MM}-\cite{rotations}.
We will refer to this algorithmic procedure as {\it rotation}.
In particular, this allows to connect systematically 
the wrapped-D5 system,  the baryonic branch discussed in~\cite{BGMPZ}
and the KS background.
Finally, the relation between the 
five-dimensional scalars of the PT ansatz 
(near a KW fixed point) and the corresponding
field theory operators is known and well understood~\cite{BGMPZ,BCPZ}.

The paper is organized as follows. 
In Section~\ref{Sec:Rotation} we review most of the material discussed above.
We start from the wrapped-D5 system,
rediscuss the class of solutions in~\cite{NPP} and apply 
to them the rotation
of~\cite{MM}-\cite{rotations}. 
In doing so, we find it 
convenient to adopt  the five-dimensional language of~\cite{BHM}.
We also briefly summarize the 
five-dimensional perspective on the KW-KT-KS solutions.
In Section~\ref{Sec:UV}, we study in detail the UV behavior of the rotated solutions, and compare them
both to the original 
unrotated solution and to the KS solutions, 
in the language of the five-dimensional
sigma-model, and in the light of the operator analysis
of all the perturbations of the KW~\cite{KW} fixed point, 
allowed within the PT ansatz~\cite{BGMPZ,BCPZ}. This allows us in particular to discuss the difference between
these various cases in terms of the operators of the 
dual field theory. We present a detailed analysis of the dual field theory
where striking coincidences between the perturbative behavior of the 
quiver field theory and the gravity solution emerge.
In Section~\ref{Sec:IR}, we examine in detail the 
behavior of the solutions in the deep IR.
We show that both the Ricci scalar and the square of the Ricci tensor are finite,
while a singularity appears in the Kretschmann scalar (the invariant built as the square of the Riemann tensor).
We also compute the expectation value of rectangular Wilson loops. The very mild nature of the IR singularity,
and the comparatively nice behavior of the rotated backgrounds in the far UV, 
allow us to follow the prescription in~\cite{MW}, and extract
from it the quark-antiquark static potential $E_{QQ}$. The results are  very similar to those in~\cite{NPR}. Linear confinement
appears at arbitrarily large quark separation $L_{QQ}\rightarrow +\infty$, 
accompanied by the non-standard feature of a first-order phase transition taking place at a finite value of $L_{QQ}$, 
for backgrounds where $\r_{\ast}$ is large enough.
The strength  of the transition depends explicitly on $\r_{\ast}$.
We conclude in Section~\ref{Sec:Outlook}, by critically discussing 
our results
and outlining 
a few possible directions for further development.

\section{A mini-review: a class of solutions interpolating within the PT ansatz\label{Sec:Rotation}}

We start by identifying the class of solutions we are going to study,
and by summarizing all the technology we need in the rest of the paper.
In doing so we make extensive use of the results and language in~\cite{HNP} and~\cite{GMNP},
transcribed into the five-dimensional formulation of the PT ansatz~\cite{PT},
following closely the notation of~\cite{BHM}.
We also briefly remind the reader what the KS~\cite{KS}, KT~\cite{KT} and  KW~\cite{KW}  solutions are.

\subsection{Wrapped-D5 system\label{Sec:wrappedD5system}}

We start from the geometry produced by stacking on top of each other $N_c$ 
D5-branes that wrap an
$S^2$ inside a CY3-fold and then taking
the strongly coupled limit of the gauge theory on this stack,
in the (type-IIB) supergravity approximation~\cite{MN,HNP}. 
We truncate type-IIB supergravity to include only gravity, 
dilaton $\Phi$ and  RR three-form $F_3$,
and define the $SU(2)$ left-invariant one-forms as 
\bea\lab{su2}
\tilde{\w}_1\,=\, \cos\psi d\tilde\theta\,+\,\sin\psi\sin\tilde\theta
d\tilde\phi\,\,,\,
\tilde{\w}_2\,=\,-\sin\psi d\tilde\theta\,+\,\cos\psi\sin\tilde\theta
d\tilde\phi\,\,,\,
\tilde{\w}_3\,=\,d\psi\,+\,\cos\tilde\theta d\tilde\phi\,\,.
\eea
We use an ansatz
that assumes the functions appearing in the background depend 
only the radial coordinate $\r$ (the range of the angles is
$0\leq \theta,\tilde{\theta}<\pi\,,\,
0\leq\phi,\tilde{\phi}<2\pi\,,\,0\leq\psi<4\pi$). We 
write the background (in Einstein frame) as
\bea
ds^2 &=& \alpha' g_s e^{\Phi(\rho)/2} \Big[ (\alpha' g_s)^{-1} dx_{1,3}^2 + ds_6^2 \Big], \nonumber\\
ds_6^2 &=& e^{2k(\rho)}d\rho^2
+ e^{2 h(\rho)}
(d\theta^2 + \sin^2\theta d\phi^2) +\nonumber\\
& &\frac{e^{2 {g}(\rho)}}{4}
\left((\tilde{\omega}_1+a(\rho)d\theta)^2
+ (\tilde{\omega}_2-a(\rho)\sin\theta d\phi)^2\right)
 + \frac{e^{2 k(\rho)}}{4}
(\tilde{\omega}_3 + \cos\theta d\phi)^2, \nonumber\\
F_{3} &=&\frac{\alpha^{\prime}g_sN_c}{4}\Bigg[-(\tilde{\omega}_1+b(\rho) d\theta)\wedge
(\tilde{\omega}_2-b(\rho) \sin\theta d\phi)\wedge
(\tilde{\omega}_3 + \cos\theta d\phi)+\nonumber\\
& & \partial_\rho b \ d\rho \wedge (-d\theta \wedge \tilde{\omega}_1  +
\sin\theta d\phi
\wedge
\tilde{\omega}_2) + (1-b(\rho)^2) \sin\theta d\theta\wedge d\phi \wedge
\tilde{\omega}_3\Bigg].
\label{nonabmetric424}
\eea
The full background is then determined by solving the equations of motion for the 
functions $(a,b,\Phi,g,h,k)$. Notice that from here on we set $\alpha^{\prime}g_s=1$.

The system of BPS equations derived using this ansatz 
can be rearranged in a convenient form, by rewriting the
functions of the background in terms of a set of functions $P(\rho),
Q(\rho),Y(\rho), \tau(\rho), \sigma(\rho)$ as~\cite{HNP}
\beqs
4 e^{2h}=\frac{P^2-Q^2}{P\cosh\tau -Q}, \;\; e^{2{g}}= P\cosh\tau -Q,\;\;
e^{2k}= 4 Y,\;\; a=\frac{P\sinh\tau}{P\cosh\tau -Q},\;\; N_c b= \sigma.
\label{functions}
\eeqs
Using these new variables, one can manipulate the BPS equations to obtain a
single decoupled second order equation for $P(\rho)$, while all other functions are
obtained from $P(\rho)$ as follows:
\bea
& & Q(\rho)=(Q_0+ N_c)\cosh\tau + N_c (2\rho \cosh\tau -1),\nonumber\\
& & \sinh\tau(\rho)=\frac{1}{\sinh(2\rho-2 \rho_0)},\quad \cosh\t(\r)=\coth(2\r-2{\r_0}),\nonumber\\
& & Y(\rho)=\frac{P'}{8},\;\;\;
e^{4\Phi}=\frac{e^{4\Phi_o} \cosh(2{\rho_0})^2}{(P^2-Q^2) Y
\sinh^2\tau},\nonumber\\
& & \sigma=\tanh\tau (Q+N_c)= \frac{(2N_c\rho + Q_o + N_c)}{\sinh(2\rho
-2{\rho_0})}.
\label{BPSeqs}
\eea
The second order equation mentioned above  reads
\beq
P'' + P'\Big(\frac{P'+Q'}{P-Q} +\frac{P'-Q'}{P+Q} - 4 
\coth(2\rho-2{\rho}_0)
\Big)=0.
\label{Eq:master}
\eeq
We will refer to Eq.~(\ref{Eq:master}) as the {\it master equation}: this is the only 
equation that needs solving in order to 
generate the large classes of solutions for the more general Papadopoulos-Tseytlin
system we are interested in.
In this paper we will always set $\r_0=0$, which amounts 
to setting to $1$ the dynamical scale 
in terms of which all other dimensionful parameters will be measured.
Also, in order to avoid a nasty singularity (`bad' according to the criteria
in~\cite{Maldacena:2000mw}) in the IR we fine-tune $Q_0=-N_c$.\footnote{As an example, the solution $P=2N_c \r$ 
gives the background of \cite{MN}. This solution will not be the focus 
of this paper.}

Finally, we revisit the definition of gauge coupling in the dual field theory.
The six-dimensional theory on
the D5-branes has a 't Hooft coupling given by the dimensionful $\lambda_6=g_s \alpha' N_c$,
and the supergravity limit is taken by keeping this fixed~\cite{Itzhaki:1998dd}.
The  branes wrap a small two-cycle $\Sigma_2$,
so that at
low energies an effectively  four-dimensional theory emerges with gauge coupling $g_4$. 
Following~\cite{VLM}, which considers a  five-brane (in the probe approximation)
extended along the Minkowski directions and the two-cycle defined by
$\Sigma_2=[\theta=\tilde{\theta},\;\;
\phi=2\pi-\tilde{\phi},\;\;\psi=\pi]$,
one arrives at \cite{NPP}
\beq
\frac{g_4^2 N_c}{8\pi^2}\,=\,\frac{N_c \coth (\r
)}{P}\,.
\label{Eq:coupling}
\eeq

We move on to discuss the result of applying the
solution-generating technique or rotation. This gives rise to the type-IIB backgrounds that are the object of study of this paper.
\subsection{Rotation: U-duality as a solution-generating technique\label{Sec:RotationUDuality}}
In the paper \cite{MM} the authors proposed a U-duality  
that takes a particular solution to Eq.~(\ref{Eq:master}) ---
hence a background of the form of Eq.~(\ref{nonabmetric424}) --- and maps 
it into another background where new fluxes are turned on.
This U-duality can be seen to be equivalent to (a particular case of)
a rescaling of the K\"ahler two-form and complex structure 
three-form characterizing the background (see 
\cite{GMNP,rotations} for details).

The effect of this solution-generating technique (that we call 
{\it `rotation'}) can be summarized by defining a basis
(below we use the definition
$\hat{h}\equiv 1-k_2^2 e^{2\Phi}$, with the parameter $k_2$ restricted to $0\leq k_2 \leq e^{-\Phi_{\infty}}$)
\bea
& &
e^{xi}= \hat{h}^{-\frac{1}{4}}e^{\frac{\Phi}{4}}dx_i,\;\;\;
e^{\r}= \hat{h}^{\frac{1}{4}}e^{\frac{\Phi}{4} +k}d\r,\;\;\;
e^{3}=\hat{h}^{\frac{1}{4}}\frac{e^{\frac{\Phi}{4} +k}}{2}
(\omega_3+\cos\theta d\varphi),\nonumber\\
& & e^{\theta}=\hat{h}^{\frac{1}{4}}e^{\frac{\Phi}{4} +h} d\theta,\;\;\;
e^{\varphi}=
\hat{h}^{\frac{1}{4}}e^{\frac{\Phi}{4} +h}\sin\theta d\varphi,\nonumber\\
& & e^{1}=\hat{h}^{\frac{1}{4}}\frac{e^{\frac{\Phi}{4} +g}}{2}(\omega_1+a d\theta ),\;\;\;
e^{2}=\hat{h}^{\frac{1}{4}}\frac{e^{\frac{\Phi}{4} +g}}{2}(\omega_2 -a\sin\theta d\varphi),
\eea
where $x_i$ are the four Minkowski directions.
The (new) generated  configuration is
\bea
& & ds_{E}^2= \sum_{i=1}^{10} (e^i)^2,\nonumber\\
& & F_3= \frac{e^{-\frac{3\Phi}{4}}}{\hat{h}^{\frac{3}{4}}}
\Big[f_1 e^{123} + f_2 e^{\theta\varphi 3}
-f_3(e^{\varphi 1 3} + e^{\theta 23})+ f_4(e^{\r 1 \theta} + e^{\r \varphi 2})
    \Big],
\nonumber\\
& & B_2=k_2\frac{e^{3\Phi/2}}{\hat{h}^{1/2}}
\Big[e^{\r 3}+ \cos\mu(e^{\theta\varphi} +e^{12}) 
+\sin\mu (e^{\varphi 1} +e^{\theta 2})   \Big], \nonumber\\
& & H_3=-k_2 \frac{e^{\frac{5\Phi}{4}}}{\hat{h}^{\frac{3}{4}}}\Big[
-f_1 e^{\theta\varphi \rho} - f_2 e^{12\r}+ f_3(e^{\theta 2 \r} +
e^{\varphi 1 \r}) - f_4(e^{\theta 1 3} - e^{\varphi 2 3})
\Big],\nonumber\\
& & F_5= k_2 \frac{d}{d\r}(\frac{e^{2\Phi}}{\hat{h}}) \hat{h}^{3/4}
e^{-k-\frac{5 \Phi}{4}}
\Big[-e^{tx1x2x3 \r}+ e^{\theta\varphi 1 2 3}    \Big],
\label{rotatedbackgroundzz}
\eea
where $\cos\mu=-\frac{P- Q\coth(2\r)}{P\coth(2\r) -Q}$, the functions $f_i, i=1,..,4$ are
\bea
& & f_1= -2 N_c e^{-k-2g},\;\;\; 
f_2 =\frac{N_c}{2}(a^2-2ab +1)e^{-k-2h},\nonumber\\
& & f_3= N_c(b-a)e^{-k-h-g},\;\;\; f_4=\frac{N_c}{2}b' e^{-k-h-g},
\eea
and we denoted
\beq
e^{ijk...l}= e^{i}\wedge e^{j}\wedge e^{k}\wedge...\wedge e^{l}.
\eeq
A necessary condition to apply this solution-generating technique
is that the quantity $e^{\Phi}$ is bounded from above (being an increasing function with $e^{\Phi(\infty)}$ its maximum value). 
This condition can be linked with the 
absence of D7-brane 
sources in the configuration of Eq.~(\ref{rotatedbackgroundzz}) (see~\cite{GMNP} for details). In most parts of this paper, we will choose 
$k_2 e^{\Phi(\infty)}=1$. This is basically keeping the sub-leading
term at infinity in an expansion of the warp factor $\hat{h}(\r)$. 
The rationale for this choice will be carefully discussed 
in the following sections.

For future reference we compare the background in 
Eq.~(\ref{rotatedbackgroundzz}) 
with the generic 
type-IIB background written in Eqs.~(3.8)-(3.11) of the paper~\cite{BHM}
(a detailed comparison will be given in Appendix \ref{detailsrotation}). 
The functions $h_1, h_2, \chi, {\cal K},a,b,\Phi, p,x,\tilde{g}$ 
in \cite{BHM} \footnote{ The function that we call $\tilde{g}$
is denoted $g$ in \cite{BHM}.} are given in terms of the functions $h,g,k,a,b,\Phi$ in Eq.~(\ref{rotatedbackgroundzz}) as
\bea
& & h_2=-\frac{k_2}{4}f_4 e^{2\Phi +h+k+g},\;\; h_2'=-k_2 e^{2\Phi}
[\frac{f_3}{2}e^{k+h+g} -\frac{f_2}{4}a e^{k+2g}],\nonumber\\
& & \chi'+h_1'= -k_2e^{2\Phi}
[f_1 e^{k+2h}   +f_3 ae^{k+h+g}-\frac{f_2}{4}a^2 e^{k+2g}    ],\;\;
h_1'-\chi'= \frac{k_2}{4}e^{2\Phi+k +2g},\nonumber\\
& & {\cal K}= -\frac{k_2}{4} e^{2\Phi+2h+2g}\Phi',\;\; 
e^{2\tilde{g}}= 4 e^{2h-2g},\nonumber\\
& & e^{-6p}=\frac{\hat{h}}{8}e^{2k+h+g+\Phi},\;\;\; 
e^{2x}=\frac{\hat{h}}{4}e^{2h+2g+\Phi},\nonumber\\
& & a\to a,\;\; b\to b,\;\; \Phi \to \Phi.
\label{changeafterrotationzz}
\eea
Let us now move on to describe the five-dimensional 
perspective for these new backgrounds, to be used later in this paper.
\subsection{Five-dimensional language}
Following the notation in~\cite{BHM}, we describe the more general PT system using an effective five-dimensional 
action that reads, up to an overall normalization,
\beqs
{\cal S}&=&\int\di^5y\sqrt{-g}\left[\frac{1}{4}R\,-\,\frac{1}{2}G_{ab}g^{MN}\partial_M\Phi^a\partial_N\Phi^b\,-\,V(\phi)\right]\,,
\eeqs
where $\Phi^a=(\tilde{g},x,p,a,b,\Phi,h_1,h_2,\chi,{\cal K})$ and $y^M=(x^{\mu},r)$.
We impose two constraints
\beqs
{\cal K}&=&M+2N(h_1 + b h_2)\,,\\
\partial_M\chi&=&\frac{(e^{2\tilde{g}}+2a^2+e^{-2\tilde{g}}a^4-e^{-2\tilde{g}})\partial_M h_1+2a(1-e^{-2\tilde{g}}+a^2e^{-2\tilde{g}})\partial_Mh_2}{e^{2\tilde{g}}+(1-a^2)^2e^{-2\tilde{g}}+2a^2}\,,
\eeqs
where ${\cal K}$ is the normalization of the $F_5$ form in ten dimensions, 
and $\chi$, $h_1$ and $h_2$ appear in the NS $B_2$ antisymmetric tensor of type IIB. The quantity 
$N$ is (up to a proportionality constant) 
the normalization of the $F_3$ form, and essentially counts how many D5-branes are present,
while $M$ would count the number of D3-branes if $N=0$.

The constraints allow to remove $\chi$ and ${\cal K}$ from the sigma-model, which is hence defined by
\beqs
G_{ab}\partial_M\Phi^a\partial_N\Phi^b
&=&
\frac{1}{2}\partial_M \tilde{g} \partial_N \tilde{g} 
\,+\,\partial_M x \partial_N x
\,+\,6\partial_M p \partial_N p
\,+\,\frac{1}{4}\partial_M \Phi \partial_N \Phi +\,
\frac{1}{2}e^{-2\tilde{g}}\partial_M a \partial_N a \\
&& +\frac{1}{2}N^2e^{\Phi-2x}\partial_M b \partial_N b 
+\frac{e^{-\Phi-2x}}{e^{2\tilde{g}}+2a^2+e^{-2\tilde{g}}(1-a^2)^2}
\left[\frac{}{}(1+2e^{-2\tilde{g}}a^2)\partial_M h_1 
\partial_N h_1\right.\\ &&\left.
+\frac{1}{2}(e^{2\tilde{g}}+2a^2+e^{-2\tilde{g}}(1+a^2)^2)\partial_M h_2 \partial_N h_2
+2a(e^{-2\tilde{g}}(a^2+1)+1)\partial_M h_1 \partial_N h_2\right]\,.\nonumber
\eeqs
The potential is
\beqs
V&=&-\frac{1}{2}e^{2p-2x}(e^{\tilde{g}}+(1+a^2)e^{-g})
+\,\frac{1}{8}e^{-4p-4x}(e^{2\tilde{g}}+(a^2-1)^2e^{-2\tilde{g}}+2a^2)\\
&&\,+\,\frac{1}{4}a^2e^{-2\tilde{g}+8p}
+\frac{1}{8}N^2e^{\Phi-2x+8p}\left[e^{2\tilde{g}}+e^{-2\tilde{g}}
(a^2-2a b +1)^2 +2 (a-b)^2\right]\nonumber\\
&&\,+\,\frac{1}{4}e^{-\Phi-2x+8p}h_2^2 
+\,\frac{1}{8}e^{8p-4x}(M+2N(h_1+b h_2))^2\,.\nonumber
\label{potentialzzz}\eeqs
The five-dimensional metric is written as (by convention the metric is mostly plus)
\beqs
\label{Eq:5Dmetric}
\di y^2 
&=& e^{2A} \eta_{\mu\nu}\di x^{\mu}\di x^{\nu}\,+\,  \di r^2\,.
\eeqs
The warp factor $A$ is determined by the Einstein equations.

In looking for solutions to the background, we assume that all the functions
have a non-trivial dependence only on the radial direction $r$.
For example, the system of wrapped-D5 in Eq.~(\ref{nonabmetric424})
is obtained, as discussed below, by setting $M=0$ and $N=N_c/4$,
in which case one can consistently 
set $h_1=h_2=\chi={\cal K}=0$, 
reducing to six the number of scalar functions
controlling the background.
The radial directions in the ten and five-dimensional languages 
are connected by the change of variable 
$2 e^{-4p} \di \r =  \di r\,$. Let us study this in more detail.
\subsection{The D5's backgrounds: master equation and rotation}

Given a solution for $P$ is found, one can
algebraically derive the background in Eq.~(\ref{nonabmetric424})
for all the active scalars of the five-dimensional model
\beqs
a&=&\frac{P}{\sinh 2 \r\left(P\coth 2\r -Q\right)},\;\;
b=\frac{2\r}{\sinh 2\r},\nonumber\\
\Phi&=&\frac{1}{4} \log \left(\frac{8 e^{4 \Phi_o} \sinh ^2(2
   \r)}{\left(P^2-Q^2\right) P'}\right),\;\; 
x=\frac{1}{8} \log \left(\frac{e^{4
  \Phi_o} \sinh ^2(2 \r)\left(P^2-Q^2\right)^3}{8192 
   P'}\right),\nonumber\\
  p&=&
  -\frac{1}{24} \log \left(\frac{e^{4 \Phi_o} \left(P^2-Q^2\right)
   \sinh ^2(2 \r) \left(P'\right)^3}{131072}\right), \;\; 
\tilde{g}=\frac{1}{2} 
\log \left(\frac{P^2-Q^2}{(Q-P \coth (2 \r))^2}\right),\nonumber\\
  h_1^{\prime}&=&0\,=\,h_2\,.
\eeqs
As a consequence,
\beqs
A=\frac{1}{6} \log \left(\frac{1}{256} e^{4 \Phi_o}
   \left(P^2-Q^2\right) \sinh ^2(2 \r)\right),\;\;\;
{\cal K}=0\,=\,\chi^{\prime}\,,
\eeqs
and the full type-IIB background in Eq.~(\ref{nonabmetric424}) is known.
For future reference, we highlight 
an important subtlety:
because $B_2=0=F_5$ in the system of Eq.~(\ref{nonabmetric424}), 
one might think that 
there are six active scalars, and hence expect a general solution 
of the BPS equations to depend on six integration constants. This is not so:
the BPS equations do not descend simply from a superpotential
for the five-dimensional description of the wrapped-D5 system, 
but rather the supersymmetric backgrounds must satisfy a system of six 
first-order equations, supplemented by a Hamiltonian constraint. 
After repackaging the resulting system in terms
of $P$ and $Q$, the general BPS solution depends on five integration constants: $Q_0$, $\r_0$, $\Phi_o$,
and the two integration constants of the general solution to the second-order equation for $P$.
As anticipated above, we fine-tune $Q_0$, so
the actual solution depends on four independent integration constants.

Provided $\Phi_{\infty}\equiv\lim_{\r\rightarrow \infty}\Phi$ is finite 
(and the dilaton is a monotonically
increasing function of $\r$, which is always true 
for the solutions we consider in this paper),
the rotation of~\cite{GMNP} allows to 
algorithmically generate the full class 
of solutions, parameterized by $0<k_2<  e^{-\Phi_{\infty}}$.
Comparing with Eq.~(\ref{changeafterrotationzz}) we obtain that in 
the five-dimensional language, the rotation acts as 
(the superscript $(r)$ indicates a `rotated' function)
\beq
	\begin{aligned}
	         a^{(r)} &= a \,,\,\,\,b^{(r)} = b\,,\,\,\Phi^{(r)} = \Phi\,,\,\,e^{2\tilde{g}^{(r)}} = e^{2\tilde{g}}\,,\\
		e^{2x^{(r)}} &= \big( 1- k_2^2 e^{2\Phi} \big) e^{2x} \,,\\
		e^{-6p^{(r)}} &= \big( 1- k_2^2 e^{2\Phi} \big) e^{-6p}\,, \\
		\partial_\rho h_1^{(r)} &= \frac{k_2 N_c}{4} e^{2\Phi} \big[ e^{2\tilde{g}} + 2 a (a-b) + e^{-2\tilde{g}} (a^2 + 1)(a^2 - 2ab +1) \big]\,, \\
		h_2^{(r)} &= \frac{k_2 N_c}{8} e^{2\Phi} \partial_\rho b \,,\\
	\end{aligned}
\eeq
and hence
\beq
	\begin{aligned}
	A^{(r)}&=A + \frac{1}{6}\log (1-k_2^2 e^{2\Phi}),\\
             {\cal K}^{(r)} &= k_2 e^{\Phi + 2x} \partial_\rho \Phi, \\
		\partial_\rho \chi^{(r)} &= \frac{k_2 N_c}{4} e^{2\Phi} \big[ e^{2\tilde{g}} + 2 a (a-b) + e^{-2\tilde{g}} (a^2 - 1)(a^2 - 2ab +1) \big]\,.
			\end{aligned}
\eeq
Notice that the combination $x+3p$ is unaffected by the rotation.

We now specify the 
type of solutions we will be mostly 
interested in in the remainder of this paper. We call them `seed' solutions
since from them, after the rotation procedure is applied,
we construct the backgrounds that are the focus of this paper.
\subsection{Seed solutions}\label{seedsectionxx}

The two-parameter family of solutions discussed 
in~\cite{NPP} is obtained by observing that
 if $P \gg Q$,  the master equation is approximately solved by
\beqs
P_0&=&c\left(\cos^3\alpha+\sin^3\alpha\left(\sinh 4\r-4\r\right)\right)^{1/3}\,.
\label{P0zzxx}
\eeqs
One can then construct the full solution for $P$ by expanding in powers of $N_c/c$,
with $P=\sum_{n=0}^{\infty} P_{n}(\r)\left(\frac{N_c}{c}\right)^{2n}$,
and iteratively solving for each $P_n$ as a function of the parameters $c$ and $\alpha$.
This procedure yields a smooth solution for $P$, provided $P>Q$ for all $\r>0$.
Ultimately, this yields the constraint
\beqs
\cot\alpha&\lsim&\exp\left[\frac{2^{4/3}c}{3N_c}\right]\,.
\label{boundzzz}
\eeqs
If $\alpha$ is small, effectively the solution for $P$ is 
approximately constant for $\r<\r_{\ast}$,
while for $\r>\r_{\ast}$ one sees that $P\simeq e^{\frac{4\r}{3}}$.

Much of this paper is devoted to analyzing the physical meaning of $\r_{\ast}$.
One finds that approximately
$4\r_{\ast}\simeq\log 2\cot^3\alpha$.
By looking at the gauge coupling defined in Eq.~(\ref{Eq:coupling}), one sees that,
provided $\r_{\ast}$ is large, 
there exists an intermediate regime in the radial direction
$\r_I<\r<\r_{\ast}$ over which 
this effective four-dimensional 
gauge coupling is finite and approximately constant~\cite{NPP}.
The scale $\r_I\sim 1$ is the value of the radial coordinate below which
the functions $a$ and $b$ (and hence the gaugino condensate) become non-trivial
(it is the scale above which $\coth 2\r \simeq 1$).

\begin{figure}[h]
\begin{center}
\begin{picture}(510,110)
\put(170,0){\includegraphics[height=3.4cm]{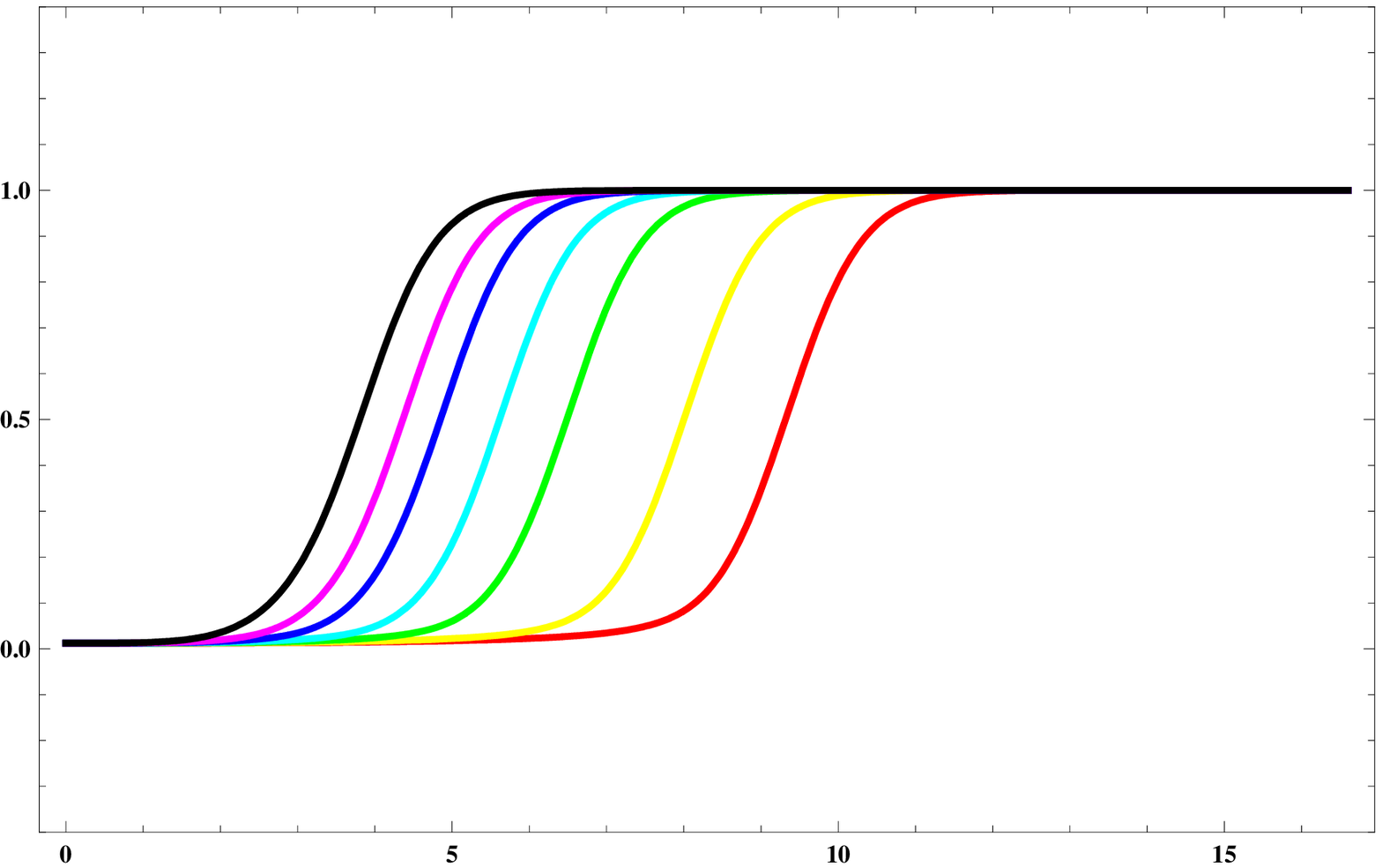}}
\put(0,0){\includegraphics[height=3.4cm]{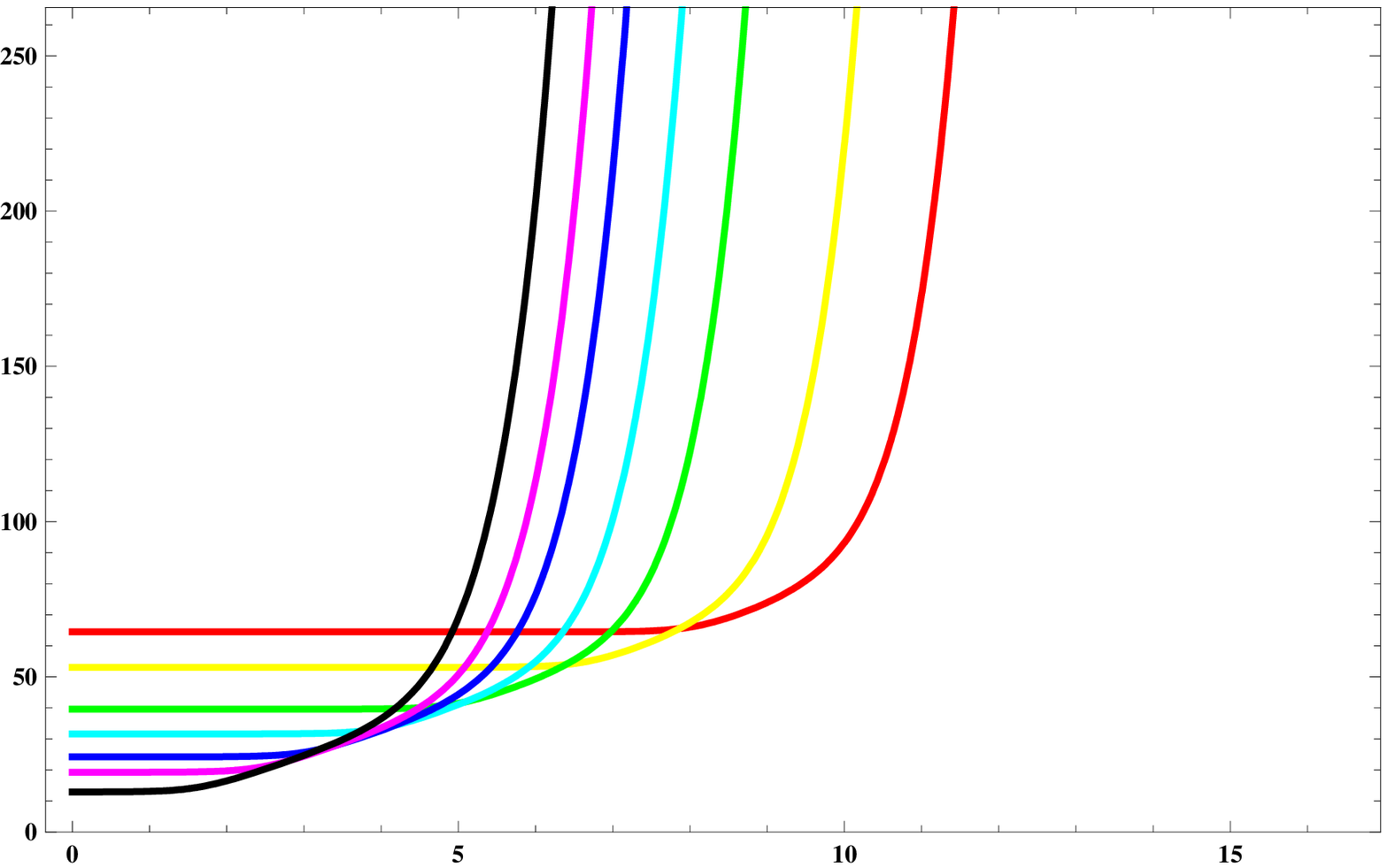}}
\put(349,0){\includegraphics[height=3.4cm]{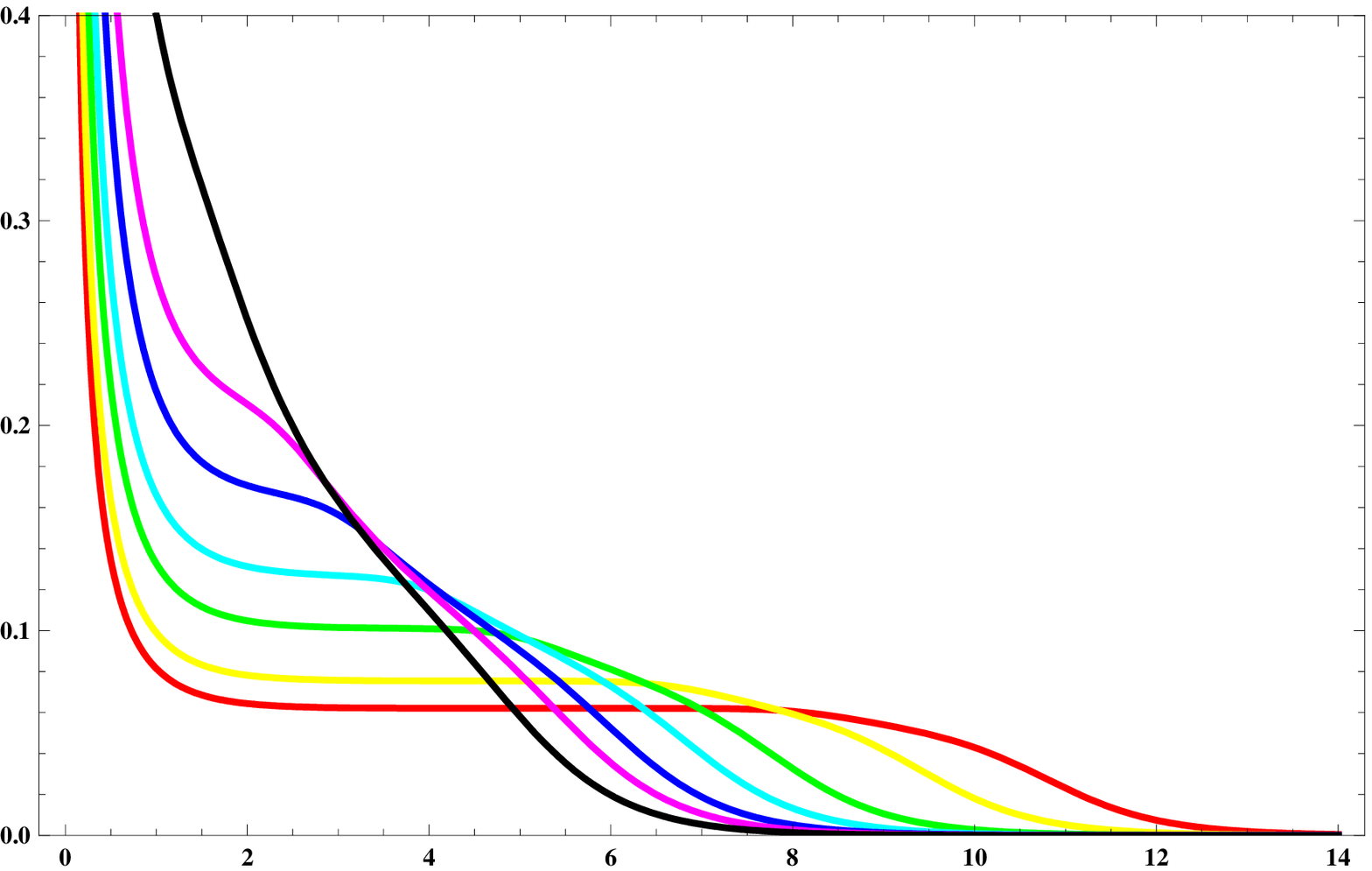}}
\put(145,-3){$\r$}
\put(315,-3){$\r$}
\put(485,-3){$\r$}
\put(161,90){$e^{2\Phi}$}
\put(-6,90){$P$}
\put(330,90){$\frac{g_4^2N_c}{8\pi^2}$}
\end{picture} 
\caption{Examples of the background functions $P$, $e^{2\Phi}$ and $\frac{g_4^2N_c}{8\pi^2}$,
obtained by solving the master equation for $N_c=4$.
Some integration constants are tuned so that the dilaton is kept
constant in the IR and UV. 
The backgrounds differ by the different value of the scale $\r_{\ast}$.
\label{Fig:walking}}
\end{center}
\end{figure}

We plot in Figure~\ref{Fig:walking} some examples of such backgrounds.
Notice that we choose the integration constants in such a way as to make the 
value of the dilaton agree in the far UV and deep IR for all solutions.  We will clarify later 
on the reason for this choice; for the time being, the figure 
has mainly illustrative purposes.

Finally, it is useful to 
remind the reader about the asymptotic expansions 
of solutions of this class~\cite{HNP},\cite{NPP}.
In the far UV, for $\r\rightarrow \infty$:
\beqs
P&=&3c_+ e^{4\r/3}\,+\,4 \frac{N_c^2}{3 c_+} \left(\r^2 - \r + \frac{13}{16}\right) e^{-4 \r/3}\,+
\,\left(-{8}c_+\r-\frac{c_-}{192c_+^2}\right)e^{-8\r/3}\,+\,{\cal O}(e^{-4\r})\,,
\label{asympzzxx}
\eeqs
where $c_{\pm}$ are the two constants characterizing all of these solutions.
In the IR, for $\r\rightarrow 0$ we have
 \beqs
 P&=&c_0\,+k_3 c_0 \r^3+\frac{4}{5}k_3 c_0 \r^5-k_3^2c_0\r^6+\frac{16(2c_0^2 k_3-5k_3 N_c^2)}{105c_0}\r^7\,+\,{\cal O}(\r^8)\,,
\eeqs
where now $c_0$ and $k_3$ are the free parameters.
One can hence write all of these solutions by specifying $N_c$ and 
any of the pairs $(c,\alpha)$, $(c_+,c_-)$ or $(c_0,k_3)$.
The relation between these is not known in analytical form,
and which parameterization to use is mostly a matter of convenience.

The subject of this paper is the class of solutions that are 
obtained from the seed solution discussed above by applying the
rotation procedure. As it will be useful in following sections, 
we move on to
describe in five-dimensional language the solutions
discovered by Klebanov-Strassler~\cite{KS}, Klebanov-Witten~\cite{KW}, 
Klebanov-Tseytlin~\cite{KT}.

\subsection{
Summary of the Klebanov-Strassler, Klebanov-Tseytlin and 
Klebanov-Witten solutions}\label{KSKTKWsummary}
We  briefly summarize in this subsection the relevant properties 
of the KS-KT-KW solutions.
The Klebanov-Strassler system can be obtained from the PT one with the constraint
\beqs
a=\tanh y, \;\;
e^{-\tilde{g}}=\cosh y\,,
\label{constraintzz}\eeqs
such that a superpotential exists:
\beqs
W&=&\frac{1}{4}e^{-2(p+x)}\left(e^{6p}\left(\frac{}{}M-2e^{2x}\cosh y +2 (h_1+b h_2) N\right)-2\right)\,.
\eeqs
This has to be understood 
in the sense that given a solution to the BPS equations,
\beqs
\partial_r A &=&-\frac{2}{3}W\,,\\
\partial_r \Phi^a&=&G^{ab}\frac{\partial W}{\partial \Phi^b}\,,
\eeqs
the resulting $A$ and $\Phi^a$ satisfy automatically the classical equations 
derived from the sigma-model.

Starting from the first-order equations of the KS system, 
one finds that
in the usual $\r$ coordinate in which $\di r = 2 e^{-4p} \di \r$, 
\beqs
y&=&-\ln \tanh (\r-\r_0)\,=\,2 {\rm arctanh}\,e^{-2(\r-\r_0)}\,.
\eeqs
Setting $\r_0=0$, as usual, yields
a second-order equation for $b$ (obtained by combining with the equation for $h_2$) 
that is solved by
\beqs
b&=&b_1\cosh 2\r +\frac{b_2+(1-b_1)2\r}{\sinh 2\r}\,.
\eeqs
Setting $b_1=0$ makes the function $b$ well behaved in the UV, while setting $b_2=0$ 
avoids the arising of a nasty singularity in the IR.
With these three choices, one has six of the background scalars:
\beqs
\Phi&=&\Phi_{\infty},\;\;
a=\frac{1}{\cosh2\r}, \nonumber\\
e^{-\tilde{g}}&=&\coth 2\r , \;\;
b=\frac{2\r}{\sinh 2\r}, \nonumber\\
h_2&=&\frac{N}{2}e^{\Phi_{\infty}} 
\partial_{\r}b\,=\,-Ne^{\Phi_{\infty}}\frac{-1+2\r \coth 2\r}{\sinh 2\r},
\nonumber\\
h_1&=&\frac{N}{2}e^{\Phi_{\infty}}\frac{2\r(1+\cosh 4\r)-\sinh 4\r}{\sinh^2 2\r}\,+\,\tilde{h}_1\,.
\eeqs
The integration constant $\tilde{h}_1$ just amounts to a rescaling $M\rightarrow \tilde{M}=M+2N \tilde{h}_1$.
The equations for $p$ and for $x$ are less friendly. By defining
\beqs
f&\equiv&x+3p+\frac{1}{2}\ln\frac{2}{3}\,,
\eeqs
one finds that
\beqs
\label{eq:KSf}
e^{2f}&=&\frac{\sinh 4\r+f_0-4\r}{\cosh 4\r -1}\,,
\label{f0zzzxx}
\eeqs
and again we set $f_0=0$ in order to avoid an IR singularity. The equation for $x$ reduces to
\beqs
\label{eq:KSeomx}
x^{\prime}+{\cal K} e^{-2x}-\frac{4}{3}e^{-2f}&=&0\,,
\eeqs
where the function ${\cal K}=M+2N(h_1+b h_2)$ is the normalization of the $F_5$ form,
and is known in closed form from the previous functions. The regular KS solution
is obtained by fine-tuning $\tilde{h}_1$ so that $\tilde{M}=0$. The equation for $x$ can be solved only numerically. Finally, the solution for the warp factor $A$ can be written in terms of $x$ as
\SP{
	A = A_0 + \frac{x + \log \sinh (2 \rho)}{3},
}
where $A_0$ is an integration constant that we put equal to zero (it can be reabsorbed into $dx_{1,3}^2$ and just sets an overall energy scale).

Because we are mostly interested in the UV expansion, some useful information
can be obtained from the (singular) KT limit, 
obtained by retaining only $x$, $p$ and $h_1$
as dynamical fields.
In this case the solution is
\beqs
\Phi&=&\Phi_{\infty},\;\;a=0\,=\,\tilde{g}\,=\,b\,=\,h_2, \nonumber\\
h_1&=&{N}e^{\Phi_{\infty}}\left(2\r-1\right)\,+\,\tilde{h}_1\;\;
x+3p=\frac{1}{2}\ln\frac{3}{2}, \nonumber\\
p&=&-\frac{4}{9}\r-\frac{1}{6}\ln\left[\left(-\frac{N^2e^{\Phi_{\infty}}}{4}+\frac{\tilde{M}}{2}+2N^2e^{\Phi_{\infty}}\r\right)e^{-\frac{8}{3}\r}\,+\,\tilde{p}\right]\,,
\label{KTzzxxcc}
\eeqs
where $\tilde{p}$ is the last integration constant in the system, while in $x+3p$ we set the integration constant $f_0=0$.
Notice that setting $\tilde{p}=0$ results in a softening of the divergence  of $p$ as a function of $\r$ (for large $\r$).

For completeness, notice that for $N=0$ there exists a 
constant solution
\beqs
\Phi&=&\Phi_{\infty},\;\;a=0\,=\,\tilde{g}\,=\,b\,=\,h_2\,=\,h_1, \nonumber\\
x+3p&=&\frac{1}{2}\ln\frac{3}{2},\;\;p=-\frac{1}{6}\log\frac{M}{2}\,.
\label{zzxxcc}\eeqs
This is the KW solution that yields 
the $AdS_5\times T^{1,1}$ background geometry mentioned earlier on.

Summarizing, the constraint $a^2+e^{2\tilde{g}}-1=0$, see 
Eq.~(\ref{constraintzz}),
allows to reduce the system to 
seven scalars, with a known superpotential.
The solution of the resulting first-order equations for the scalars depends on seven integration constants
$(\Phi_{\infty}, \r_0, b_1, b_2, \tilde{h}_1, f_0, \tilde{p})$, besides $N$ and $M$.
While $\Phi_{\infty}$  has little to no physical effect on the resulting solution,
and $\r_0=0$ simply defines the (dynamical) scale of the theory, by setting the end-of-space in the radial direction,
one must set $b_1=b_2=f_0=\tilde{p}=0$ together with $\tilde{h}_1=-M/(2N)$ in order to avoid singular behaviors in the IR and in the UV.
As a result, the general (regular) KS solution depends on the two harmless, 
independent integration constants $\Phi_{\infty}$ and $\r_0$.

\section{Short-distance physics: towards a systematic field theory interpretation\label{Sec:UV}}

This section is mostly devoted to the study of the UV asymptotic behavior of the rotated solutions.
By doing so, we can interpret the integration constants in terms of the operators deforming the KW fixed points.
By comparing the rotated solution with the 
unrotated solution and with the KS solution,
we can precisely identify what is the difference between these three classes of backgrounds,
in terms of couplings and VEVs of field theory operators.

\subsection{General analysis\label{Sec:GA}}

First, we summarize some general results that hold for all the backgrounds compatible with the 
PT ansatz.
Because the background with $AdS_5\times T^{1,1}$ geometry is dual to a conformal theory,
it is sensible to expand the potential around the KW fixed point(s).
In doing so, one finds that the general solution differs from the conformal one(s) by
the presence of terms that scale with power $\Delta$, that is,
for a generic field $\varphi\sim z^{\Delta}$ as $z\rightarrow 0$, with 
$z=e^{- \frac{2}{3}\r}$, in the radial coordinate used in the previous section.
This power is either the physical dimension of an operator of dimension $\Delta$ that is developing a VEV,
or the dimension of the coupling of an operator of dimension $4-\Delta$ that is added to the dual theory.
The allowed values of $\Delta$ can be classified in full generality, by requiring that the background satisfies the
PT ansatz. 
We start this analysis by recalling what are 
the allowed values of $\Delta$, and 
what background fields they are associated with.

Perturbations driven by the scalar fields 
$\Phi$ and $h_1$ correspond to the scaling dimensions $\Delta=0,4$ (which can be interpreted in terms of
a marginal deformation and its conjugate VEV).
 Notice  that while the dual of $\Phi$ is exactly marginal, 
 the dual of $h_1$ is not, an observation that we will recall and use 
later on.\footnote{Notice that what we mean by exactly marginal here is
 only the fact that the leading-order expansion in small $z$ 
contains a constant, but not a logarithm.}
With the field $a$  
are associated scaling dimensions $\Delta=1,3$, while to $\tilde{g}$ scaling dimension 
$\Delta=2$. The  system of $h_2$ and $b$ mixes, and the resulting scaling dimensions are $\Delta=-3,1,3,7$.
Finally, the mixed system of $x$ and $p$ corresponds to scaling dimensions $\Delta=-4,-2,6,8$. 
Summarizing, the dual field theory 
can be described in terms of a conformal theory, perturbed by the presence of VEVs and
couplings of a set of eight possible operators: we have 
operators  of dimensions 
2,6,7 and 8 (one operator for each dimension), 
two operators of dimension 3 and two of dimension 4.

At the microscopic level, the field theory dual of the KW background 
is based on an ${\cal N}=1$ gauge theory with gauge group $SU(M)\times SU(M)$,
containing chiral superfields $A_{1,2}\sim(M,\bar{M})$ and $B_{1,2}\sim(\bar{M},{M})$.
All the corresponding field theory operators of the dual field theory can be found in~\cite{BCPZ}
and we summarize them schematically in Table~\ref{Fig:dimensions}.
Because we will always work with BPS equations, only at most
half of the admissible scaling dimensions are going to appear in the UV expansions of the solutions.
By inspection, it turns out that we expect at most the presence of four couplings: the two marginal
couplings are related to the two gauge couplings, and the coupling of the dimension-7 and dimension-8 operators are allowed.
Four possible VEVs are also present: for the two dimension-3 operators, for the dimension-2 operator and for the dimension-6 operator~\footnote{In the literature, the dimension-3 and dimension-2 VEVs are associated with the deformation and the resolution of the conifold, respectively,
 while the dimension-6 VEV has been discussed for example in~\cite{PZT}.}.

All the solutions we discuss differ by which of these couplings and VEVs are  non-zero (and {\it independent}).
In the KS case, 
as we discussed in Section~\ref{KSKTKWsummary} 
and summarized below Eq.~(\ref{zzxxcc}), 
we are setting to zero five integration constants, plus imposing a constraint that
reduces to seven the number of scalars.
This means that the only allowed couplings are the two marginal ones,  
but with coefficients that are related to one another.
Ultimately, the fact that one of the two is never 
exactly marginal, unless $N=0$,  is what makes KS not asymptotically AdS in the far UV.
 Also, the VEV of a combination of the dimension-3 operators is present (the gaugino condensate).
This can be verified explicitly in the expansions in 
Appendix~\ref{Sec:expansions},
where we retained for completeness $\tilde{M}\neq 0$, and expanded for small $z$,
with $\r=-\frac{3}{2}\log z$.

\begin{table}
\begin{tabular}{|c|c|c|c|c|c|c|c|c|}
\hline\hline
$\Phi^a$ & $\Delta$ & ${\cal O}$~\cite{BCPZ} & BPS & KS & $k_2=0$ & $0<k_2<e^{-\Phi_{\infty}}$ & $k_2=e^{-\Phi_{\infty}}$ & Butti et al.~\cite{BGMPZ}\cr
\hline
$a$ & 1,3&$\Tr (W_1^2-W_2^2)$&3&3&3&3&3&3\cr
$\tilde{g}$ &2&$\Tr (A\bar{A}-B\bar{B})$&2&&&&2&2\cr
$\Phi$ &0,4&$\Tr (F_1^2+F_2^2)$&0&0&0&0&0&0\cr
$h_1$ &0,4&$\Tr (F_1^2-F_2^2)$&0&&&0&&\cr
$x$, $p$ &-4,-2,6,8&$\Tr W^2\bar{W}^2$&-4,6&&-4,6&-4,6&6&\cr
$b$, $h_2$ &1,3&$\Tr( W_1^2+W_2^2)$&3&&&&&\cr
&-3,7&$\Tr (A\bar{A}+B\bar{B}) W^2$&-3&&&&&\cr
\hline\hline
\end{tabular}
\caption{Field-theory operator analysis, based on expanding  in the UV near the KW fixed points.
The columns show the five-dimensional fields, the scaling dimensions of the perturbations they allow,
the corresponding field-theory operators in terms of the two gauge groups, the scaling dimensions selected by the
BPS equations. The last five columns show which couplings or VEVs
correspond to the  {\it independent} integration constants that can be dialed, 
labelled by the corresponding scaling-dimension of the gravity-dual scalar.
Notice that some of the couplings/VEVs that are not explicitly highlighted are
present, but their UV-boundary values are not independent, in particular the dimension-2 VEV is present in all the solutions constructed starting
from the master equation.}
\label{Fig:dimensions}
\end{table}

Let us move on to study the new solutions.
First of all, we notice that two important combinations of the scalars are
unaffected by the rotation.
One is
\beqs
a^2+e^{2\tilde{g}}-1&=&\frac{2Q}{P\coth 2\r -Q}\,.
\eeqs
Setting this to zero would amount to imposing the constraint that defines the
KS system. But this is not allowed, because Q cannot vanish: all the rotated solutions belong on
the baryonic branch of the KS system.
This is also indicated by the fact  that the dimension-2 VEV is turned on, as can be seen from the expansion of $\tilde{g}^{(r)}$
from Appendix~\ref{Sec:expansions}.
In the limit in which  $P\gg Q$, equivalently $N_c/c\rightarrow 0$
--- see Eq.~(\ref{asympzzxx}) --- the violation of the constraint becomes parametrically small,
and  this is the regime in which the approximation $P\simeq P_0$, 
see Eq.~(\ref{P0zzxx}), becomes accurate.

\begin{figure}[b]
\begin{center}
\begin{picture}(460,550)
\put(0,0){\includegraphics[height=20cm]{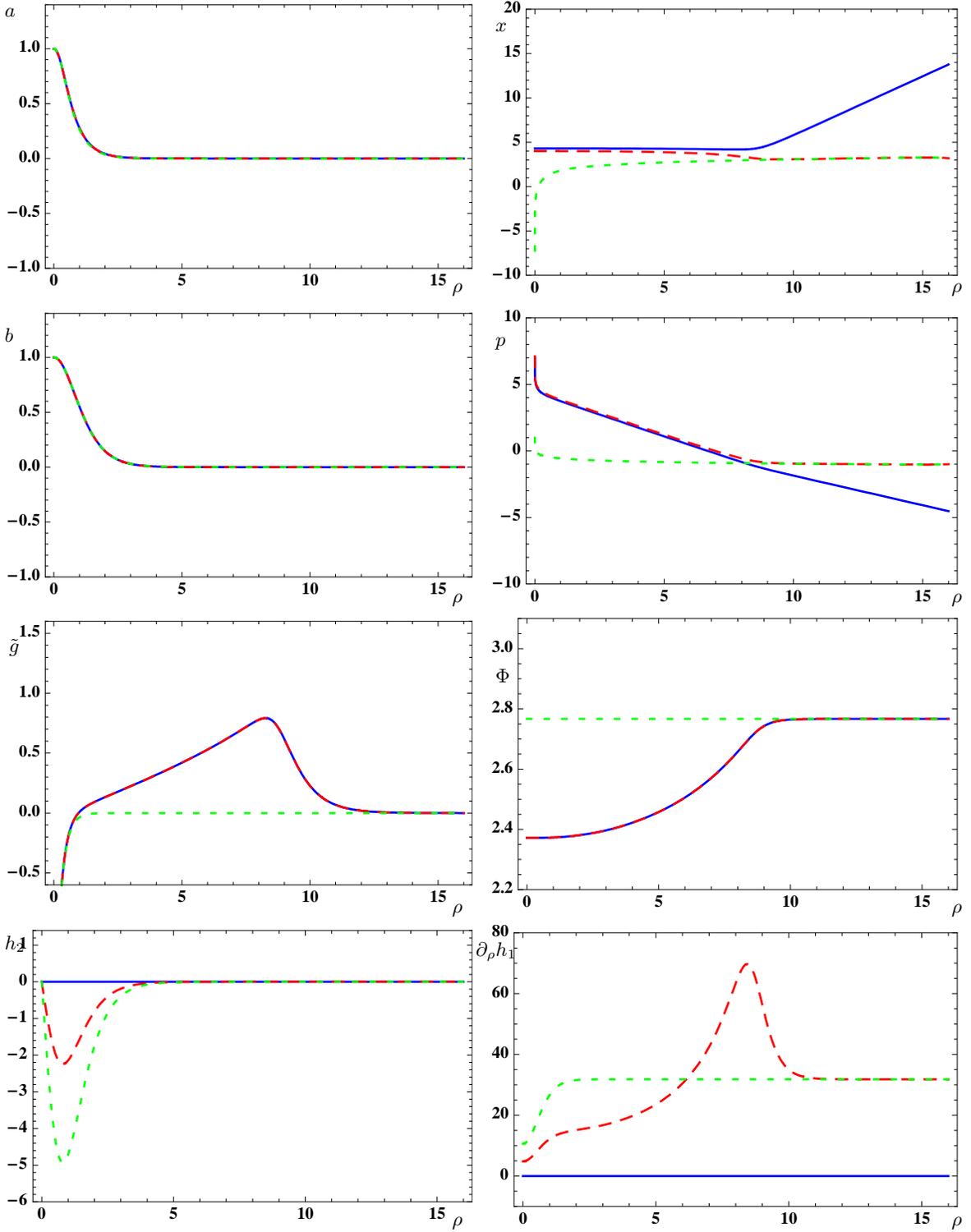}}
\put(-1,560){$a$}
\put(-1,410){$b$}
\put(1,267){$\tilde{g}$}
\put(-1,130){$h_2$}
\put(225,553){$x$}
\put(225,408){$p$}
\put(225,253){$\Phi$}
\put(216,127){$\partial_{\r}h_1$}
\put(435,3){$\r$}
\put(435,147){$\r$}
\put(435,290){$\r$}
\put(435,433){$\r$}
\put(205,3){$\r$}
\put(205,147){$\r$}
\put(205,290){$\r$}
\put(205,433){$\r$}
\end{picture} 
\caption{Three exact BPS backgrounds in the PT system, obtained numerically.
In blue (continuum line) a numerical solution for the wrapped-D5 system, obtained by solving the master equation.
In red (long-dashed line) the result of rotating the solution in blue, and fine-tuning $k_2$. In green (short-dashing)
 a regular KS solution, obtained by matching (where possible) 
the boundary conditions.}
\label{Fig:plotcompare}
\end{center}
\end{figure}

The second interesting invariant is $x+3p$, which in terms of the variables entering the master equation is given by
\beqs
e^{4x+12p}&=&\frac{4(P^2-Q^2)}{(\partial_{\r}P)^2}\,.
\eeqs
In the KS case, the solution for $x+3p$ depends on 
the integration constant $f_0$,
that is set to zero---see around Eq.~(\ref{eq:KSf}). 
In the case of the seed solutions in Section \ref{seedsectionxx}, 
this is not the case: the constant $c_-$ appears
in the coefficient of corrections scaling as $z^6$ (see again 
Appendix~\ref{Sec:expansions}). 
This means that when allowing  a non-vanishing value of $c_-$
we are turning on the VEV of the dimension-6 operator, 
with arbitrary strength.

Another important invariant of the rotation is the dilaton $\Phi$.
As a consequence of the fact that the constraint yielding KS is violated, the dilaton  has non-trivial dynamical equations,
and hence a non-trivial profile.  Again from the expansion in 
Appendix~\ref{Sec:expansions}
one sees that indeed the corrections are proportional to $N_c/c_+$.

The fact that the dilaton is non-trivial, and goes to a finite value in the UV, means that it is possible to fine-tune $k_2=e^{-\Phi_{\infty}}$.
This is of crucial importance, let us explain why.
First of all, notice that whenever an irrelevant operator is 
inserted, it makes little sense to perform 
the expansion
as in Appendix~\ref{Sec:expansions}.  One should first 
find a regime in which the background is at least
in some sense close to conformal, and expand from there.

In order to do so, we consider the UV expansion of $P$ (setting $c_-=0$ for simplicity),
replace in the expression for $x^{(k_2=0)}-p^{(k_2=0)}$, and (formally) expand first for small $c_+$, and then for
small $z$. The result is trustable only at the leading order, which yields 
\beqs
x^{(k_2=0)}-p^{(k_2=0)}&=&
\frac{10 {c_+}^2 (30 \log (z) (3 \log (z)+2)+37)}{3 {N_c}^2 z^4
   (12 \log (z) (3 \log (z)+2)+13)^2}\,.\label{e6pkt}
\eeqs
The choice of $x-p$ is just dictated by convenience, similar results holding for any 
combination of $x$ and $p$ (aside from $x+3p$).
The conclusion of this exercise
 is that in the seed solutions, as well as in their rotation with generic values of $k_2$,
the constant $c_+^2/N_c^2$ controls the coupling of the dimension-8 operator, and is the analog of $\tilde{p}$ in the KS system.
This is ultimately what renders pathological the UV behavior of the wrapped-D5 backgrounds with $P\simeq c_+e^{\frac{4}{3}\r}$,
which would correspond to field theories that need a UV completion, 
because their UV dynamics is dominated  by the higher-dimensional operator.

The fine-tuning of $k_2$ allows to adiabatically switch off 
this higher-order operator, as we explain now.
We start with  two minor remarks, which are important for technical reasons.
Let us try to identify (at least at leading-order) the expansions of 
the KS solutions
--- see Eqs.~(\ref{KSexpansionszz})-(\ref{KSexpansionfinal}) --- with those of the 
rotated and fine-tuned solutions --- see 
Eqs.~(\ref{bbexpansionzz})-(\ref{bbexpansionfinal}).
In order to do so, one sees that by choosing
\beqs
k_2&=&e^{-\Phi_{\infty}}\,=\,(18c_+^3)^{1/4}e^{-\Phi_o}\,,
\eeqs
together with $N=N_c/4$, $c_-=0$ and $N_c/c_+=0$,\footnote{Notice that this has to be understood as a limit procedure, in which one
keeps $N_c$ fixed, and dials $c_+$ to large values, hence producing backgrounds that, after rotating, 
approximate to the KS ones~\cite{GMNP}.} one makes $x+3p$, $a$, $b$, $\Phi$, $\tilde{g}$, $h_1$ and $h_2$ agree with KS in the far UV (at leading-order).
Interestingly, for $x$ to actually agree one needs also $\tilde{M}=M+2N\tilde{h}_1=0$. This last observation
will help us understand the field theory interpretation of the rotation itself and we will make extensive use of it in the following subsections.
The dilaton being an invariant 
and well behaved in the far UV, we can use the expansion for $\Phi$
from Appendix~\ref{Sec:expansions}, from which one sees that by 
fine-tuning $k_2$
one ensures that the rotation factor 
$\hat{h}=1-k_2^2e^{2\Phi} \propto z^4$, 
which cancels the $1/z^4$ term, for example in the expansion of Eq.~(\ref{e6pkt}).
In practice, this means that the coupling of the  
marginal operator in $h_1$ (related to $k_2$) is fine-tuned 
against the marginal operator in $\Phi$ 
(related to $e^{\Phi_{\infty}}$) 
in such a way as to switch off the dimension-8 operator,
while preserving the dimension-2 VEV (related to the function $\tilde{g}$).

Finally, let us summarize what couplings and VEVs are present in each of the cases.
The seed solutions depend, as we said, 
on four integration constants $\r_0$, $\Phi_o$, $c_+$ and $c_-$.
The quantity $\r_0$ corresponds to a VEV for 
the dimension-3 operator (gaugino condensate), 
in the same sense as in the KS solution.
The normalization of the dilaton corresponds to a marginal coupling, while the absence of the $h_i$ fields 
is related to the fact that there is only one gauge group, and hence one gauge coupling in the dual theory.
The constant
$c_+^2/N_c^2$ corresponds to a deformation 
due to the dimension-8 operator,
while the quantity $c_-/c_+^3$ to the VEV of the dimension-6 operator.
The generic rotated solution differs by the fact that the second gauge group is now present,
and hence a second quasi-marginal deformation proportional to $h_1$ (hence
proportional to $k_2$) is 
driving the flow.
The fine-tuned rotated solution corresponds to a peculiar choice 
such that the coupling of the dimension-8 
operator is switched off adiabatically 
(i.~e. keeping the dimension-2 VEV fixed), while at the same time relating
among each other in a specific way the couplings of the two marginal operators.
All of these solutions live on the baryonic branch, 
because the dimension-2 VEV is present (though
its boundary value is fixed, as we said, in such a way as to avoid a nasty singularity in the IR).
The whole analysis is summarized by Table~\ref{Fig:dimensions}.

Concluding, the main difference between the KS solutions and 
the rotated and fine-tuned 
solutions is the insertion of two VEVs: 
a dimension-2 one, which brings the background on the baryonic branch,\footnote{We did not check the existence of
a massless normalizable glueball associated with the breaking of baryonic symmetry.} 
and a dimension-6 one. 
Hence in the UV the rotated and fine-tuned solutions are going to
be almost indistinguishable from KS. Big differences will emerge 
for $\r<\r_{\ast}$, due to the VEVs.
This is illustrated graphically in Fig.~\ref{Fig:plotcompare}.
What the figure shows is the background value of the eight scalars for three solutions.
First of all, we plot the original seed solution~\cite{NPP}, which belongs to the wrapped-D5 system,
as shown by the vanishing of $h_{1,2}$.
The UV of such a solution is bad, as shown by the divergence of $p$ and $x$, ultimately due 
to the presence of the dimension-8 operator.
$\Phi$ not being constant, but approaching a constant in the UV,
 one can apply the rotation and fine-tune $k_2$ so as to remove 
the dimension-8 operator from the dual field theory,
hence smoothening the far-UV behavior of $x$ and $p$. At the same time, this 
 induces non-trivial profiles for $h_{1,2}$.
 We then compare to the KS solution, chosen so as to match the rotated solution in the far UV
 (in particular, by setting $\tilde{M}=0$).
 Above $\r_{\ast}\simeq 9$ the two are almost indistinguishable.
 However, below $\r_{\ast}$  the VEVs are playing an important role. The KS solution
has very different $\tilde{g}$, $\Phi$, $x$ and $p$, while $a$, $b$, $h_1$ and $h_2$ are
qualitatively very similar.

A very final comment concerns the relation of these classes of solutions to the baryonic branch in~\cite{BGMPZ}.
Indeed,  the constraint $a^2+e^{2\tilde{g}}-1=0$ is always violated
and hence the solutions never really agree with KS. Far in the UV, they rather agree with the solutions in~\cite{BGMPZ}.
The main difference with respect to~\cite{BGMPZ} is the presence of the dimension-6 VEV, which results in the background being very different 
in the deep IR, where a (mild) singularity appears, which 
is absent in~\cite{BGMPZ}.
If one were to evolve from the UV the rotated solutions with $c_-=0$ towards the IR,
the singularity at the end-of-space in the IR would disappear and one would exactly describe the baryonic branch,
in the same sense as in~\cite{GMNP}.
  
  \subsection{The rotation and its field theory interpretation}
  
In this subsection we propose a field-theory 
interpretation of the rotation procedure,
that  integrates and complements the discussions in~\cite{KS,MN,GMNP,MM,AD}.
We start by highlighting a set of seemingly puzzling facts about the
backgrounds we built. Some of what we say here repeats previous
results, but we find it convenient to collect together all the useful information we have. 
  
In short, the unrotated solutions differ from the KS
  one in four respects.
  Two of these are well known and admit a clean field-theory interpretation.
  There are neither $B_2$ nor $F_5$ in the wrapped-D5 system, and as
  a consequence one of the (quasi-) marginal deformations,
  signaled by $h_1$, is absent. In the dual field theory language,
  this means that there is only one gauge group with adjoint matter
in contrast with two gauge groups and bifundamental matter as in the 
quiver theories dual
  to the KW and KS backgrounds.
  There is a dimension-8 operator deforming the theory. As a result, the dual field theory is not
  UV complete, not even in the generalized sense of KS.
  These first two differences are affected by the rotation, which depending on the value
  of $k_2$ amounts to switching back on the second marginal 
deformation (and hence the dual field theory
 is a quiver). 
In particular, fine-tuning $k_2$ to its maximal value leaves us with a KS-like
quiver, in which only one tunable parameter controls both marginal 
deformations, and at the same time, in this 
limit the dimension-8 deformation is exactly switched off.
  
Aside from this, there are 
two differences between the wrapped-D5 background 
and the KS background which are not affected
  by the rotation, and that yield two puzzling results. 
First of all,  the quantity $e^{2\tilde{g}} +a^2 -1$ 
(that when vanishing yields the KS system) is non-vanishing,
  and hence we are always describing backgrounds that, after rotation, 
belong on the baryonic branch
  (they are hence more closely related to the backgrounds in~\cite{BGMPZ} 
than to the ones in~\cite{KS}). So, we find the first puzzle:
\begin{itemize}
\item{Puzzle: the UV expansion of the coefficient of this VEV (or $\tilde{g}$)
is proportional to $N_c/c_+$,
  while the coefficient of the dimension-8 deformation is 
proportional to $c_+^2/N_c^2$.
  Such a precise relation between two superficially 
independent coefficients demands an explanation.}
\end{itemize}
Remember that in the process of solving the master equation
we restricted ourselves to a subset of the possible solutions, by 
fine-tuning $Q_0=-N_c$ in order to avoid a 
nasty singularity --- see the discussion below Eq.~(\ref{Eq:master}).
This is the technical reason that makes the VEV and the coupling related.
But when such a kind of 
fine-tuning is needed on the gravity side of the correspondence 
in order to avoid a singularity, it is often the case that the 
fine-tuning has a clean explanation
in terms of the dual field theory, and this is the first thing we would like to understand.

\begin{figure}[h]
\begin{center}
\begin{picture}(500,150)
\put(50,0){\includegraphics[height=4.6cm]{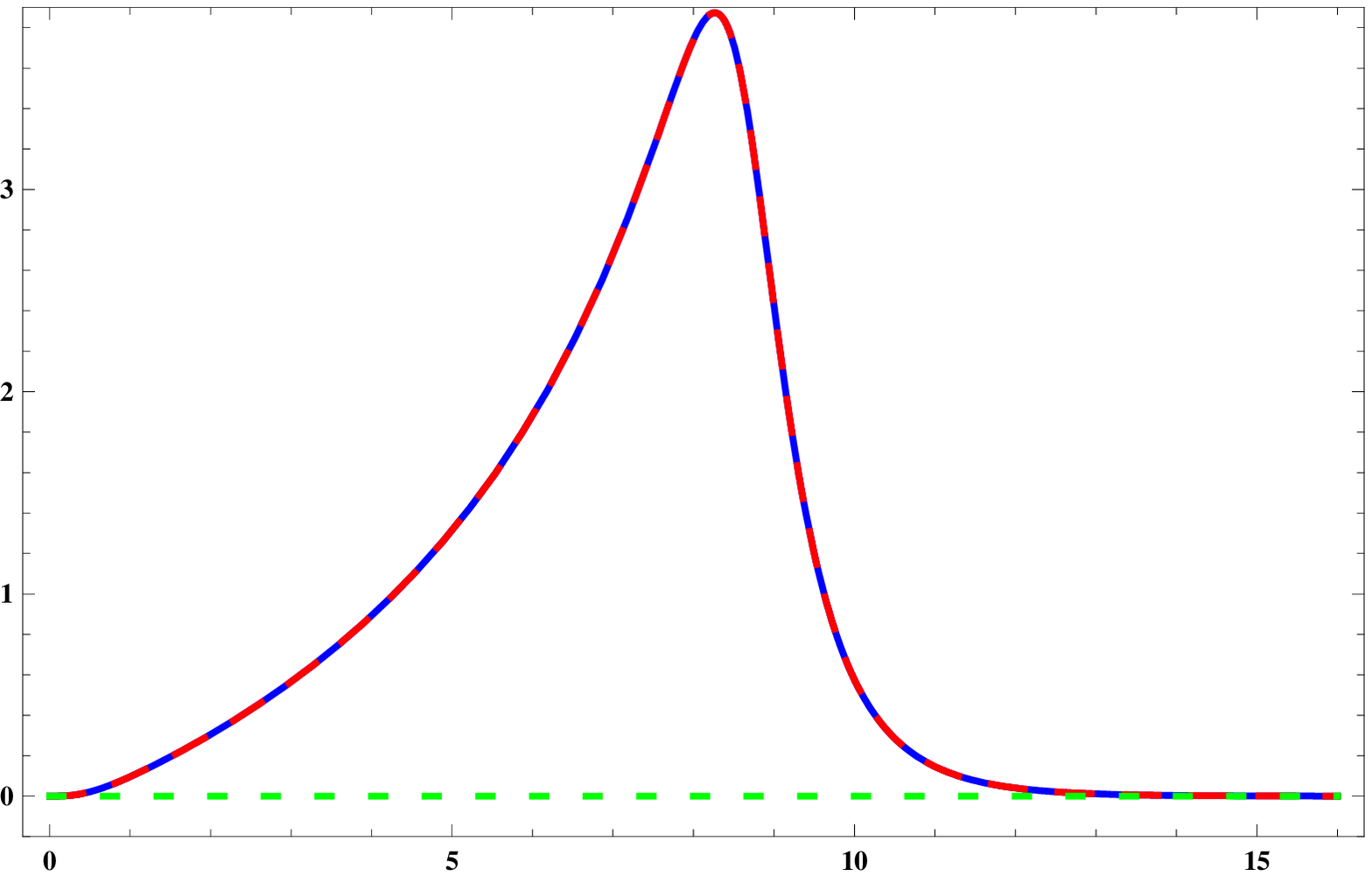}}
\put(300,0){\includegraphics[height=4.6cm]{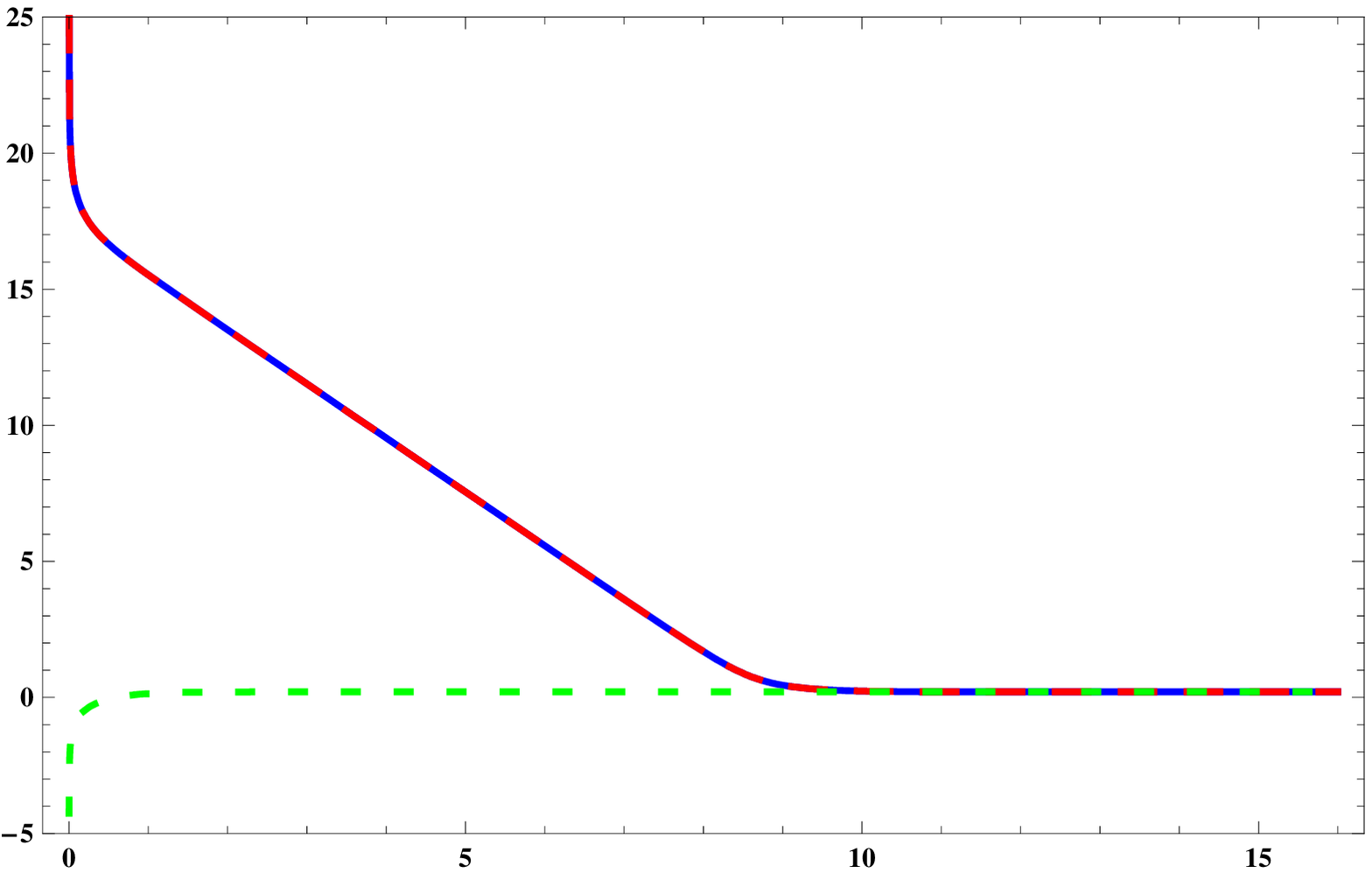}}
\put(1,120){$a^2+e^{2\tilde{g}}-1$}
\put(270,120){$x+3p$}
\put(475,0){$\r$}
\put(225,0){$\r$}
\end{picture} 
\caption{The (rotation-invariant) combinations $a^2+e^{2\tilde{g}}-1$ and $x+3p$,
as a function of the radial direction $\r$, for the same backgrounds as in Fig.~\ref{Fig:plotcompare}, with the same
color-coding.}
\label{Fig:invariants}
\end{center}
\end{figure}

The second puzzling fact has to do with the relation between
the behavior of three independent background functions, all of which are unaffected by the rotation.
We plot in Fig.~\ref{Fig:invariants} the two invariant quantities $a^2+e^{2\tilde{g}}-1$ and $x+3p$
for the same backgrounds as in Fig.~\ref{Fig:plotcompare}. 
First of all, $x+3p$ agrees with KS in the far UV, but differs for $\r<\r_{\ast}\simeq 9$.
This is simply the effect of the presence of  the dimension-6 VEV, which changes the IR, but not the UV dynamics.
\begin{itemize}
\item{Puzzle: the puzzle comes from the fact that on the 
solutions we are interested in,
the invariant combination $a^2+e^{2\tilde{g}}-1$ 
assumes a non-trivial profile at the same scale $\r_{\ast}$
at which the function $x+3p$ sensitive to the
dimension-6 VEV is taking over the dynamics.

}
\end{itemize}
While this could probably be explained in terms of the very 
non-trivial behavior of the RG evolution in the field theory
language (in the gravity language, the fact that the  BPS equations for the background scalars are coupled),
what is surprising is that below the scale $\r_{\ast}$ this combination is suppressed,
and  vanishes (exactly) at the end-of-space in the IR.

The third rotation-invariant quantity we referred to is the dilaton. 
The coefficient of the $z^4$ term in the UV expansion of the dilaton depends only on $c_+/N_c$,
and yet the dilaton profile changes significantly at the scale $\r_{\ast}$ controlled by $c_-$.
Deep in the IR the dilaton becomes again practically constant (see Fig.~\ref{Fig:plotcompare}).
All of this in spite of the fact that $c_-$ appears nowhere in the UV expansion of the dilaton itself.
Again, this might just be the effect of operator mixing. And yet, it demands a more precise explanation.

Probably connected to the second puzzle, we make an observation 
that anticipates one of the results of the next section.
In the presence of $c_-\neq 0$, the background is singular. 
This might suggest that what we are doing by turning on $c_-$ is not allowed in the dual gauge theory:
after all, we are tampering with a dimension-6 VEV, and hence the vacuum structure, without  changing the couplings (dynamics),
and it is hence not surprising that we run into troubles in the deep IR.
But this is too simplistic an explanation: 
the singularity we obtain is surprisingly mild, yielding finite Ricci scalar $R$ and $R_{\mu\nu}R^{\mu\nu}$.
Only the invariant $R_{\mu\nu\rho\sigma}R^{\mu\nu\rho\sigma}$ shows the singularity.
In the presence of various matter fields, 
`mild' singularities like this are not typical.
As a result of this, many physical low-energy 
quantities can (and will, in the next sections) be computed without 
obvious obstacles, contrary to what is expected in the presence of a singular background.
This suggests that the singularity is resolvable, probably by relaxing the constraints yielding
the PT system, and considering a more general truncation such as those in~\cite{CF},
\cite{BGGHO}.
We postpone the (non-trivial) question of how to resolve the IR singularity in the presence of $c_-$ 
to a future study,  in the hope that this would shed some light also on the other aspects of this puzzle,
and go back to the first of the two puzzles, taking the attitude that these solutions, although singular,
admit a sensible field-theory interpretation.

The rotated solution with 
$k_2$ fine-tuned to its 
maximal value $k_2 e^{\Phi_{\infty}}=1$ automatically enforces the constraint $\tilde{M}=0$.
We can summarize the second puzzle by saying that there is some non-trivial 
relation between the dimension-2 and dimension-6 VEV, that demands an explanation
in the context of the field theory based on the $SU(M)\times SU(M+N)$ quiver.
We will devote the following subsection to making more precise all the
elements of this puzzle, while postponing its resolution to a 
dedicated field theory study, in which the precise role of dimension-2 and dimension-6 VEVs will be 
studied in detail, in the context of more general classes of solutions to the master equation for $P$ than those addressed in this paper.

\subsubsection{Higher-order operators\label{Sec:HOO}}

We want to understand why the coefficients of the UV expansion of the dimension-2 VEV
and of the coupling of the dimension-8 operator for the solutions to the wrapped-D5 system are not independent,
and why by 
fine-tuning the parameter $k_2$ (which controls 
the extended gauge symmetry of the dual of the KS system 
with respect to the dual of the wrapped-D5 system) 
one ends up switching off the dimension-8 operator, without affecting the VEV.
What kind of field theories do we know of, in which the coefficients of a set of higher-order operators are
precisely related to the value of a VEV? Two  examples are the chiral Lagrangian of QCD
and the electro-weak effective action obtained  by integrating out the heavy gauge bosons from the Standard Model.
Let us digress and remind the reader about the basic properties of the latter.

Suppose that one wants to compute the amplitude of a given
flavor-changing neutral current process
involving hadrons (i.e. one or more of the five lightest quarks), within the Standard Model.
In principle what one could do is simply to compute all the relevant Feynman diagrams
at some order in perturbation theory. But this is not a good idea, for two reasons. First of all,
because one needs to compute also the relevant matrix elements of hadrons, for currents built out of quarks,
and this is a strongly-coupled problem that requires input from the lattice (or from some other non-perturbative tool). 
But even more importantly, because the diagramatics would become far too difficult,
due to the fact that even perturbative-QCD
effects are large enough that they must be included, often at the next-to-next-to-leading-order level.
In particular, perturbation theory does not like calculations that involve largely separated mass scales
(such as the masses of the $W$-boson and of the $b$-quark), because of potentially large logarithms appearing from the
brute-force evaluation of the loop diagrams,
and this requires to RG-improve the perturbative calculations.

A systematic and organized way of proceeding exists, and yields sensible results
in many phenomenologically relevant applications (see for example~\cite{Buras} for a pedagogical 
review on the subject).
The basic idea is to break down the calculation in three stages.
First one uses perturbation-theory methods to compute the relevant amplitudes 
in the original theory (the Standard Model with gauge group $SU(3)\times SU(2)\times U(1)$),
up to some loop order. 
This intrinsically assumes that all the couplings are small, and that it makes sense to compute
in terms of quarks and gluons, which is the case  provided this is done
at  the electro-weak scale.
Then one uses these amplitudes to match (at the electro-weak scale) 
onto the coefficients of an effective theory, which is obtained by suppressing the heavy degrees of freedom ($W$ and $Z$ gauge bosons,
top quark and possible Higgs fields) and writing an effective Lagrangian with the unbroken $SU(3)\times U(1)$,
and which contains a complete basis of higher-order operators involving the light degrees of freedom (quarks, leptons,
photon and gluons), in which the coefficients are chosen in such a way as to yield the same amplitudes.
The second stage consists of using the RG equations of the effective theory in order to evolve the coefficients
 from the electro-weak scale down to some relevant physical scale (the mass of the $B$ meson, for example).
Finally, one uses the input from the lattice, computing the matrix elements at the same low-energy scale
(and within the same renormalization scheme), and finally obtain the phenomenologically relevant
amplitudes to be compared to the data (which are scheme and scale independent).

So much for this digression.
The point is that the effective Lagrangian used in the second stage of this procedure is a generalization and refinement 
of the Fermi theory, supplemented by the interactions of the unbroken gauge group.
At leading-order, the coefficients of the higher-dimensional operators 
are precisely related to the VEV in the original theory  $v_W$ (the Standard Model Higgs VEV
responsible for electro-weak symmetry breaking) by the 
Fermi constant  $G_F$ as in
\beqs
\frac{G_F}{\sqrt{2}}&=&\frac{g_W^2}{8M_W^2}\,=\,\frac{1}{2v_W^2}\,,
\eeqs
where $M_W$ is the mass of the $W$ gauge bosons, and $g_W$ the $SU(2)$ gauge coupling.
Deep in the IR, this theory is  equivalent to the Standard Model, up to some finite order in the perturbative expansion.
However, on the one hand adopting the effective field theory language
  makes it easier (in practice) to keep into account precisely the dynamics of the unbroken gauge group.
On the other hand, the effective theory
 contains higher-order operators, and if one were to evolve its RG equations from the electro-weak scale towards the UV
(rather than towards the IR, as one should) one would run into big troubles.
While more manageable when dealing with phenomenological questions, this effective theory is not UV complete,
it is valid (and useful) only up to a UV cutoff, set by the masses of the heavy states that have been integrated out in the process of constructing it.

Let us now close the digression and get back to our problem.
The analogies should be evident. In the wrapped-D5 system we have a VEV
and a higher-dimensional operator which are strongly correlated. The dual field theory
has gauge group $SU(N_c)$, but is UV incomplete, because of the higher-order operator,
which takes over the dynamics above the scale set by $\r_{\ast}$.
The process of rotating for a generic value of $k_2$ changes the dual gauge theory
into some quiver theory, which has a larger gauge group, more degrees of freedom and more couplings. 
The presence of more gauge 
degrees of freedom translates into the fact that, because the deep IR is very similar,
the coefficient of the higher-order operator is modified as a function of $k_2$.
However, there exists a special value of $k_2$ (and hence a special choice of dual quiver 
gauge theory) for which this process ends up switching off the higher-dimensional operator,
making the theory healthier when extrapolated towards the far UV.

Hence our proposal for the interpretation of what is going on.
\begin{itemize}
\item{
The UV-complete dual field theory of the backgrounds obtained by rotating with a fine-tuned value of $k_2$
is a quiver, similar to the one of Klebanov-Strassler.
This theory undergoes a chain of  Seiberg dualities  (the cascade), as in KS. However,
 the cascade does not proceed all the way down to its latest stages. 
 Rather, a non-trivial VEV Higgses the gauge symmetry.
 The VEV interrupts the cascade at a stage that is controlled by $\r_{\ast}$.
In the process, most of the vector and chiral multiplets acquire a mass and decouple.
The wrapped-D5 system provides the gravity dual of the effective field theory description 
valid after integrating out these degrees of freedom, below the scale fixed by $\r_{\ast}$.
Ultimately, the unbroken $SU(N_c)$ gauge group leads to confinement and to the formation of the gaugino condensate.}
\end{itemize}

One can integrate out the heavy degrees of freedom from the quiver gauge theory, and 
in this way obtain a new gauge theory in which the field content is the one of the dual description of the wrapped-D5
system. However, the result is an effective theory which contains  higher-order operators,
with coefficients determined by the symmetry-breaking VEV.
The gravity dual of this is the wrapped-D5 (unrotated) background.
It yields (almost) the same physics in the deep IR as the original quiver, but it is now UV incomplete.
Rotated backgrounds with generic values of $k_2$ correspond to partial UV completions,
in which an incorrect number of degrees of freedom has been added (the gauge group is not 
large enough), and hence the dimension-8 operator cannot be completely removed.
But notice that in doing so one keeps the dimension-2 and dimension-6 VEV unchanged,
which is reflected in the fact that (in the five-dimensional language) 
$a^2+e^{2\tilde{g}}-1$ and $x+3p$ are unaffected by the rotation.

In short, what we are suggesting is that the UV completion of the backgrounds 
obtained in the wrapped-D5 system (and having UV asymptotics with $P\sim e^{4\r/3}$)
can be constructed by rotating according to~\cite{MM,GMNP} and 
fine-tuning the rotation parameter $k_2$. This
yields the dual of a quiver (in our particular case, the theory is in
 the Higgs phase).
Hence the rotated and unrotated backgrounds are not dual to 
two different unrelated field theories,
but to two theories one of which is the low-energy effective description of the other.

A final cautionary remark, mostly technical in nature.
By close inspection of the five-dimensional scalars that enter the metric ($a$, $\tilde{g}$,
$x$ and $p$), one can notice that the geometry in the IR is not exactly the same,
before and after the rotation, and hence one might question whether our interpretation 
really holds.
However, by looking more carefully at all the background functions, it turns out 
 the only effect of the rotation in the metric deep in the IR can be reabsorbed by rescaling
the four-dimensional coordinates, and by rescaling at the same time $\alpha^{\prime} g_s$ (or $N_c$, as done in~\cite{GMNP}).
I.e., this is just the effect 
of an inaccurate matching at the cutoff scale, which can be trivially fixed.
The only substantial difference is that the functions $h_1, h_2$ 
that enter in the  background values of the fields $B_2$ and $F_5$ are non-trivial
 below the scale $\r_{\ast}$. We will come back to this later on, but we anticipate here that 
 this is  not a reason for concern, because it turns out that both $F_5$ and $H_3$ are strongly suppressed below $\r_{\ast}$.

\subsubsection{Perturbative results: a summary}

In this subsection we collect from the literature a set of perturbative results
that are relevant in order to provide a complete field theory interpretation of 
the backgrounds we discussed.
We  start this discussion by reminding
the reader of the so-called {\it planar equivalences}. 
Since 1998, it has been proposed
that orbifold and orientifold projections of {\it parent}
supersymmetric field theories (for example
${\cal N}=4$ SYM, ${\cal N}=1$ SYM, et cetera, with gauge group $SU(N_c)$) 
to less symmetric {\it daughter} theories
shared the same planar diagrams in the large $N_c$ limit.
This implies that the perturbative expansions are coincident.
While at this level the equivalence is kinematical, the 
non-perturbative nature of the equivalence was suggested to be valid.
This equivalence relates the connected correlation functions and VEVs 
of corresponding {\it neutral} operators\footnote{By {\it neutral}
we refer to operators  in the parent theory that are gauge invariant, 
single trace
and invariant under the discrete symmetries that define the projection.} 
in both theories. In order for this planar equivalence to be valid 
it is needed that  the discrete symmetries  that define the projection
do not undergo spontaneous symmetry breaking.
For a summary of this line of research, see the papers \cite{planar}.

In this paper we will only {\it suggest} a planar 
equivalence between the
${\cal N}=1$ single node $SU(N_c)$ supersymmetric theory with an infinite tower of massive 
excitations (that arise when compactifying with the D5-branes), the ${\cal N}=1^*$ SYM theory around a particular Higgs vacuum
and the two-node KS quiver with bifundamentals.\footnote{Something like this planar equivalence may exist, but
it must be  slightly different from the ones studied in
\cite{planar} and references therein.} Nevertheless,  sometimes we may use the language developed in flows
of ${\cal N}=1^*$ SYM (with 
three chiral multiplets $\Phi_i,i=1,2,3$) to refer to the KS quiver
with bifundamentals $A_\alpha, B_\beta$, with $\alpha,\beta=1,2$.
In the following, we will  give some details on key results
that highlight this connection.

In Section~\ref{Sec:HOO} 
of this paper, we
interpreted the rotation procedure
as a  `conspiration' between a
quasi-marginal coupling and a dimension-2 operator getting a VEV,
so that an 
irrelevant operator $O_8$
(that without this tuning would be present and driving the UV dynamics)
is actually not 
present. This proposal was exemplified by what 
happens in the Standard 
Model and its low-energy effective  field 
theory, the Fermi theory, both of them being weakly coupled.
But we would like to emphasize that our 
proposal in section~\ref{Sec:HOO}
is for some dynamics that
takes place at strong coupling.

A weakly coupled version of the connection we proposed was presented by 
Maldacena and Martelli in
\cite{MM}.
Interestingly, their proposal goes from the KS quiver field theory 
into the one node with adjoints field theory, 
while ours (just like the rotation on the string/gravity side) 
goes from the one-node QFT (with the irrelevant inserted $O_8$)
into the quiver, that acts as the correct UV completion and decouples
this $O_8$. 

The authors of \cite{MM}
consider the KS quiver and study its 
{\it perturbative}
dynamics (by this, we mean that the K\"ahler potential is trivial, 
hence no gravity background can be a good approximation to the dynamics).
We need to solve the D-term equations that read
(see also~\cite{Dymarsky:2005xt})
\bea
& & \sum_\alpha A_\alpha A_\alpha^{\dagger}
- \sum_{\beta} B_\beta^\dagger B_\beta=
\frac{U}{M}1_{M},\nonumber\\
& & \sum_\alpha A_\alpha^\dagger A_\alpha- \sum_{\beta}
 B_\beta B_\beta^\dagger= 
\frac{U}{M+N_c}1_{(N_c+M)},\nonumber\\
& & U= Tr(\sum_\alpha A_\alpha A_\alpha^{\dagger}
- \sum_{\beta} B_\beta^\dagger B_\beta),
\label{dtermeqxx}\eea
where we have used that the quiver is $SU(M)\times SU(M+N_c)$ and that
$A_\alpha, B_\beta$ transform as bifundamentals in each group
($\alpha,\beta=1,2$).
It was shown in~\cite{Dymarsky:2005xt} that there are two types of solutions to these equations. 
Those where both $A_\alpha, B_\beta$ are nonzero and 
that correspond to mesonic branches
(where ${\cal M}\sim AB$) and those in which either $A=0$ or $B=0$
that correspond to baryonic branches and that arise only if $M=q N_c$ (where $q$ is an integer).
The authors of \cite{Dymarsky:2005xt} complement their analysis with the 
non-perturbative induced superpotentials, solve also the F-term equations
 and point
out how the moduli space changes from the classical solution (the conifold)
into non-singular deformed conifolds. 

Coming back to the perturbative analysis in the paper \cite{MM},
the authors proceed by expanding around a 
particular ({\it perturbative})
baryonic solution---presented
in section 4.2 of  \cite{Dymarsky:2005xt},
and find that the gauge group is Higgsed from 
$SU(M+N_c)\times SU(M)\to SU(N_c)$ 
(we emphasize, 
with $M=q N_c$, otherwise such a perturbative
baryonic solution does not exist). Also, they 
obtain that the {\it perturbative} mass spectrum of this quiver
in this particular vacuum is nearly coincident with the
{\it perturbative} mass spectrum found by the authors of 
\cite{AD}, that we now revisit.

Andrews and Dorey~\cite{AD} studied
the KK (with twisting) decomposition of the six-dimensional 
field theory with sixteen supercharges and gauge group $SU(N_c)$, 
that is the theory living on $N_c$ D5-branes that wrap a 
holomorphic two-cycle in the resolved conifold. 
After a careful analysis, they
obtained a spectrum for a four-supercharge 
$SU(N_c)$ field theory that consists
of a massless vector multiplet and a  tower
of massive vector and chiral multiplets. Degeneracies 
and masses at each level are given in \cite{AD}. 
Again, we stress that this is a classical calculation.
In the same work, the authors of \cite{AD} studied
the F-flatness condition coming from the ${\cal N}=1^*$ SYM superpotential
\beq
{\cal W}(\Phi_i)= Tr\Big[i\sqrt{2} \Phi_1[\Phi_2,\Phi_3] +\eta(\Phi_1^2 +
\Phi_2^2+ \Phi_3^2)   \Big]\to 
[\Phi_i,\Phi_j]=i\sqrt{2}\eta \epsilon_{ijk}\Phi_k,
\eeq
that after a rescaling of the fields $\Phi_i$ leads to the $SU(2)$ algebra.
The solutions to the equations are any representation of $SU(2)$.
Expanding around the Higgs vacuum $\Phi_i=J_i^{(N_c)}$ 
that breaks $U(N_c)\to U(1)^{N_c}$, 
they find that the 
vacuum defines a fuzzy sphere (and, in the limit $N_c\to \infty$, a sphere).
Carefully expanding the Lagrangian for ${\cal N}=1^*$ and 
keeping only quadratic terms around the vacua 
$\Phi_i= J_i^{(N)} +\delta\Phi_i$ 
they also find the mass spectrum, that  matches 
(for finite value of $N_c$ 
with a  truncated version of)
the one of  the compactified D5-brane theory discussed above. Also, they showed that at leading order in the fluctuations, the Lagrangians match.
In other words, they are showing how the four-dimensional ${\cal N}=1^*$ SYM
theory deconstructs the six-dimensional theory on the five-branes.

For future reference, we note that if the choice of Higgs vacuum is
\beq
\Phi_i= 1_{N_c} \times J_i^{(q+1)}
\eeq
then the gauge symmetry is broken according to
\beq
U((q+1)N_c)\to U(N_c)
\eeq
and the mass spectrum of the weakly coupled ${\cal N}=1^*$ SYM contains a tower of massive chirals and massive vectors (aside from the massless
vector multiplet). The heaviest state is a vector multiplet with mass
\beq
M^2= \eta^2 q(q+1)
\eeq 
followed by a  chiral of mass $M^2=\eta^2 (q+1)^2$. 
The masses and degeneracies for vector and chiral multiplets
are
\bea
& & M_v^2 =\eta^2k(k+1),\;\;\; deg=(2k+1)N_c^2,\nonumber\\
& & M_{ch}^2=\eta^2 k^2,\;\;\; deg= 4k N_c^2,\;\;\; k:1,...,q\,.
\label{massesdeg}
\eea

In summary, Higgsing the ${\cal N}=1^*$ SYM theory around one of its classical vacua
exactly reproduces the truncated perturbative mass spectrum of the 
compactified theory on the D5-branes.
Higgsing the KS quiver around one of its perturbative baryonic solutions roughly
reproduces the perturbative spectrum 
of the theory on the compactified D5-branes.
The coincidences are notable. The three theories are linked and this suggests
a relation between ${\cal N}=1^*$ and the KS quiver, perhaps along the lines of
\cite{Hollowood:2004ek}  (it would be nice to realize this in string theory).
We emphasize that both in our strongly coupled version 
and in Maldacena-Martelli \cite{MM}
weakly coupled connection, 
the Higgsing  plays a central role. It is indeed the way of connecting
a quiver theory with a single node theory.
It should be interesting to make more formal the idea of a 
possible planar equivalence between these three theories.

We would like to make a brief comment about the phenomenon of Higgsing
in these backgrounds. In the paper \cite{Aharony:2000pp}, Aharony
proposed that, aside from a sequence of Seiberg dualities, 
the Higgs mechanism 
could be the reason why the decrease in ranks of the KS cascade takes place.
It was argued that at every position where one usually performs a 
Seiberg duality, there is a source that, once crossed, Higgses the gauge groups.
This proposal found a clean realization in the solutions with 
sources (flavor branes) 
of \cite{GMNP}, where one can see 
that the warp factor is the superposition of both phenomena 
(the cascade and the
Higgsing).
Here, we are proposing that even in the absence of sources, the 
Higgsing interpretation may be adequate. Indeed, the 
equivalence between the two pictures (Seiberg duality and Higgsing)
was argued in more generality in 
\cite{Hollowood:2004ek}.

\subsubsection{About the vacuum structure of the dual theory}

We conclude the field theory analysis 
by discussing the physics connected with the second puzzle we highlighted earlier on,
in particular with the roles played by the dimension-2 and dimension-6 VEVs.
Because we can think of the process of rotating and fine-tuning 
(the choice of $k_2$)
as yielding the UV completion of the dual to the wrapped-D5 system, 
and hence as a way of describing in different terms the same long-distance physics,
we will here concentrate on the rotated and fine-tuned  solutions, 
the results extending 
to the whole class under consideration. 

We start by listing some important  properties of the backgrounds we studied in this paper.
\begin{itemize}

\item The solution is controlled by the coefficients $c_+$ and $c_-$ appearing in the UV expansion, which correspond to
dimension-2 and dimension-6 VEVs in the field theory.

\item The physical meaning of the freedom we have in choosing backgrounds with different $\r_{\ast}$
is connected with the scale at which the cascade stops and the gauge group is Higgsed, and with the parameter $q$ controlling the breaking $SU(q N_c+ N_c)\times SU(q N_c)\rightarrow SU(N_c)$.

\item The rotated and fine-tuned solutions always have $\tilde{M}=0$.

\item There is a non-trivial correlation between the behavior of $\Phi$, $a^2+e^{2\tilde{g}}-1$ 
and $x+3p$ at and below
the scale $\r_{\ast}$, but not above it (as suggested by the UV expansions). Also, the combination
$a^2+e^{2\tilde{g}}-1$ vanishes at the end-of-space in the IR.

\item There are physical, measurable differences between backgrounds in which both dimension-2 and dimension-6 VEVs  are present (the seed solution and the rotated one in this paper),
the backgrounds in which the dimension-6 VEV is absent (such as Butti et al.~\cite{BGMPZ}), and the KS background, in which both VEVs vanish. We will explicitly show this fact later on, by computing the
expectation value of the rectangular Wilson loop.

\item The singularity in the IR does not seem to be
 a reason for major concern, as it 
is not preventing us from a consistent field theory interpretation,
but rather the calculation of physical quantities seems to proceed unaffected by it.
For this reason, in this paper we took the pragmatic approach of analyzing the background 
in field-thery terms, assuming that the singularity is resolvable, and that the possible
resolution does not affect the observables we are interested in.

\item  There seems to be an emerging general picture, in field theory terms, 
explaining what the relation is between the rotated backgrounds, their relatives within the PT ansatz,
and the deformations of ${\cal N}=4$ super-Yang-Mills. This picture is, for the time being, based on
circumstantial evidence and
striking analogies at the perturbative level, elements of which appear to manifest themselves also in the gravity dual at the non-perturbative level.

\end{itemize}

Ultimately, we would like to understand if there is a comprehensive field theory picture that 
explains all of the above. This requires conducting a more systematic study of the dual field theory,
which we postpone to a future study. 
Such a study requires including also solutions of the master equation that we did not 
include in the present paper, such as those in which $P\simeq 2 N_c \r$ (as in~\cite{MN}), 
at least for some range of $\r$ (as in~\cite{ENP,NPR}
for example).
In doing so, we should be able to ask whether the physics associated with the dimension-3, dimension-2 and dimension-6 
VEVs which control the non-trivial features of these backgrounds
can be interpreted as a genuinely multi-scale dynamical model.

We conclude with a geometric observation, possibly connected 
with the roles of the dimension-2 VEV 
represented by $(a^2+e^{2\tilde{g}}-1)$ and the dimension-6 VEV in $(x+3p)$.
Let us take the (string-frame) metric of the PT system. 
Consider now a pair of three-cycles $\Sigma_3=[\theta,\varphi,\psi]$ and 
$\tilde{\Sigma}_3=[\tilde{\theta},\tilde{\varphi},\psi]$  
in the internal manifold. 
The resulting induced metrics on each of the cycles $\Sigma_3, \tilde{\Sigma}_3$ are
\beqs
\Sigma_3:~\di s^2_3&=&e^{{\Phi}/2+x-\tilde{g}} \left(a^2+e^{2 \tilde{g}}\right)\left(\frac{e^{-6 p-2x+\tilde{g}}}{\left(a^2+e^{2 \tilde{g}}\right)} ({\di \psi}+{\di \phi} \cos \theta
   )^2+ 
   \left({\di \theta}^2+{\di \phi}^2 \sin ^2\theta \right)\right)\,,\\
  \tilde{\Sigma}_3:~ \di s^2_{\tilde{3}}&=&
   e^{{\Phi}/2+x-\tilde{g}} \left(e^{-6 p-2x+\tilde{g}} ({\di \psi}+{\di \tilde{\phi}} \cos
   \tilde{\theta})^2+
   \left({\di \tilde{\theta}}^2+{\di \tilde{\phi}}^2 \sin
   ^2\tilde{\theta}\right)\right)\,.
   \eeqs
Both are proportional to the metric on the squashed three-sphere, which can be written in terms of the 
three angles $0\leq\theta< \pi$,  $0\leq \phi <2\pi$ and $0\leq \psi <4\pi$ 
as
\beqs
\di s^2&=&\di \theta^2+\sin^2\theta \di \phi^2 + \alpha \left(\di \psi +\cos\theta \di \phi\right)^2\,,
\eeqs
and which reproduces the metric on the 3-sphere for $\alpha=1$.

Notice that in the KS system, 
in which the VEVs of baryon and antibaryon operators coincide,
one has $a^2+e^{2 \tilde{g}}=1$, so that the two three-dimensional surfaces have the same geometry.
This is an effect of the global $Z_2$ symmetry of the KS system.

Let us focus now on the cycle $\tilde{\Sigma}_3$ or $\di s^2_{\tilde{3}}$. 
Expressed in terms of $P$ and $Q$, the squashing factor becomes
\beqs
\alpha_{{2}}\,=\,e^{-6p-2x+2\tilde{g}}&=&\frac{\partial_{\r}P}{2(P\coth 2\r-Q)}\,,
\eeqs
with  $Q=N_c(2\r \coth2\r -1)$.

When $P\sim e^{\frac{4\r}{3}}$ (in the case of this paper, when $\r>\r_{\ast}$), then $\alpha_2\simeq \frac{2}{3}$, which 
is the familiar geometric factor characterizing $T^{1,1}$. This is not a surprising result.
More interesting is the fact that when $P\simeq c_0$---as in our background in the deep IR---
then $\alpha_2\sim e^{4(\r-\r_{\ast})}$ is exponentially suppressed,
and one obtains that $\di s^2_{\tilde{3}}$ is a three-sphere which becomes 
extremely squashed near the origin of the space.
By comparison, in the linear-dilaton solution of~\cite{MN} one has $P=2N_c \r$, and hence $\alpha_2=1$,
so that  $\di s^2_{\tilde{3}}$ is exactly a three-sphere.

The Wilson-loop expectation value was calculated in~\cite{NPR} focusing on backgrounds in which
$P\sim c_0$ for $\r<\bar{\r}$, and $P\simeq 2 N_c \r$ for $\r>\bar{\r}$. A first-order phase transition 
was shown to appear when $\bar{\r}>\r_I$. As we will see, this behavior is present also in the case
discussed in this paper, suggesting that the phase transition has to do with the squashing of the sphere, and ultimately with the dimension-6 VEV.
Again, a dedicated study of the field theory, and a systematic comparison among all possible classes of regular solutions to the master equation
is necessary, in order to elucidate this point.

We will now study a couple of quantities that will reinforce 
the field theory interpretation put forward above.

\subsection{Central charge, Maxwell charge and the scale $\r_{\ast}$\label{Sec:other}}
In this subsection we perform a study of a set of non-trivial field-theory
quantities that can be computed in the rotated background.
The main purpose of this subsection is to illustrate the difference between the rotated
backgrounds in the ranges $\r<\r_{\ast}$ and $\r>\r_{\ast}$.
In doing so, we are going to test the proposal for the field-theory interpretation 
of the previous subsection.
In particular, we will show that the dual theory below $\r_{\ast}$ has a smaller number of effective degrees of freedom,
and a smaller gauge group, compared to what is expected in the KS case or along the baryonic branch of KS described in~\cite{BGMPZ}.

\subsubsection{Central charge}

\begin{figure}[htpb]
\begin{center}
\begin{picture}(470,160)
\put(0,0){\includegraphics[height=140pt]{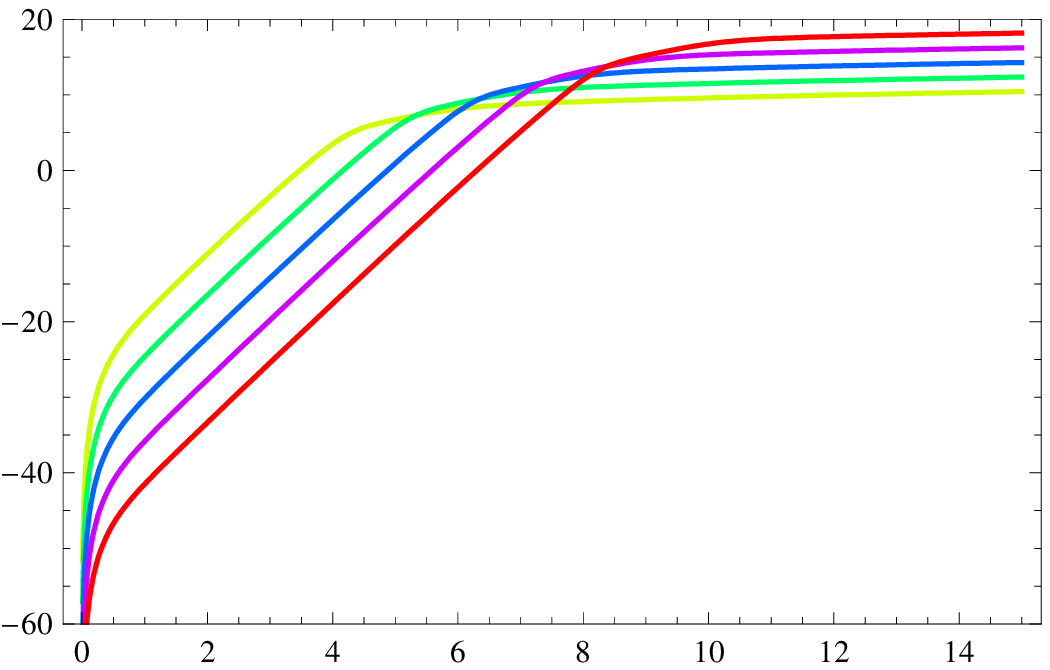}}
\put(240,0){\includegraphics[height=140pt]{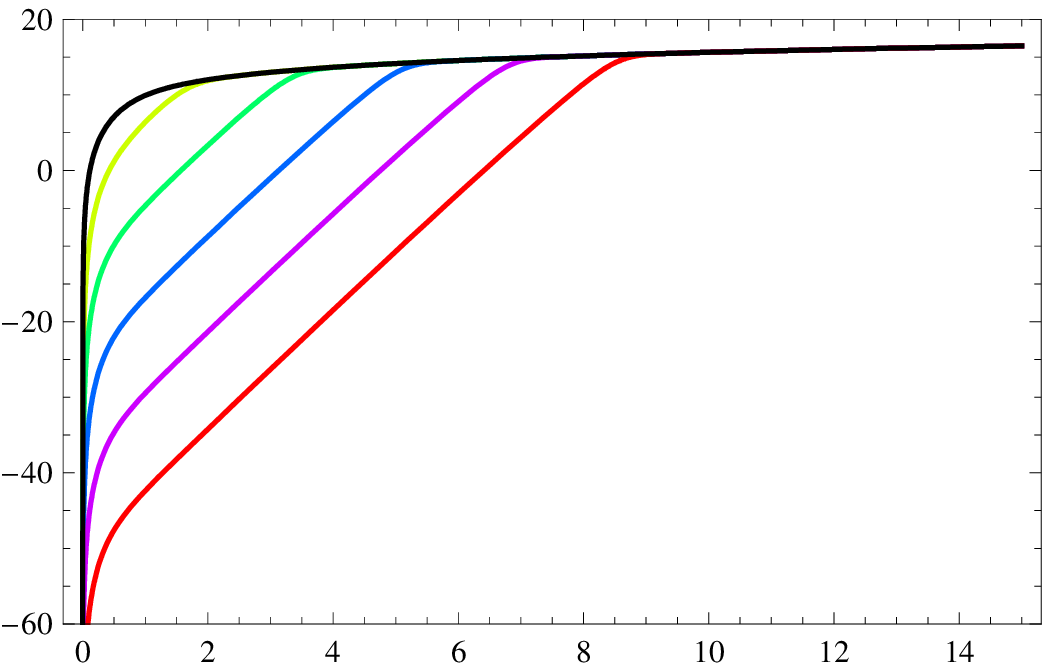}}
\put(5,145){$\log c$}
\put(245,145){$\log c$}
\put(222,7){$\rho$}
\put(462,7){$\rho$}
\end{picture}
\caption{The left panel shows the central charge as a function of the radial coordinate $\rho$ for a few of the rotated solutions. The right panel shows the same, but for deformations of Klebanov-Strassler given by different values of $f_0$ (the black line corresponds to $f_0 = 0$, i.e. the original solution of Klebanov-Strassler).}
\label{Fig:CentralCharge}
\end{center}
\end{figure}

The holographic central charge is given by \cite{CentralCharge}
\SP{
	c \sim \frac{1}{(\partial_r A)^3}.
}
Using Einstein's equations, one can show that this quantity is a monotonically decreasing function as one flows towards the IR. In terms of the ten-dimensional variables
\SP{
	c \sim \frac{(1-k_2 e^{2\Phi})^2 e^{2\Phi+2h+2g+4k}}{\Big( \partial_\rho \big[ 4\Phi + 4h +4g + 2k + \log (1-k_2 e^{2\Phi}) \big] \Big)^3}.
}
We plot this quantity for a few of the rotated solutions as well as deformations of 
Klebanov-Strassler in Figure~\ref{Fig:CentralCharge} (by `deformations of KS'
we mean solutions where the integration constant $f_0$ in  eq.(\ref{f0zzzxx}) 
is non-zero).

The presence of the scale $\r_{\ast}$ has, both in our rotated solutions and in the case of the singular modification of KS, 
the effect of drastically reducing the central charge at and below the scale $\r_{\ast}$. In field-theory terms, this means that 
a large number of degrees of freedom freeze below this scale, their dynamics becoming 
trivial and decoupled.

Whilst the central charge is only well defined at fixed points, we believe this analysis
points to the particular behavior we proposed. Let us now analyze another observable.

\subsubsection{Maxwell charge}

\begin{figure}[htpb]
\begin{center}
\begin{picture}(470,160)
\put(0,0){\includegraphics[height=140pt]{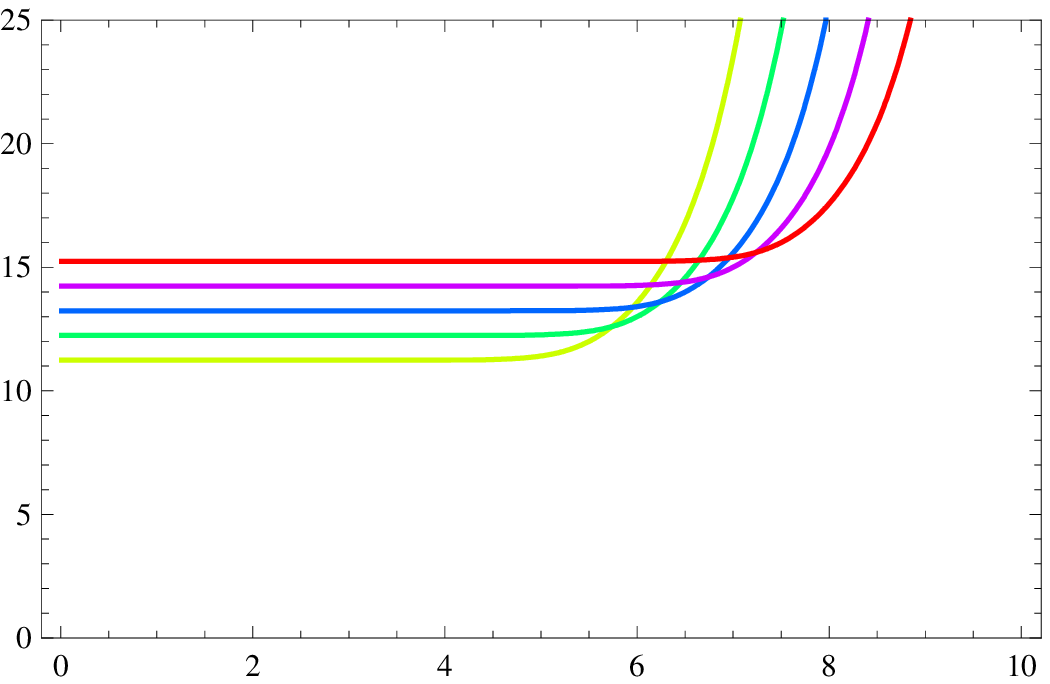}}
\put(240,0){\includegraphics[height=140pt]{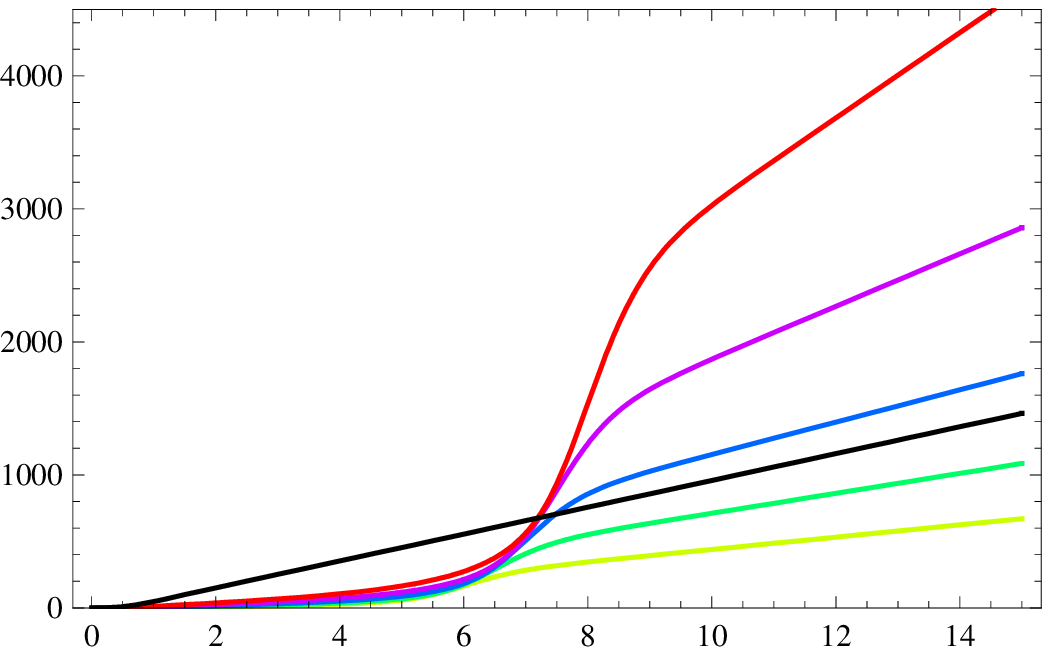}}
\put(5,145){$P$}
\put(245,145){$Q_{Maxwell, D3}$}
\put(218,8){$\rho$}
\put(470,8){$\rho$}
\end{picture}
\caption{The right panel shows the Maxwell charge $Q_{Maxwell, D3}$ for Klebanov-Strassler (black line), and a few different rotated solutions, the $P$ of which is shown in the left panel.}
\label{Fig:MaxwellCharge}
\end{center}
\end{figure}

We define a Maxwell charge 
associated with the D3-branes as 
\SP{
	Q_{Maxwell, D3} = \frac{1}{16\pi^4}\int_{\Sigma_5}F_5 = \frac{4}{\pi} {\cal K},
}
with the manifold $\Sigma_5 =[\theta,\varphi,\tilde{\theta},\tilde{\varphi},\psi]$.
We will use this Maxwell charge as an indicator 
of the `number of degrees of freedom' in the quiver field theory as originally suggested
in \cite{KS}.
Notice that in the five-dimensional language 
$Q_{Maxwell, D3}$ is nothing but the function ${\cal K}$ in Eq.~(\ref{changeafterrotationzz}).
For the rotated solutions this becomes equal to
\SP{
	Q_{Maxwell, D3} = \frac{k_2 e^{2(\Phi+g+h)} \partial_\rho \Phi}{\pi} = \frac{k_2 e^{2 \Phi_o}}{2 \pi} \frac{Q \left(N_c \cosh (2 \rho) -\sigma \right)}{\sqrt{Y
   \left(P^2-Q^2\right)}},
}
whereas for Klebanov-Strassler (and deformations corresponding to non-zero $f_0$, but $\tilde{M}=0$) it is equal to
\SP{
	Q_{Maxwell, D3} = -\frac{4}{\pi}\frac{e^{\Phi_{\infty}} N^2 (2 \rho  \coth (2 \rho )-1) (4 \rho -\sinh (4 \rho ))}{\sinh^2(2 \rho )}.
}
Note that the deformation parameter $f_0$ does not enter into this expression. In Figure~\ref{Fig:MaxwellCharge}, $Q_{Maxwell, D3}$ is plotted as a function of $\rho$ for a few of the rotated solutions, as well as for Klebanov-Strassler.

If one is to interpret this quantity in terms of diluted D3 
in the background, or equivalently as giving a rough estimation for
the rank of the
gauge group of the dual quiver theory,\footnote{The Page charge, as used in \cite{Benini:2007gx}, gives the same 
result as in the KS background
$Q_{Page, D3}=0$.} what this shows is the expected behavior of the cascade for $\r>\r_{\ast}$.
However, below $\r_{\ast}$ this rank suddenly drops virtually to zero. This supports the suggestion that the formation of the
condensates results in the Higgsing of the theory, in which the last steps of the duality cascade are replaced by the spontaneous breaking
$SU(M)\times SU(M+N)\rightarrow SU(N)$.
In the ten-dimensional language, the fact that ${\cal K}$ is 
strongly suppressed below $\r_{\ast}$
means in turn that $B_2$, $H_3=dB_2$ and $F_5$ are also suppressed 
compared to $F_3$. In this sense, 
approximating them with zero, and looking and the wrapped-D5 system instead, is another way
of thinking of the latter as an effective field theory. It contains many less degrees of freedom, and is hence much simpler and convenient,
while at the same time this is just a leading-order approximation, which yields quite accurate results below the 
cutoff connected with $\r_{\ast}$, but above which a completion is needed.

Concluding this short subsection: both the central charge and the Maxwell charge, computed for the rotated backgrounds
 behave in two different ways for $\r>\r_{\ast}$ and for $\r<\r_{\ast}$. In the former case, a slow evolution is compatible with
 the duality cascade. In the latter case, there is a sudden drop, with the rank of the gauge group of the dual theory and its  
number of effective degrees of freedom falling towards their minimal values.
Both these two derived quantities seem to support our interpretation of the background, according to which
$\r_{\ast}$ is the cutoff scale below which the dual theory is in the 
Higgsed phase.

We now move on to study IR aspects of our new backgrounds.
\section{Long-distance physics: IR asymptotics and Wilson loop\label{Sec:IR}}

This section is devoted to the long-distance physics of the solutions.
As we said, in the deep-IR region there is little difference between
the backgrounds 
that belong to the class of 
the wrapped-D5 system and their rotated counter-parts.
However, the presence of the dimension-6 VEV is going to result in rather important
differences compared to the KS solution~\cite{KS} 
and to the baryonic branch solutions in~\cite{BGMPZ}.
We focus the study on the rotated and fine-tuned solutions, for simplicity.

\subsection{IR asymptotics and curvature singularity\label{Sec:Curvature}}

The original `seed' solutions discussed in 
Section \ref{seedsectionxx} have singularities in the IR. 
This is also true after the rotation has been applied to 
them. Various criteria have been given in the literature, 
as to when an IR singularity is good (and it is believed 
that the supergravity background captures the relevant 
physics) or bad (in which case the 
supergravity description  breaks down and a resolution of the singularity in supergravity
or even the full 
string theory is needed). One such criterion is the one due to 
Gubser ~\cite{Gubser:2000nd}, which 
states that the five-dimensional potential $V$ --- see Eq.~(\ref{potentialzzz}) --- evaluated on a particular solution to the equations 
of motion, should be bounded 
from above in order for the singularity to be good. It is argued that 
this is a necessary condition for the existence of near-extremal 
generalizations of the background, in which the singularity is hidden 
behind the horizon of a black hole. 
Another criterion, given in~\cite{Maldacena:2000mw}, 
states in its strong form that the 
$g_{tt}$ component of the metric as function 
of the radial coordinate should not increase 
as one approaches the IR singularity. This is motivated by 
the interpretation of the radial coordinate 
as corresponding to the energy scale of the dual 
field theory: excitations in the bulk that are closer to 
the IR singularity should correspond to excitations of 
lower and lower energy as seen from the boundary.

In order to better understand the nature of the IR singularities for the rotated solutions, we will now study three invariant objects related to the curvature. These are the Ricci scalar $R$, the Ricci tensor squared $R_{\mu\nu}R^{\mu\nu}$, and the Riemann tensor squared $R_{\mu\nu\tau\sigma}R^{\mu\nu\tau\sigma}$. Using that $P$ can be expanded in the IR as
\SP{
	P = c_0+c_0 k_3 \rho ^3+\frac{4}{5} c_0 k_3 \rho ^5-c_0
   k_3^2 \rho ^6+\frac{16 k_3 \left(2
   c_0^2-5 N_c^2\right)}{105 c_0} \rho ^7 +\mathcal O \left(\rho
   ^8\right),
}
where $c_0$ and $k_3$ are integration constants, 
one finds that these three objects have IR expansions given by
\SP{
	R = \frac{16 \left(\frac{2}{3}\right)^{5/8}
   e^{-\frac{\Phi_o}{2}} N_c^2 \left(\sqrt{6}
   \sqrt{c_0^3 k_3}+8 k_2^2 e^{2 \Phi_o}\right)}{3
   c_0^{15/8} k_3^{5/8} \left(\sqrt{6} \sqrt{c_0^3
   k_3}-8 k_2^2 e^{2 \phi
   _0}\right){}^{3/2}}-\frac{128
   \left(\left(\frac{2}{3}\right)^{5/8}
   e^{-\frac{\Phi_o}{2}} N_c^2 \left(\sqrt{6}
   \sqrt{c_0^3 k_3}+8 k_2^2 e^{2 \phi
   _0}\right)\right)}{9 \left(c_0^{15/8} k_3^{5/8}
   \left(\sqrt{6} \sqrt{c_0^3 k_3}-8 k_2^2 e^{2 \phi
   _0}\right){}^{3/2}\right)} \rho^2 + \mathcal O \left(\rho ^4\right),
}
\SP{
	R_{\mu\nu}R^{\mu\nu} =& -\frac{512 \left(\sqrt[4]{\frac{2}{3}} e^{-\Phi_o}
   N_c^4 \left(8 \sqrt{6} k_2^2 e^{2 \Phi_o}
   \sqrt{c_0^3 k_3}-93 c_0^3 k_3-992 k_2^4 e^{4 \phi
   _0}\right)\right)}{81 \left(c_0^{15/4} k_3^{5/4}
   \left(\sqrt{6} \sqrt{c_0^3 k_3}-8 k_2^2 e^{2 \phi
   _0}\right){}^3\right)}+ \\& \frac{8192
   \sqrt[4]{\frac{2}{3}} e^{-\Phi_o} N_c^4
   \left(8 \sqrt{6} k_2^2 e^{2 \Phi_o} \sqrt{c_0^3
   k_3}-93 c_0^3 k_3-992 k_2^4 e^{4 \phi
   _0}\right)}{243 c_0^{15/4} k_3^{5/4}
   \left(\sqrt{6} \sqrt{c_0^3 k_3}-8 k_2^2 e^{2 \phi
   _0}\right){}^3} \rho^2 +\mathcal O\left(\rho ^4\right),
}
\SP{
	R_{\mu\nu\tau\sigma}R^{\mu\nu\tau\sigma} = \frac{160 \sqrt[4]{\frac{2}{3}} e^{-\Phi_o}
   \sqrt[4]{\frac{c_0}{k_3^5}}}{3
   \left(\sqrt{6} \sqrt{c_0^3 k_3}-8 k_2^2 e^{2 \phi
   _0}\right)} \rho^{-8} +\frac{1024 \sqrt[4]{\frac{2}{3}}
   e^{-\Phi_o} \sqrt[4]{\frac{c_0}{k_3^5}}}{\left(72 k_2^2 e^{2 \Phi_o}-9 \sqrt{6}
   \sqrt{c_0^3 k_3}\right)} \rho^{-6}+ \mathcal O\left(\rho
   ^{-5}\right).
}

\begin{figure}[htpb]
\begin{center}
\begin{picture}(470,320)
\put(0,160){\includegraphics[height=140pt]{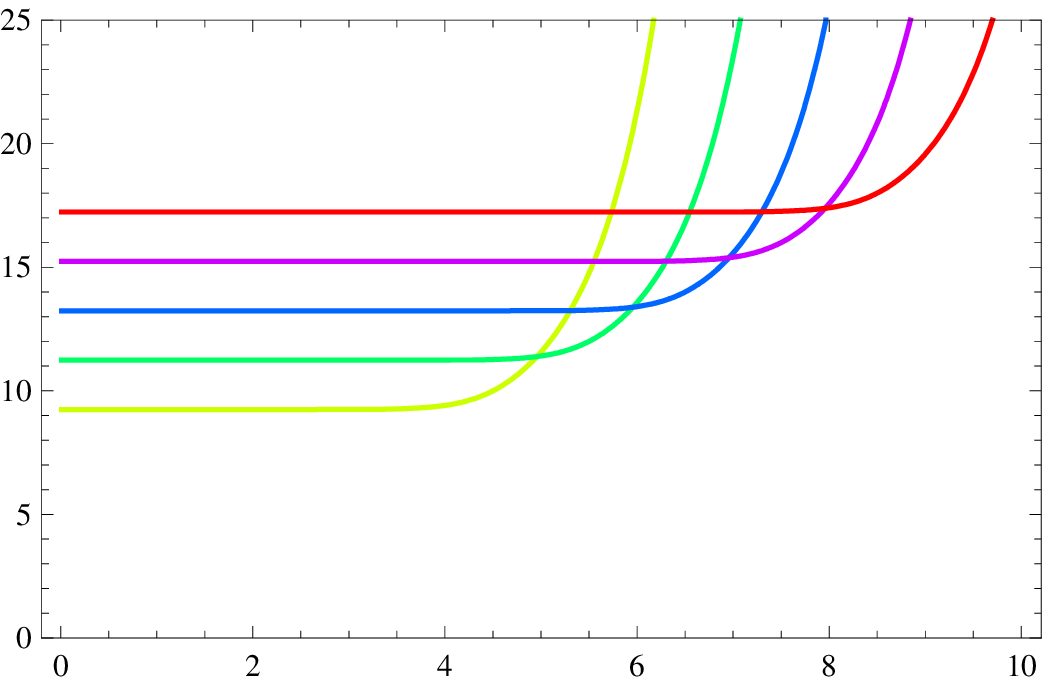}}
\put(240,160){\includegraphics[height=140pt]{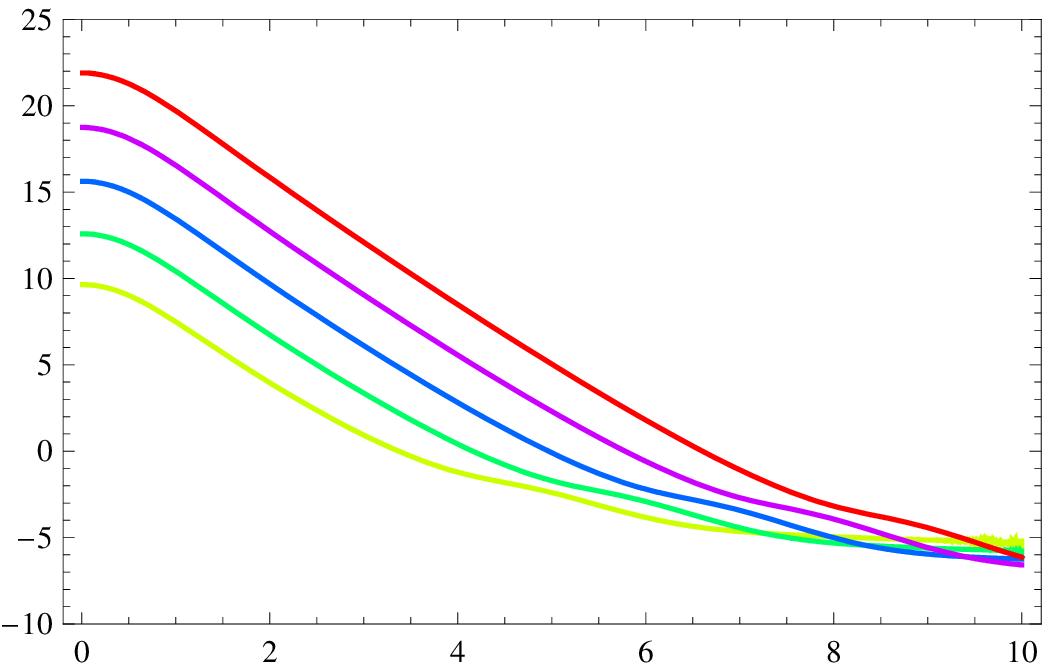}}
\put(0,0){\includegraphics[height=140pt]{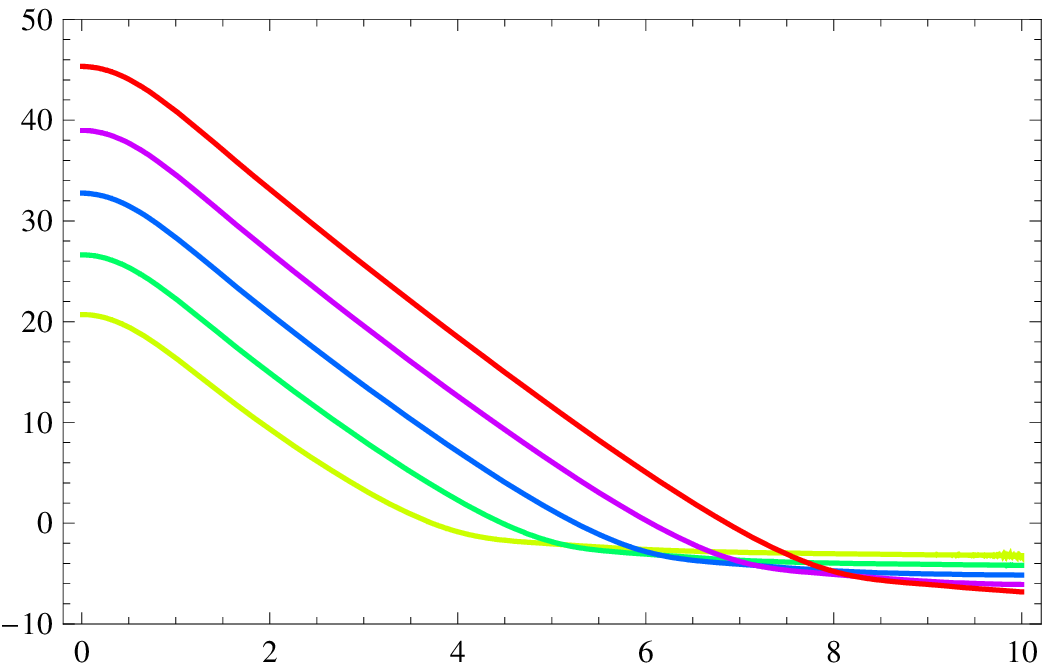}}
\put(240,0){\includegraphics[height=140pt]{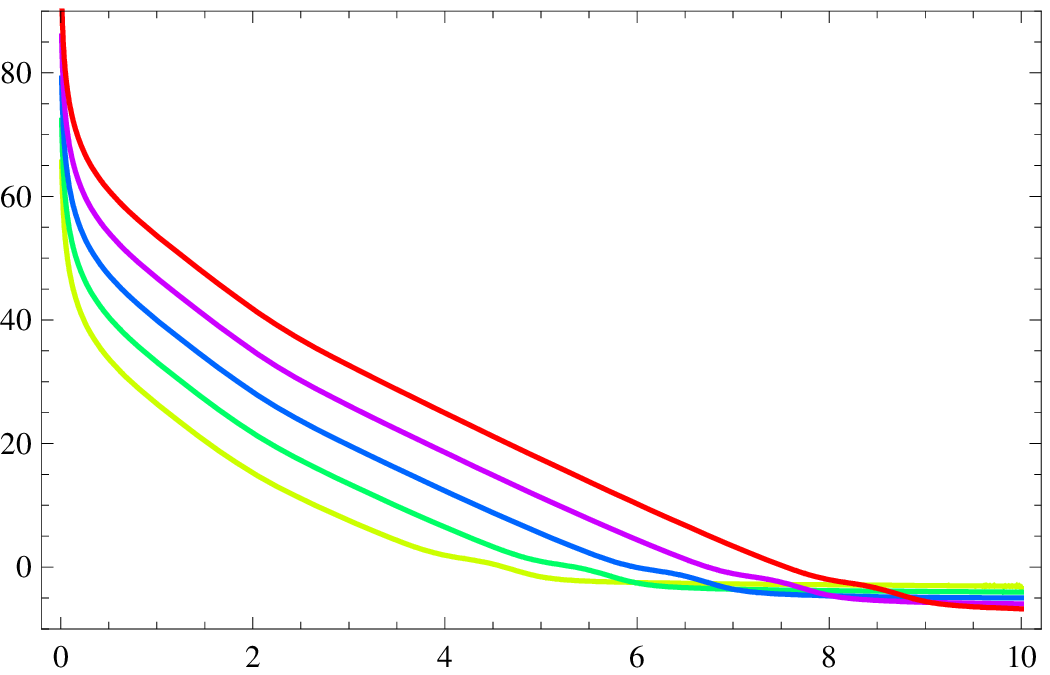}}
\put(5,145){$\log R_{\mu\nu}R^{\mu\nu}$}
\put(5,305){$P$}
\put(240,145){$\log R_{\mu\nu\tau\sigma}R^{\mu\nu\tau\sigma}$}
\put(245,305){$\log R$}
\put(220,7){$\rho$}
\put(460,7){$\rho$}
\put(222,167){$\rho$}
\put(462,167){$\rho$}
\end{picture} 
\caption{The Ricci scalar $R$, the Ricci tensor squared $R_{\mu\nu}R^{\mu\nu}$, and the Riemann tensor squared $R_{\mu\nu\tau\sigma}R^{\mu\nu\tau\sigma}$ for a few rotated solutions given by the $P$ of the upper left panel.}
\label{Fig:CurvatureRotated}
\end{center}
\end{figure}

As can be seen, $R$ and $R_{\mu\nu}R^{\mu\nu}$ stay finite in the IR 
(another simple invariant that is finite is $\sqrt{|g|}$), 
whereas $R_{\mu\nu\tau\sigma}R^{\mu\nu\tau\sigma}$ 
diverges as $\rho^{-8}$. This agrees with the numerically 
obtained plots shown in Figure~\ref{Fig:CurvatureRotated}.

It did not escape our attention that many other backgrounds 
(like the negative-mass Schwarzchild solution) follow the same pattern. What is 
characteristic of the present example is the presence of many matter 
fields accompanied by the mildness of the singularity.

For  more detail about these invariants in related solutions, see Appendix~\ref{curvatureccc}.
We move now to the study of an important observable, the Wilson loop.
\subsection{Wilson loops, confinement and phase transition}

We compute here the expectation value of the Wilson loop, and hence the quark-antiquark potential,
following the prescription in~\cite{MW}. For convenience, and because in the IR all the solutions discussed here
are very similar to those in~\cite{NPR}, we follow the notation introduced there.
We probe the background with an open string, the end-points of which are fixed on a D3-brane that 
extends in the Minkowski directions, and that is located at some very large value $\r_2$ of the 
radial direction.
The string configuration (given the distance in the Minkowski directions $ L_{QQ}$ between the end-points) 
is computed by minimizing 
the classical Nambu-Goto action. The result is a classically stable configuration in which the string hangs 
in the radial direction down to a minimal value $\hat{\r}_o$ at its middle-point.
In practice, one varies $\hat{\r}_o$, and for each possible value computes the string configuration,
the separation $L_{QQ}$ between the end-points of the string on the UV brane, and the energy $E_{QQ}$
as the classical action evaluated on the solution. Solving for $\hat{\r}_o$ yields the relation $E_{QQ}(L_{QQ})$,
and one can then verify that when 
$\hat{\r}_o$ approaches the end-of-space $\r_0=0$,
then $L_{QQ}$ diverges, and $E_{QQ}=\sigma L_{QQ} + {\cal O} (1/L_{QQ})$, yielding confinement.

In order to do so, one has to write the ten-dimensional metric in string frame.
We do not allow the string to explore the internal space, hence only the $g_{tt}$, $g_{xx}$ and $g_{\r\r}$
components of the metric are used.
One then defines
\beqs
F^2&=&g_{tt}g_{xx}\,,\\
G^2&=&g_{tt}g_{\r\r}\,,\\
V_{\rm eff}^2(\r,\hat{\r}_o)&=&\frac{F^2(\r)}{F^2(\hat{\r}_o)G^2(\r)}\left(F^2(\r)-F^2(\hat{\r}_o)\right)\,,\\
L_{QQ}(\hat{\r}_o)&=&2\int_{\hat{\r}_o}^{\r_2}\di \tilde{\r}\frac{1}{V_{\rm eff}(\tilde{\r},\hat{\r}_o))}\,,\\
E_{QQ}(\hat{\r}_o)&=&2\int_{\hat{\r}_o}^{\r_2}\di \tilde{\r}\sqrt{\frac{F^2(\tilde{\r})G^2(\tilde{\r})}{F^2(\tilde{\r})-F^2(\hat{\r}_o)}}\,.
\eeqs

In order for this calculation to make sense, it is necessary that in the UV
\beqs
\lim_{\r\rightarrow +\infty} V_{\rm eff}(\r,\hat{\r}_o)&=&+\infty\,,
\eeqs
which encodes the fact that appropriate boundary conditions must be satisfied by the string in order to end on the D3-brane.
In order for $L_{QQ}\rightarrow +\infty$ when $\hat{\r}_o\rightarrow 0$, then one must have 
\beqs
\lim_{\hat{\r}_o\rightarrow 0}V_{\rm eff}(\r,\hat{\r}_o)&\propto&\r^{\gamma} + \cdots\,,
\eeqs
with $\gamma\geq 1$, where an expansion in small $\r$ is 
understood, see the discussion around Eq.~(25) in the paper~\cite{NPR}.

If both these conditions are satisfied, then one can ask if the theory confines, and what is the 
value of the string tension. The linear potential is recovered for $\gamma=1$.
The result is
\beqs
\sigma&=&\lim_{L_{QQ}\rightarrow +\infty}\frac{\di E_{QQ}}{\di L_{QQ}}\,=\,\lim_{\r\rightarrow 0} F(\r)\,.
\eeqs

Within the PT ansatz, the string-frame metric is
\beqs
\di s^2&=&e^{\frac{\Phi}{2}}\di s^2_E\,=\,e^{2p-x} \di y^2\,+\,\cdots\nonumber\\
&=&e^{2p-x+\frac{\Phi}{2}+2A}\di x_{1,3}^2\,+\,e^{2p-x+\frac{\Phi}{2}}\di r^2\,+\,\cdots\nonumber\\
&=&e^{2p-x+\frac{\Phi}{2}+2A}\di x_{1,3}^2\,+\,e^{-6p-x+\frac{\Phi}{2}+\log 4}\di \r^2\,+\,\cdots\,,
\eeqs
where we omitted the internal part of the metric.
As a result, all the necessary information is contained in
\beqs
F^2&=&e^{4p-2x+{\Phi}+4A}\,,\\
G^2&=&e^{-4p-2x+{\Phi}+2A+\log 4}\,.
\eeqs

\subsubsection{The rotated solutions}

We start from the wrapped-D5 system, with no rotation in place, in which case
the exact relation $A=\frac{\Phi}{4}+\frac{x}{2}-p$ holds.
Hence in this case one has
\beqs
F^2_0&=&e^{2\Phi}\,,\\
G^2_0&=&e^{-6p-x+\frac{3}{2}\Phi+\log 4}\,=\,F_0^2 \frac{P^{\prime}}{2}\,,
\eeqs
where the prime refers to a derivative with respect to $\r$.

In order to see whether these solutions satisfy the UV conditions allowing for the probe-string calculation to be done, we study the asymptotic behavior of $V_{\rm eff}$.

The dilaton approaches a constant in the far UV, while $P^{\prime}\propto e^{4\r/3}$, hence
\beqs
V_{{\rm eff}\,(k_2=0)}^2\,\propto\, e^{-\frac{4}{3}\r}\rightarrow 0\,.
\eeqs
This is not compatible with the boundary conditions of the open string on the D3-brane at infinity.
We will not discuss these solutions any further.

The action of the rotation for generic $k_2$ is,
\beqs
F^2_{k_2}&=&\left(1-k_2^2F_0^2\right)^{-1}\,F^2_0\,,\\
G^2_{k_2}&=&G^2_0\,,
\eeqs
hence it does affect $F^2$ but not $G^2$.
Most importantly, the whole calculation requires to know 
and specify only $\Phi$ and $P^{\prime}$.

The effect of the rotation is, as we said, to change the asymptotic behavior of $F$, but not $G$.
In practice, at asymptotically large values of $\r$ and having 
fine-tuned $k_2$:
\beqs
V_{{\rm eff}\,(r)}^2 \propto \left(\frac{N_c^2}{c_+^2}e^{-\frac{8}{3}\r}\frac{}{}\right)^{-2}\,e^{-\frac{4}{3}\r}\,\propto\,e^{4 \r}\,.
\eeqs
In these expressions, we neglected the $\log z$ dependence of the ${\cal O}(z^4)$ terms in the expansion of $\Phi$.
Hence, within this class of solutions, only those in which the 
fine-tuning of $k_2$ has been implemented can be probed with 
the string and the procedure in~\cite{MW} carried out.
Notice that because of the rotation we must include corrections of order $N_c^2/c_+^2$, or else the effective potential is not well defined.

\subsubsection{IR expansions}

In order to study the effective potential in the deep IR region, we need to expand 
 first. Remember that, as we saw, only in the fine-tuned case does this calculation make
 sense.
Within the wrapped-D5 system, this expansion yields
\beqs
P&=&c_0\,+k_3 c_0 \r^3+\frac{4}{5}k_3 c_0 \r^5-k_3^2c_0\r^6+\frac{16(2c_0^2 k_3-5k_3 N_c^2)}{105c_0}\r^7\,+\,{\cal O}(\r^8)\,,\\
Q&=&N_c\left(\frac{4}{3}\r^2-\frac{16}{45}\r^4+\frac{128}{945}\r^6\,+\,{\cal O}(\r^8)\right)\,,
\eeqs
and after some algebra:
\beqs
F^2_0&=&e^{2\Phi}\,=\,
4 \sqrt{2\frac{e^{4 {\Phi_o}}}{3{c_0}^3
   {k_3}}} \left(1+\frac{16 {N_c}^2}{9c_0^2} \r^4\right) \,,\\
   G^2_0&=&2 e^{2 \Phi_o} \sqrt{\frac{6k_3}{c_0}} \r^2\,.
\eeqs
This means that for $\hat{\r}_o\rightarrow 0$ we find that $V_{\rm eff} \propto \r$.
We can interpret this as linear confinement,
with the string tension given by
\beqs
\sigma_{(r)}&=&
\left(1-e^{-2\Phi_{\infty}}\sqrt{\frac{32e^{4 {\Phi_o}}}{3{c_0}^3
   {k_3}}}  \right)^{-1/2} \left(\frac{32e^{4 {\Phi_o}}}{3{c_0}^3
   {k_3}}\right)^{1/4}\,.
\eeqs

In order to learn something, we need to connect the IR and UV expansions.
To do so, we can make use of $P_0$, but we must keep in the expression leading-order corrections in $N_c^2/c_+^2$.
This was done in part in~\cite{NPP}, where the IR expansion of the solution was modified to (at ${\cal O}(N_c^2/c_+^2)$)
\beqs
P&=&c\cos\alpha\left(1\,+\,\left(\frac{2^5}{3^2}\tan^3\alpha+\frac{2^4N_c^2\sin^3\alpha}{3^2c^2\cos^5\alpha}\log^2(2\cot^3\alpha)\right) \r^3\right)+\cdots\,,\nonumber
\eeqs
which yields the identifications
\beqs
c_0&=&c \cos\alpha\,,\\
k_3&=&\frac{2^6}{3^2}\frac{1}{(2\cot^3\alpha)}\left(1\,+\,\frac{N_c^2\log^2(2\cot^3\alpha)}{2c_0^2}\right)\,.
\eeqs
The UV expansion of $P$  yields 
the identifications
\beqs
c_+&=&\frac{c \sin\alpha}{2^{1/3}3}\,,\\
-\frac{c_-}{192 c_+^3}&=&2\cot^3\alpha\,.
\eeqs
By combining these results one then concludes that (at least for small values of $N_c/c_+$)
\beqs
\frac{3 c_0^3 k_3}{32}e^{-4\Phi_o}&=&
e^{-4\Phi_{\infty}}\left(1+\frac{ {N_c}^2 \log
   ^2\left(-\frac{{c_-}}{192
   {c_+}^3}\right)}{18c_+^2(-\frac{{c_-}}{192 c_{+}^3})^{2/3}}\right)\,,
\eeqs
and the string tension is
\beqs
\sigma_{(r)}&\simeq&e^{\Phi_{\infty}}\left(\frac{6 c_+\left(-\frac{c_-}{192c_{+}^3}\right)^{1/3}}{N_c\log \left(-\frac{c_-}{192c_{+}^3}\right)}\right)\,.
\eeqs

Going a step further, again from~\cite{NPP} one has $\log \left(-\frac{c_-}{192c_{+}^3}\right)=\log (2\cot^3\alpha)\sim 4\r_{\ast}$,
provided $\r_{\ast}\gg 1$,  and if we replace:
\beqs
\sigma_{(r)}
&\simeq&e^{\Phi_{\infty}}\left(\frac{3 c_+}{2 N_c\r_{\ast}}e^{\frac{4}{3}\r_{\ast}}\right)\,.
\eeqs
Notice how this implies that for large values of $\r_{\ast}$ the string tension would increase, if we were to keep the other parameters
fixed.

As we recalled around Eq.(\ref{boundzzz}), one has to require 
 that $P>Q$ at the scale $\r_{\ast}$~\cite{NPP}. 
Hence one has to impose the approximate bound
\beqs
\frac{N_c}{c_+}&\lsim&\frac{3 e^{\frac{4}{3}\r_{\ast}}}{2^{2/3}\r_{\ast}}\,,
\eeqs
which, replacing in the expression we gave for the string tension, yields
\beqs
\sigma_{(r)}&\gsim&2^{-1/3}e^{\Phi_{\infty}}\,.
\eeqs
The actual numerical coefficient should not be trusted, aside from the fact that it is ${\cal O}(1)$.
However, this exercise shows that by tuning appropriately the parameters in our class of 
solutions (besides tuning $k_2$), one can get a whole family of solutions which confine and have the same string tension,
while differing by the value of $\r_{\ast}$, and that this value is controlled by the value of the dilaton at infinity.

\subsubsection{Numerical study}

The class of solutions we are looking at can be characterized in terms of six parameters.
We study numerically the Wilson loops restricting to a one-parameter family of rotated solutions
selected in the following way.
The generating function $P$, which solves the master equation, depends in general on
three parameters: $c_+$, $c_-$ and  $N_c$.  The overall scale is fixed by $\r_0$, which we choose to vanish $\r_0=0$.
The rotated solution depends explicitly on $k_2$.
The calculation of the Wilson loop depends also on the
value of the dilaton $\Phi_o$.
We keep $N_c=4$ fixed (an arbitrary numerical choice, that does not affect any of the 
physical results), and vary $c_+$ and $\Phi_o$ in such a way that
the dilaton in the far UV and deep IR is kept fixed, while 
fine-tuning $k_2=e^{-\Phi(\infty)}$,
so that the calculation is sensible.
In doing so, we generate a one-parameter family of solutions to the equations
that differ only by the value of $\r_{\ast}$ 
(or $c_-$), the value of the radial coordinate 
below which $P$ is approximately constant 
and above which it is dominated by terms proportional to $e^{4\r/3}$.
Furthermore, we choose  $P$ so that $P(\r_{\ast})\simeq Q(\r_{\ast})$, 
in such a way as to maximize the effects of the $N_c/c_+$
corrections which (after the rotation) 
take us away from the KS solutions.
The resulting function $P$, the 't Hooft coupling $\frac{g_4^2N_c}{8\pi^2}$ and the dilaton $\Phi$ are shown in Fig.~\ref{Fig:walking}

\begin{figure}[h]
\begin{center}
\begin{picture}(520,550)
\put(255,290){\includegraphics[height=4.5cm]{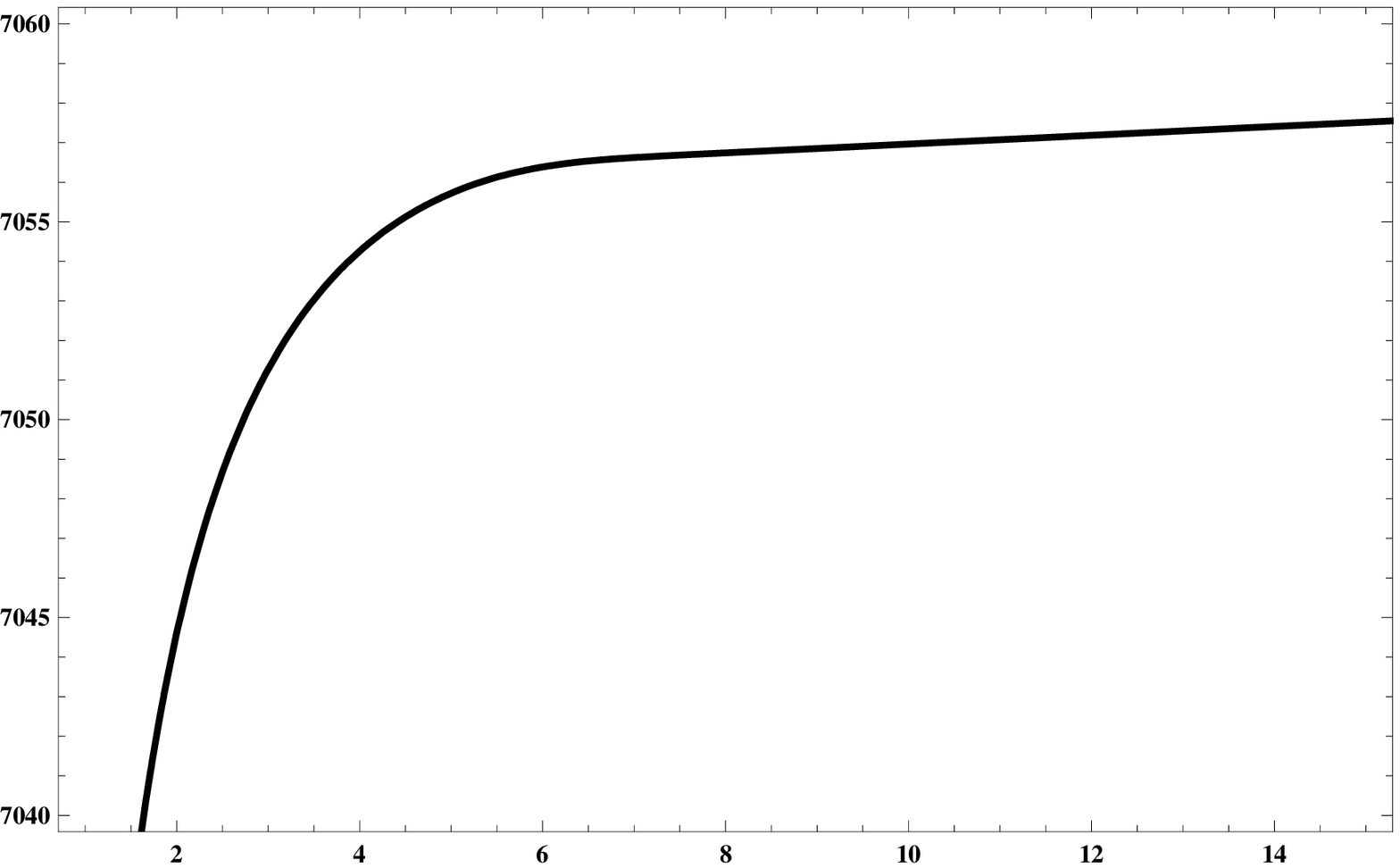}}
\put(13,145){\includegraphics[height=4.5cm]{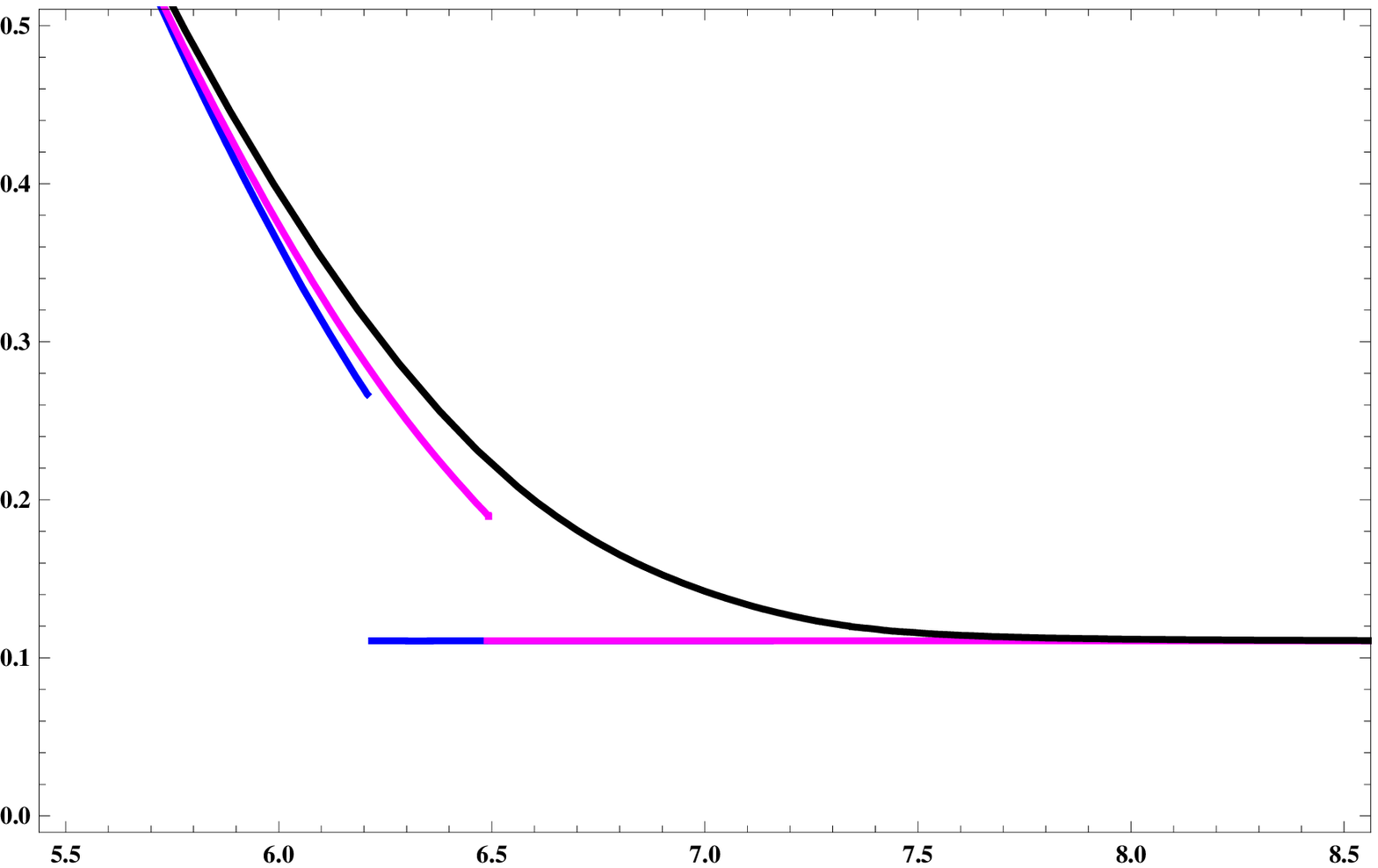}}
\put(253,435){\includegraphics[height=4.5cm]{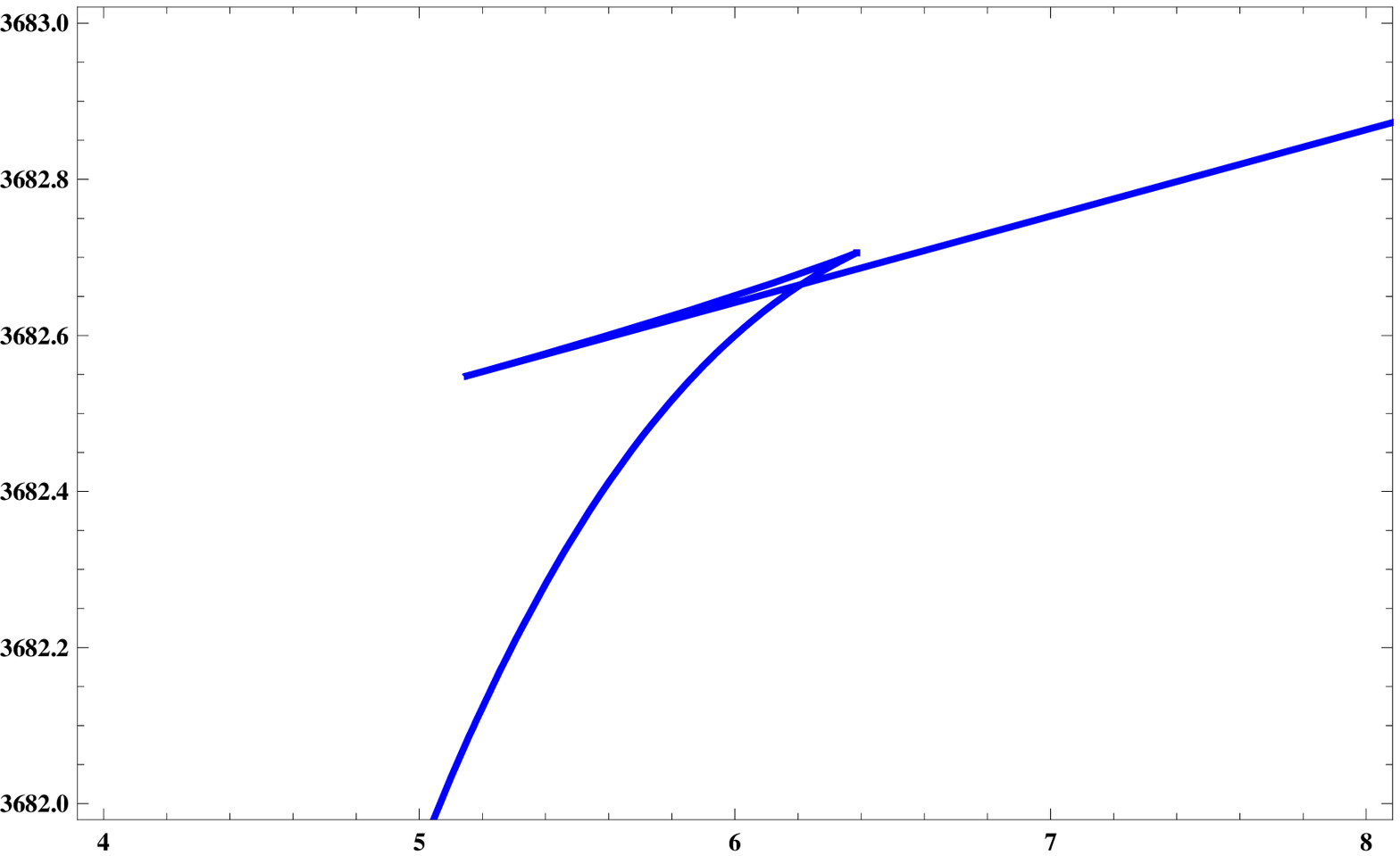}}
\put(0,290){\includegraphics[height=4.5cm]{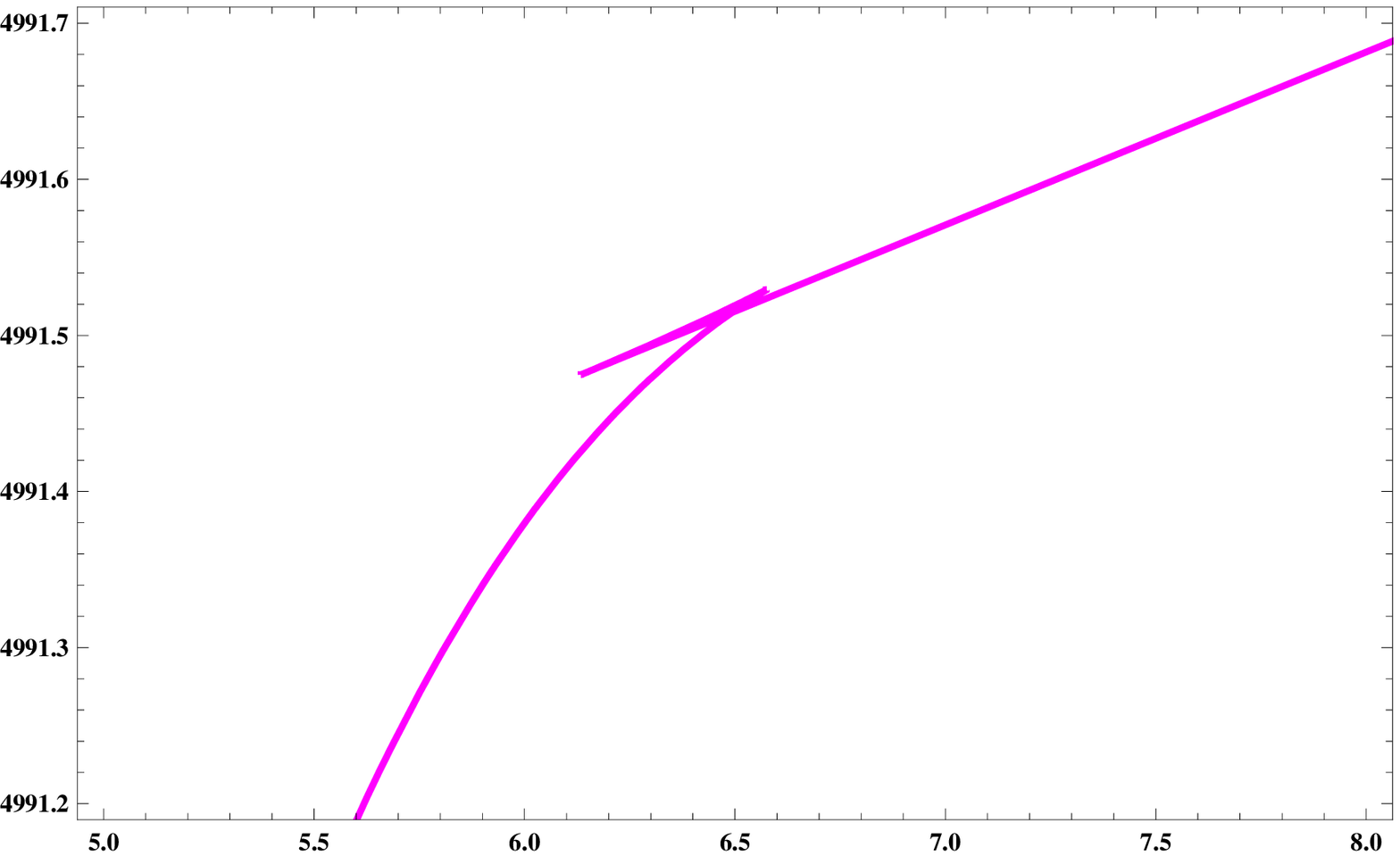}}
\put(260,145){\includegraphics[height=4.5cm]{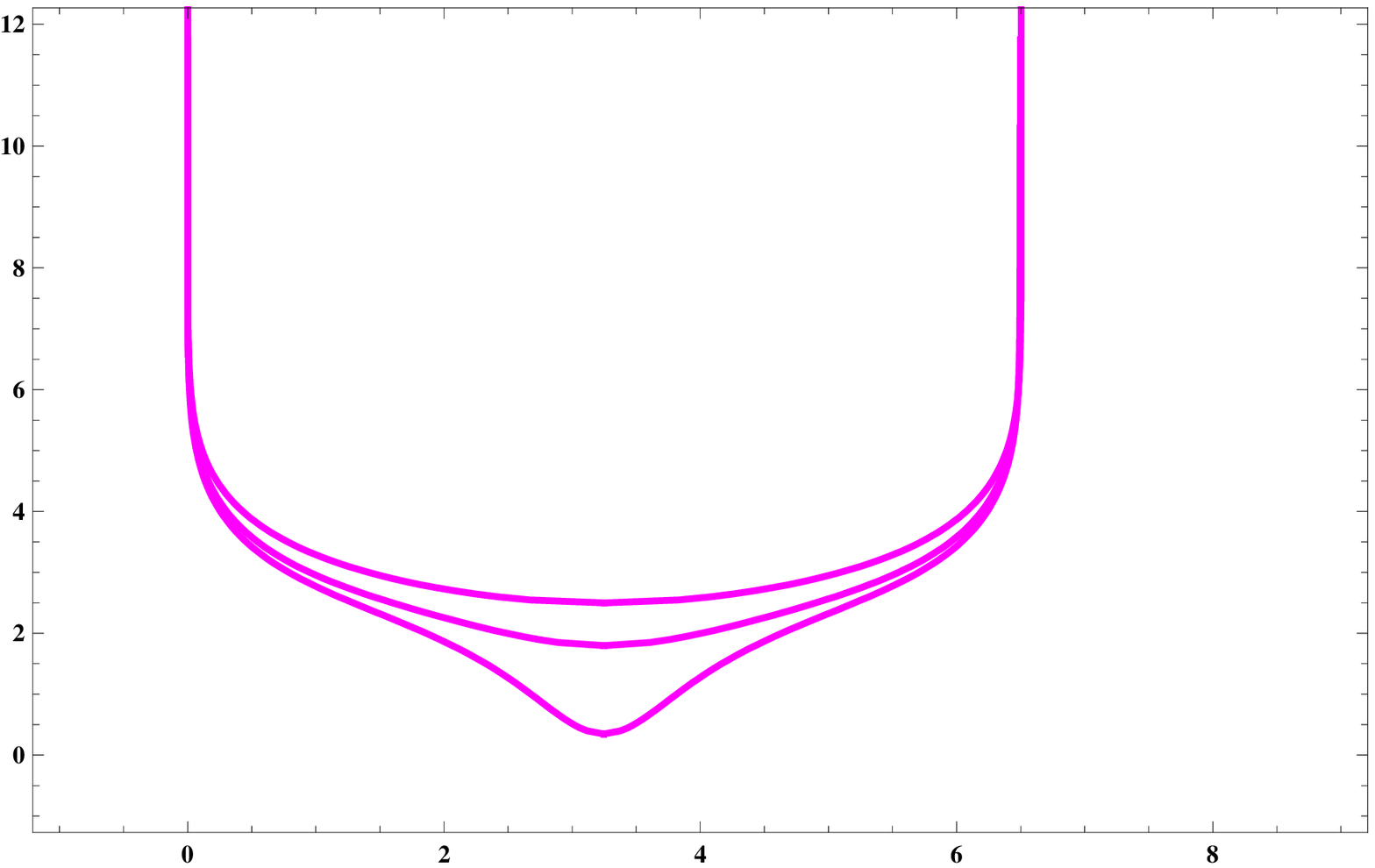}}
\put(10,435){\includegraphics[height=4.5cm]{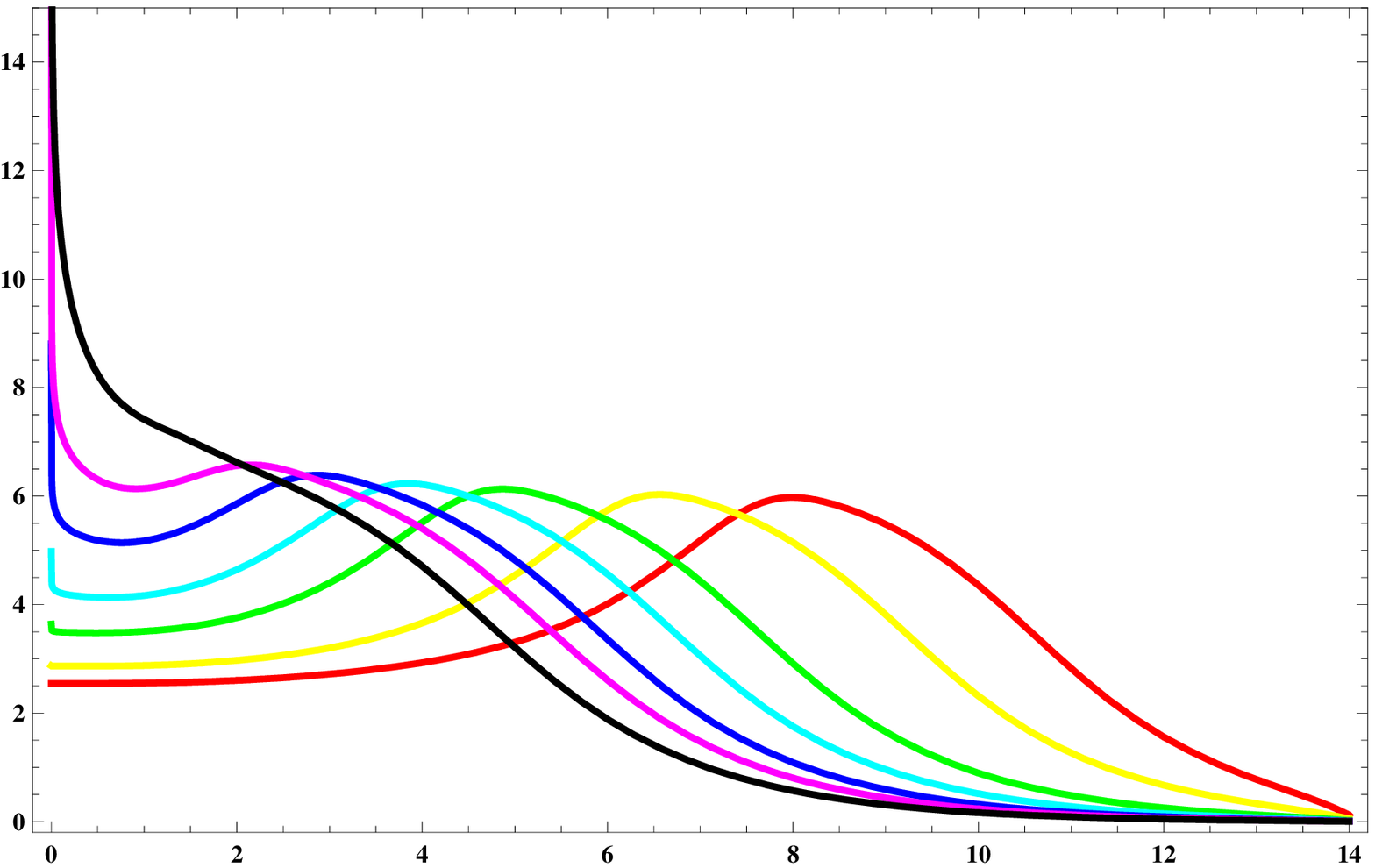}}
\put(7,0){\includegraphics[height=4.5cm]{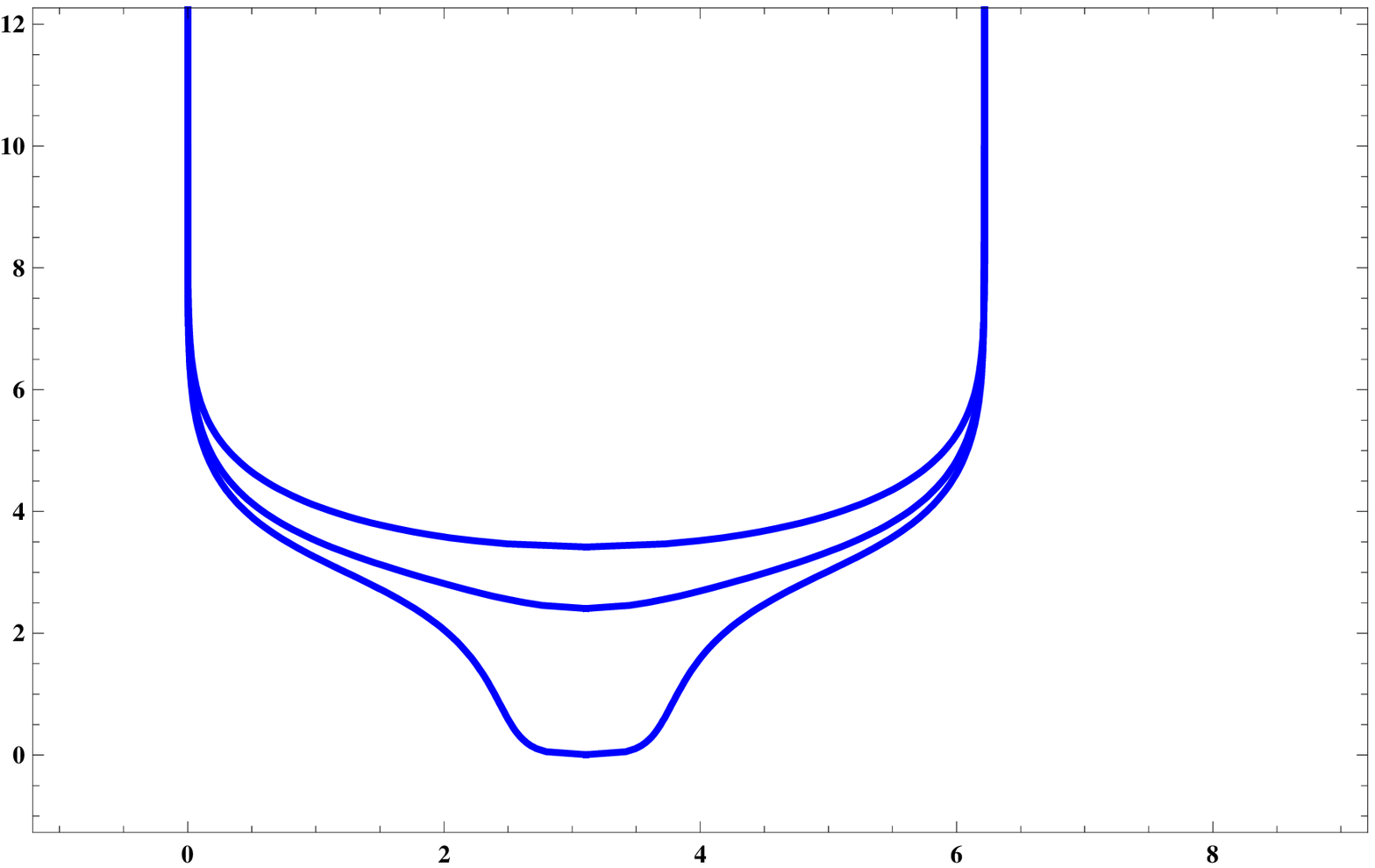}}
\put(260,0){\includegraphics[height=4.5cm]{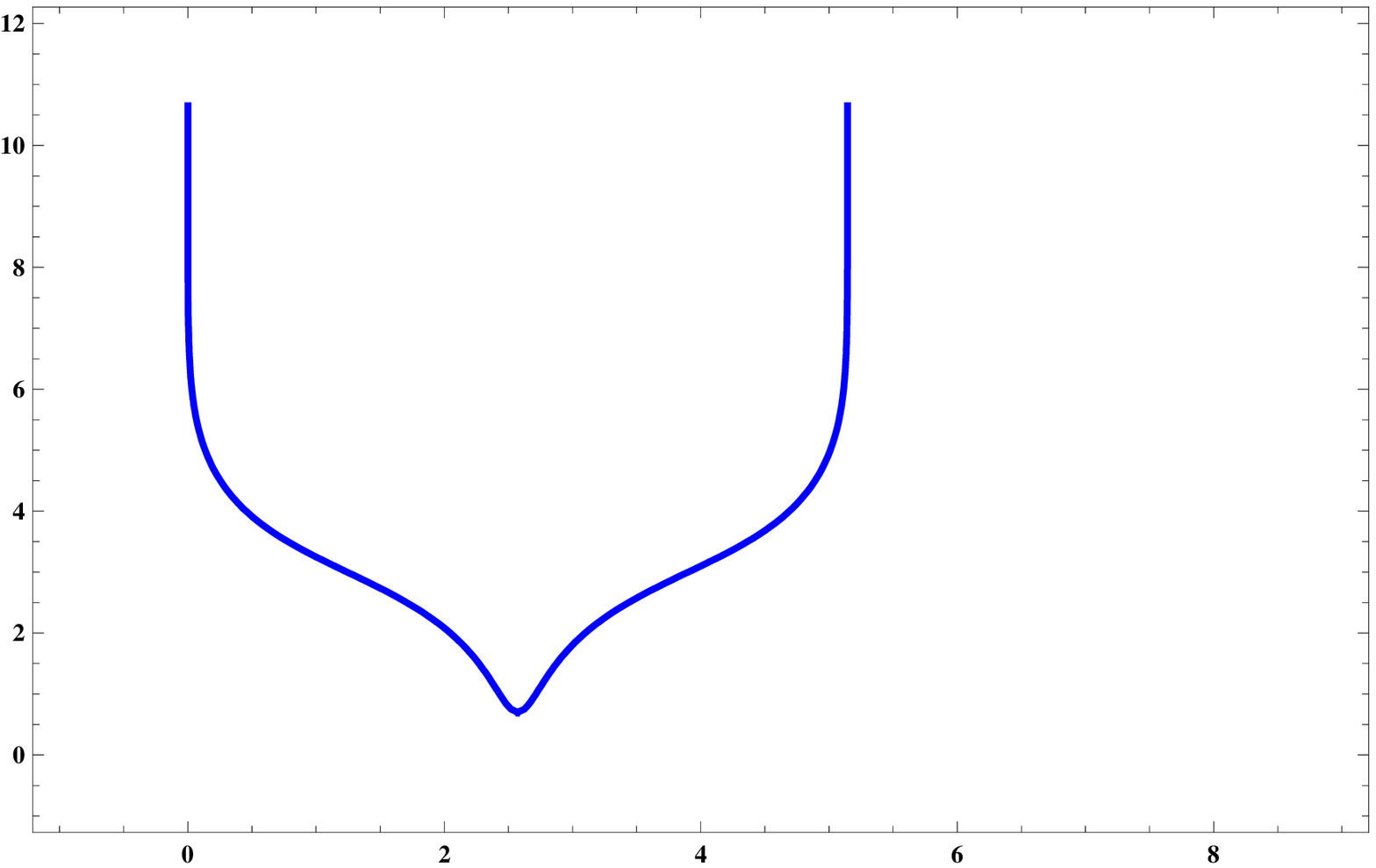}}
\put(454,-2){$x$}
\put(455,144){$x$}
\put(200,-2){$x$}
\put(200,142){$L_{QQ}$}
\put(450,286){$L_{QQ}$}
\put(244,405){$E_{QQ}$}
\put(255,262){$\r$}
\put(-0,116){$\r$}
\put(-10,260){$\frac{\di E_{QQ}}{\di L_{QQ}}$}
\put(255,116){$\r$}
\put(194,286){$L_{QQ}$}
\put(200,432){$\hat{\r}_o$}
\put(435,432){$L_{QQ}$}
\put(-10,405){$E_{QQ}$}
\put(-10,550){$L_{QQ}$}
\put(244,550){$E_{QQ}$}
\end{picture} 
\caption{The result of the Wilson-loop numerical analysis. The color coding in the figure is
such that the same color always corresponds to quantities computed on the same 
background, and agrees with Fig.~\ref{Fig:walking}. 
The top-left panel shows  $\hat{\r}_o(L_{QQ})$. Notice that the resulting function is invertible only
for the solution with smallest $\r_{\ast}$ (black). The fact that backgrounds with large values of $\r_{\ast}$
do not extend to $L_{QQ}\rightarrow +\infty$ is due only to the limited numerical precision.
The next three plots show a detail of  $E_{QQ}(L_{QQ})$ for the three backgrounds with smallest values of $\r_{\ast}$.
Notice that the result is multi-valued for two of them, while in the last case the function is invertible
and the transition between Coulomb and confined phase is smooth. 
The fifth plot shows the discontinuity of $\frac{\di E_{QQ}}{\di L_{QQ}}$ at the transition, which is absent for small values of $\r_{\ast}$,
together with the fact that the string tension is identical in all the cases considered.
The next two plots show the shape of the probe string for three 
choices of 
$\hat{\r}_o$ for two of the backgrounds  (notice the color coding),
chosen so that $L_{QQ}$ is the same, and coincides with the
critical value at which the phase transition is taking place. Hence,
$E_{QQ}$ is  the same for the upper and lower curves, while the intermediate have 
a higher energy $E_{QQ}$.
The last panel shows one unstable solution, highlighting the cuspy shape at its middle already observed and discussed
in detail in~\cite{NPR}, and which is
connected with the fact that this is an unstable classical solution of the equations derived from the Nambu-Goto action.
\label{Fig:Wilsonresults}}
\end{center}
\end{figure}

With all of this in place, we perform numerically the calculation of $L_{QQ}$ and $E_{QQ}$.
Numerically, our solutions extend towards the UV only up to $\r\lsim 15$. We hence vary $0<\hat{\r}_o\lsim12$,
keeping explicitly a UV cutoff $\r_2=14$, and
compute $L_{QQ}$ and $E_{QQ}$. We plot the result, for the numerically chosen backgrounds,
in Fig.~\ref{Fig:Wilsonresults}.

The result is that there is a first-order phase transition, as a function of $L_{QQ}$.
In order to visualize the strength, we plot in  Fig.~\ref{Fig:Wilsonresults} the derivative of the 
energy with respect to the quark separation, $\di E_{QQ} / \di L_{QQ}$,
computed only on the minimum-$E_{QQ}$ configurations. 
The result illustrates two things. First of all, the solutions have the exact same value of the string tension
($\di E_{QQ} / \di L_{QQ}$ is a universal constant at large $L_{QQ}$), as a result of the tuning we did on the IR value of the dilaton.
Second, the discontinuity in $\di E_{QQ} / \di L_{QQ}$ depends on $\r_{\ast}$, becoming larger
when $\r_{\ast}$ is large.

Some comments are in order.
Qualitatively, these results are hardly any different from those in~\cite{NPR}, reflecting the fact that
the IR of the geometry of these classes of solutions is very similar.
In particular, the last panel of Fig.~\ref{Fig:Wilsonresults} shows an interesting fact: those configurations of the string that
solve the classical equations but correspond to a maximum of the energy (the choices of $\hat{\r}_o$ for which $L_{QQ}$ is an increasing 
function of  $\hat{\r}_o$) have a peculiar shape. The solutions are completely smooth, however they do show a 
fast turning at their tip, which is ultimately responsible for the fact that the configuration has a comparatively high energy, and is unstable. 

Let us comment more about the phase transition we observe.
The exact value of the critical $L_{QQ}$ at which the transition takes place appears to depend
on the solution. This is a numerical artifact: both the energy $E_{QQ}$ and the quark separation $L_{QQ}$
have been computed while keeping the same value of the UV cutoff $\r_2$, and hence there is some intrinsic uncertainty
due to a possible overall shift in the two, which is purely UV dependent. One should not give any special meaning to it.
More interestingly, we can define the following dimensionless quantity
\beqs
s&\equiv&\frac{\left.\frac{\di E_{QQ} }{ \di L_{QQ}}\right|_- - \left.\frac{\di E_{QQ} }{ \di L_{QQ}}\right|_+}
{\lim_{L_{QQ}\rightarrow+\infty}\frac{\di E_{QQ} }{ \di L_{QQ}}}\,,
\eeqs
and use it to classify how strong the transition is.
In the examples in the plots this quantity appears to be ${\cal O}(1)$, pointing to the fact that already for $\r_{\ast}\sim 3$
we are in the presence of a strong first-order transition. The larger $\r_{\ast}$, the larger $s$ becomes.

An interesting observation: this phase transition exists only provided $\r_{\ast} > \r_I \sim 1$.
Below some finite value of $\r_{\ast}$, $E_{QQ}(L_{QQ})$ is single valued, and the transition between Coulomb phase and 
confined phase is completely smooth.
We should emphasize that this `double turn around'
phenomenon discussed in~\cite{NPR}, is present in our solution and
also in many other systems with two independent scales. 
See for example the papers in \cite{AFA}.

\section{Conclusions\label{Sec:Outlook}}

We conclude the paper by summarizing  the results we obtained, and 
our interpretation. And finally, we summarize what are the open problems, and possible ways to test and extend our 
results and their interpretation.

We started our analysis from a rather general, four-parameter class of
type-IIB backgrounds obtained by solving the master equation 
(\ref{Eq:master}) 
characterizing the
wrapped-D5 system (only gravity, dilaton and $F_3$ form are non-trivial).
The main features of such backgrounds are that: the theory confines, and the gaugino condensate appears,
in the deep IR ($\r\rightarrow 0$), but the dynamics above this scale is characterized by two very different behaviors,
with a smooth transition at a value $\r_{\ast}\gg 0$ of the radial direction. 
We applied to the backgrounds a solution-generating technique (rotation)
that (as a function of a new parameter $k_2$) allows to algebraically construct
backgrounds that are more general and fall within the PT ansatz.
In this way, a flux for $F_5$,  $B_2$ and $H_3=dB_2$ is induced.
We exhibited explicitly the action of the rotation both using
the ten-dimensional language and the five-dimensional  language
obtained by consistent truncation of the KK decomposition.

We then studied the backgrounds obtained by the rotation,
and compared them to the original ones, to the KS backgrounds and 
to the baryonic branch of KS.
Performing the study of the UV asymptotic behavior of all of these solutions, 
we concluded that the most important differences with KS are: 
the presence of a dimension-2 VEV that brings all our solutions on the baryonic branch,
the presence of a dimension-8 operator, whose dynamics makes the dual models UV incomplete
(unless $k_2$ is fine-tuned against the asymptotic value of the dilaton, in which case the dimension-8 operator disappears),
and the presence of what appears as a dimension-6 VEV.

We provided a simple field-theory interpretation of the rotation and of the backgrounds it 
relates. The unrotated background, with  $F_5=H_3=0$, provides a simple
gravity dual to the low-energy effective field theory description of the system, which consists of
a one-site ${\cal N}=1$, $SU(N_c)$  gauge theory coupled to adjoint matter.
This description is good up to the cutoff scale indicated by $\r_{\ast}$,
above which the dynamics is driven by the dimension-8 effective operator.
The rotation allows to adiabatically switch off the higher-dimensional operator of the dual theory,
while keeping the VEV(s) fixed and by fine-tuning $k_2$ one finds that the 
resulting background {\it roughly} interpolates between the KS solution for $\r>\r_{\ast}$ and the 
original `seed' wrapped-D5 background for $\r<\r_{\ast}$.
In particular, for $\r>\r_{\ast}$ the field-theory dual is essentially
the cascading $SU(M)\times SU(M+N_c)$ quiver as in KS.

We observed that the rotated backgrounds automatically
implement the constraint $\tilde{M}=0$.

This, put together  with the observation that
at the perturbative level  the $SU(M)\times SU(M+N_c)$ 
quiver field theory and the ${\cal N}=1^{\ast}$ deformation
of ${\cal N}=4$ SYM result in a spectrum that (at large $N$) 
deconstructs the (fuzzy) sphere \cite{MM},\cite{AD},
leads us to suggest that at the scale $\r_{\ast}$ the gauge group is undergoing the Higgsing
$SU(q N_c)\times SU(qN_c +N_c)\rightarrow SU(N_c)$. 
The perturbative analog to the (non-perturbative) 
scale separation $\r_{\ast}\gg 1$ comes from the fact
that the spectrum of heavy vector multiplets contains a (finite) tower of states with
lowest states with mass $M^2 \sim \eta^2$ and highest states with mass $M^2\sim q^2 \eta^2$.
We hence interpreted the freedom in the choice of $\r_{\ast}$ in terms of the freedom of the choice of Higgs vacuum in the dual theory (i.~e. in the choice of $q$).
We collected quite a large set of elements supporting this interpretation
by studying the properties of the supergravity background.

We then studied the long-distance properties of the theory, in order to test our interpretation.
Most striking is the fact that in the presence of a substantial hierarchy $\r_{\ast}\gg 1$,
the calculation of the Wilson loops yields a non-trivial result for the static quark-antiquark potential,
exhibiting the features of a strong first-order transition at intermediate distances,
while at long distances the theory confines in the traditional sense.
The former behavior disappears when the coefficient of the dimension-6 operator is tuned to very small values. 
This agrees with what was found in~\cite{NPR}, 
in a class of solutions that are very similar to 
the seed solution studied here in the IR (for $\r<\r_{\ast}$), 
but in which $P$ is linear, rather than exponential, in the UV
for ($\r>\r_{\ast}$). This indicates that the phase transition, 
if physical, has to do with the coefficient of the dimension-6 VEV.

Interestingly, the Wilson loop yields a perfectly healthy long-distance behavior,
as expected in a confining theory well captured by a supergravity dual, in spite of the fact that
the generic background with non-vanishing dimension-6 VEV is singular in the IR.
This surprising behavior is probably connected with the fact that 
the singularity is unusually mild for systems with many matter 
fields like ours: both the
Ricci scalar $R$ and the invariant $R_{\mu\u}R^{\mu\nu}$ are finite, while the singularity is manifest 
in $R_{\mu\nu\rho\sigma}R^{\mu\nu\rho\sigma}$.

We conclude by suggesting some future research programs.
First of all, a coherent picture starts to emerge, that unifies all the best-known solutions 
belonging to the PT ansatz, in which many features resemble what is expected in 
the case of mass deformations of ${\cal N}=4$ (see also~\cite{PW,LS,PS,KMPP,HPRW}).
Making a satisfactory connection with the linear-dilaton solution of wrapped-D5 system~\cite{MN},
in which $P=\hat{P}=2 N_c \r$,
requires an additional step, in which a larger class of solutions $P$ of the 
master equation must be analyzed
in detail. This is due to the fact that the rotation procedure cannot be applied to this 
specific class of backgrounds, in which the dilaton diverges in the UV.
We suggested an interesting idea, on a geometric basis, which could help provide a field theory explanation for the presence of the dimension-6 VEV,
particularly in relation to the phase transition we found by computing the Wilson loop expectation value.
A more systematic analysis of the (weakly coupled) dual field theory would also be useful along these lines.

An immediate test of all that we said would be to compute the spectrum of scalar glueballs
of the dual (confined) field theory, by studying the fluctuations of the five-dimensional sigma-model
in full generality.
The technology for doing so exists and is well understood~\cite{BHM,EP}, although
some subtlety connected with holographic renormalization does require a careful analysis.
This study would allow to answer unambiguously two open questions.
One is whether the scale separation between the heaviest and lightest massive vector multiplets 
survives also at strong coupling, or if it is only a perturbative result.
Another is to explicitly verify whether the spectrum is also at  strong coupling 
deconstructing the (non-fuzzy, at large $N$) sphere, and whether this is affected by the presence
and magnitude of the dimension-6 VEV.

The second open question relates to the IR singularity and its resolution.
It is not known whether this singularity has to be taken as signaling the fact that the background is
intrinsically pathological, whether it admits a resolution in terms of stringy physics (signaling 
that the singularity is actually due to a physical effect), or whether  it is just an unfortunate
result of the restrictions imposed in the derivation of the background equations,
which admits a resolution within supergravity.
On the basis of the arguments summarized earlier on, it seems plausible 
that the backgrounds we studied, while singular, provide a sensible quantitative description,
in supergravity terms, of the physics of the dual field theory, but this needs to be tested.
A good place to start is with the spectrum, in order to see if there are pathologies due to
the singularity itself. Aside from this, 
it should also be interesting to reanalyze the results
of the papers \cite{various}, in the more complete setup presented here.

A final comment about applications.
One phenomenological application of this research is related to the finding of an
isolated, anomalously light scalar in the spectrum of glueballs of a 
backgroud that shares many similarities with those analyzed here.
It is tempting to identify this scalar as a light four-dimensional 
pseudo-dilaton,
and if so, the conditions under which it appears are directly
relevant to the strongly coupled physics of electro-weak symmetry breaking.
However, the properties of this scalar are not fully understood
and it would hence be useful to see if its existence is a generic feature,
and to find a robust physical explanation for it.

\vspace{1.0cm}
\begin{acknowledgments}
We would like to thank some colleagues for 
interesting discussions: Adi Armoni, Anatoly Dymarsky, Johanna Erdmenger, Veselin Filev, 
Nick Halmagyi, Prem Kumar.

This work was completed while CN was a DAAD fellow in the Max-Planck Intitute
of Theoretical Physics (Munich), he thanks the hospitality extended by the MPI. 
The work of MP is supported in part  by
the Wales Institute of
Mathematical and Computational Sciences and
by the STFC Grant
ST/G000506/1. The work of JG is supported in part by MICINN and FEDER under grant FPA2008-01838, by the Spanish Consolider-Ingenio 2010 Programme CPAN (CSD2007-00042) and by Xunta de Galicia (Conselleria de Educacion grant INCITE09 206 121 PR). DE is grateful for the generous support of the people of India for research in the basic sciences.

\end{acknowledgments}

\appendix
\section{Relation to PT ansatz and details about the rotation}\label{detailsrotation}

Here we would like to make clear the relation between the backgrounds described in Section~\ref{Sec:Rotation} and the PT ansatz, as well as present in some more detail the effect of the rotation. The PT ansatz is given by\footnote{We are following the notation of~\cite{BHM}. However, in order
to reproduce the conventions used in Eq.~(\ref{nonabmetric424}), we make use of the fact that the ten-dimensional equations are symmetric
under a simultaneous change of sign of all the RR forms $F_1$, $F_3$ and $F_5$. Furthermore, the 
equations for the functions determining the background are also symmetric under the simultaneous change of sign of $a$, $b$ and $h_2$.
With respect to~\cite{BHM}, we apply both these changes of sign to all the functions appearing in the background.}
\beqs
	ds^2&=&e^{2p-x} ds_5^2 + (e^{x+ \tilde g} + a^2 e^{x- \tilde g}) (e_1^2 + e_2^2) + e^{x- \tilde g} \left( e_3^2 + e_4^2 - 2a (e_1 e_3 + e_2 e_4) \right) + e^{-6p-x} e_5^2, \\
	ds_5^2&=&dr^2 + e^{2A} dx_{1,3}^2, \\
	F_3&=& N \left[ e_5 \wedge \left(  e_4 \wedge e_3 + e_2 \wedge e_1 - b (e_4 \wedge e_1 - e_3 \wedge e_2) \right) + dr \wedge \left( \partial_r b ( e_4 \wedge e_2 + e_3 \wedge e_1) \right) \right], \\
	H_3&=&-h_2 e_5 \wedge (e_4 \wedge e_2 + e_3 \wedge e_1 ) + d r \wedge \Big[ \partial_r h_1 (e_4 \wedge e_3 + e_2 \wedge e_1 ) - \\ & & \partial_r h_2 (e_4 \wedge e_1 - e_3 \wedge e_2 ) + \partial_r \chi (-e_4 \wedge e_3 + e_2 \wedge e_1 ) \Big], \\
	F_5&=&\tilde F_5 + \star \tilde F_5, \ \ \tilde F_5 = -{\cal K} e_1 \wedge e_2 \wedge e_3 \wedge e_4 \wedge e_5.
\eeqs
Here, $\{p,x,g,a,b,h_1,h_2,\chi,{\cal K}\}$ are functions of the radial coordinate $r$, and we have defined the one-forms
\beqs
e_1&=&-\sin\theta\di \phi\,,\\
e_2&=&\di \theta\,,\\
e_3&=&\cos\psi\sin\tilde{\theta}\di \tilde{\phi}\,-\,\sin\psi\di\tilde{\theta}\,,\\
e_4&=&\sin\psi\sin\tilde{\theta}\di \tilde{\phi}\,+\,\cos\psi\di\tilde{\theta}\,,\\
e_5&=&\di \psi+\cos\tilde{\theta}\di \tilde{\phi}+\cos{\theta}\di {\phi}\,.
\eeqs

Now, consider the wrapped-D5 system described in Section~\ref{Sec:wrappedD5system}. The ansatz for the metric there falls into the more general case
\SP{
ds^2 = e^{2 \Delta} \left( dx_{1,3}^2 + ds_6^2 \right),
}
where $ds_6^2$ is of the same form as before and $\Delta$ is a function of $\rho$. Comparing with the PT ansatz, we obtain the following one-to-one map between variables ($N = N_c/4$)
\SP{\label{eq:onetoonemap}
	\Delta &=	A+p-\frac{x}{2}, \\
	g &= -A-\frac{\tilde g}{2}-p+x+ \log 2, \\
	h &= -A+\frac{\tilde g}{2}-p+x, \\
	k &= -A-4 p+ \log 2, \\
	d\rho &= \frac{1}{2} e^{4p} dr.
}

The solution-generating technique outlined in Section~\ref{Sec:RotationUDuality} starts with a solution of the kind described in Section~\ref{Sec:wrappedD5system}, i.e. a background describing the wrapped-D5 system, and then, applying a set of transformations, generates a new (rotated) solution which has the following form (the superscript $(r)$ refers to the rotated solution):
\SP{
	ds^{(r)2} &= e^{\Phi/2} \Big[ \left(1 - k_2^2 e^{2\Phi} \right)^{-1/2} dx_{1,3}^2 + \left(1 - k_2^2 e^{2\Phi} \right)^{1/2} ds_6^2 \Big], \\
	\Phi^{(r)} &= \Phi, \\
	F^{(r)}_3 &= F_3, \\
	H^{(r)}_3 &= - k_2 e^{2\Phi} *_6 F_3, \\
	F^{(r)}_5 &= -k_2 (1 + *_{10}) {\rm vol}_{(4)} \wedge d \left(e^{-2\Phi} - k_2^2 \right)^{-1}.
}
More explicitly, the transformations are given by
\SP{
	a^{(r)} &= a, \ \ b^{(r)} = b, \ \ \Phi^{(r)} = \Phi, \\
	e^{2\Delta^{(r)}} &= \left(1 - k_2^2 e^{2\Phi} \right)^{-1/2} e^{\Phi/2}, \ \ e^{2g^{(r)}} = \left(1 - k_2^2 e^{2\Phi} \right) e^{2g}, \\
	e^{2h^{(r)}} &= \left(1 - k_2^2 e^{2\Phi} \right) e^{2h}, \ \ e^{2k^{(r)}} = \left(1 - k_2^2 e^{2\Phi} \right) e^{2k}, \\
	\partial_\rho \chi^{(r)} &= \frac{k_2 N_c}{16} e^{-2 (g+h-\Phi )} \left(8 a e^{2 (g+h)} (a-b)+\left(a^2-1\right) \left(a^2-2 b
   a+1\right) e^{4 g}+16 e^{4 h}\right), \\
	\partial_\rho h_1^{(r)} &= \frac{k_2 N_c}{16} e^{-2 (g+h-\Phi )} \left(8 a e^{2 (g+h)} (a-b)+\left(a^2+1\right) \left(a^2-2 b
   a+1\right) e^{4 g}+16 e^{4 h}\right), \\
	h_2^{(r)} &= \frac{k_2 N_c }{8} e^{2 \Phi } \partial_\rho b = -\frac{k_2 e^{2 (g+\Phi )} \left(e^{2 g} \left(a^2-1\right)+4 e^{2 h}\right) a}{4 \sqrt{e^{4
   g} \left(a^2-1\right)^2+16 e^{4 h}+8 \left(a^2+1\right) e^{2 (g+h)}}}, \\
	{\cal K}^{(r)} &= \frac{k_2}{4} e^{2 (g+h+\Phi )} \partial_\rho \Phi =  \frac{k_2 N_c}{16} \frac{e^{2 \Phi } \left(e^{2 g} \left(a^2-1\right)+4 e^{2 h}\right) \left(e^{2 g}
   \left(a^2-2 b a+1\right)+4 e^{2 h}\right)}{\sqrt{e^{4 g}
   \left(a^2-1\right)^2+16 e^{4 h}+8 \left(a^2+1\right) e^{2 (g+h)}}}.
}
Using the one-to-one map given in Eq.~\eqref{eq:onetoonemap}, one easily shows that in terms of the variables that appear in the PT ansatz, the rotation takes the form (note that for the unrotated solution $A =	\frac{\Phi}{4} - p + \frac{x}{2}$)
\SP{
	a^{(r)} &= a, \ \ b^{(r)} = b, \ \ \Phi^{(r)} = \Phi, e^{2\tilde g^{(r)}} = e^{2 \tilde g}, \\
	e^{2A^{(r)}} &= \left(1 - k_2^2 e^{2\Phi} \right)^{1/3} e^{2A}, \ \	e^{-6p^{(r)}} = \left(1 - k_2^2 e^{2\Phi} \right) e^{-6p}, \ \ e^{2x^{(r)}} = \left(1 - k_2^2 e^{2\Phi} \right) e^{2x}, \\
	\partial_{r^{(r)}} \chi^{(r)} &= \frac{k_2 N_c}{8} \left(1 - k_2^2 e^{2\Phi} \right)^{-2/3} e^{4 p+2 \Phi -2 \tilde{g}} \left(a^2+e^{2 \tilde{g}}-1\right) \left(a^2-2 b
   a+e^{2 \tilde{g}}+1\right), \\
	\partial_{r^{(r)}} h_1^{(r)} &= \frac{k_2 N_c}{8} \left(1 - k_2^2 e^{2\Phi} \right)^{-2/3} e^{4 p+2 \Phi -2 \tilde{g}} \left(2 a e^{2 \tilde{g}} (a-b)+e^{4
   \tilde{g}}+\left(a^2+1\right) \left(a^2-2 b a+1\right)\right), \\
	h_2^{(r)} &= \frac{k_2 N_c}{4} e^{2 \Phi -4 p} \partial_r b = -\frac{k_2 e^{x+\frac{3 \Phi}{2}-\tilde{g}} \left(a^2+e^{2 \tilde{g}}-1\right) a}{\sqrt{a^4+2 \left(-1+e^{2 \tilde{g}}\right) a^2+\left(1+e^{2 \tilde{g}}\right)^2}}, \\
	{\cal K}^{(r)} &= 2 k_2 e^{-4 p+2 x+\Phi } \partial_r \Phi = \frac{k_2 N_c e^{x+\frac{3 \Phi }{2}-\tilde{g}} \left(a^2+e^{2 \tilde{g}}-1\right) \left(a^2-2
   b a+e^{2 \tilde{g}}+1\right)}{4 \sqrt{a^4+2 \left(-1+e^{2 \tilde{g}}\right)
   a^2+\left(1+e^{2 \tilde{g}}\right)^2}},
}
where
\SP{
	dr^{(r)} = 2 e^{-4p^{(r)}} d\rho = e^{4 (p - p^{(r)})} dr = \left(1 - k_2^2 e^{2\Phi} \right)^{2/3} dr.
}

\section{UV asymptotic expansions\label{Sec:expansions}}

In this appendix we expand by 
brute force the expressions for the eight  background scalars
for the various classes of solutions discussed in the paper, by
defining $\r=-\frac{3}{2}\log z$, and by expanding for small $z$.

For the KS solutions (having fixed $b_1=0=b_2=f_0=\tilde{p}$):
\beqs
a^{(KS)}&=&2 z^3\,+\,{\cal O}(z^9),\label{KSexpansionszz}\\
b^{(KS)}&=&-6 z^3 \log z\,+\,{\cal O}(z^9)\,,\\
\Phi^{(KS)}&=&\Phi_{\infty}\,,\\
  x^{(KS)}&=&
  \frac{1}{2} \log \left(\frac{3}{8} \left(-12 \log (z) N^2e^{\Phi_{\infty}}-N^2e^{\Phi_{\infty}}+2
   \tilde{M}\right)\right)\nonumber\\ && 
  + \frac{2}{125} \left(\frac{3 \left(7 N^2e^{\Phi_{\infty}}-10 \tilde{M}\right)}{12 \log (z)
   N^2e^{\Phi_{\infty}}+N^2e^{\Phi_{\infty}}-2 \tilde{M}}+150 \log (z)-5\right) z^6
  \,+\,{\cal O}(z^9)\,,\\
  p^{(KS)}&=&-\frac{1}{6} \log \left(-3 \log (z)
   N^2e^{\Phi_{\infty}}-\frac{N^2e^{\Phi_{\infty}}}{4}+\frac{\tilde{M}}{2}\right)+\nonumber\\ && 
  \frac{z^6 \left(-31 N^2e^{\Phi_{\infty}}+70 \tilde{M}-30 \log (z) \left(60 \log (z) N^2e^{\Phi_{\infty}}+23 N^2e^{\Phi_{\infty}}-10
   \tilde{M}\right)\right)}{125 \left(-12 \log (z) N^2e^{\Phi_{\infty}}-N^2e^{\Phi_{\infty}}+2
   \tilde{M}\right)}
  \,+\,{\cal O}(z^9)\,,\\
  \tilde{g}^{(KS)}&=&-2z^6\,+\,{\cal O}(z^9)\,,\\
  \partial_{\r}h_1^{(KS)}&=&
 2 e^{\Phi_{\infty}} N + 
 16 e^{\Phi_{\infty}} N z^6 (1 + 3 \log z )
  \,+\,{\cal O}(z^9) \,,\\
  h_2^{(KS)}&=&2 N e^{\Phi_{\infty}} z^3 (1+3 \log z)\,+\,{\cal O}(z^9)\,.
\label{KSexpansionfinal}
\eeqs
In particular
\beqs
x^{(KS)}+3p^{(KS)}&=&\frac{1}{2}\log \frac{3}{2}\,+\,z^6(1+6\log z)\,+\,{\cal O}(z^9)\,.
\eeqs

Turning back to the wrapped-D5 system, and  making use of the
UV asymptotic expansion for $P$, one finds that
\beqs
\lim_{\r\rightarrow +\infty} e^{-4\Phi}&=&18 c_+^3 e^{-4\Phi_o}\,,
\eeqs
and hence we  fine-tune $k_2=e^{-\Phi_{\infty}}=(18 c_+^3)^{1/4}e^{-\Phi_o}$.
The resulting rotated and fine-tuned solution, expanded in the same way as the 
KS solution, yields
\beqs
	         a^{(r)} &=&{2} z^3 \left(1-\frac{3 {N_c} (3 \log (z)+1)
   z^2}{9{c_+}}+\frac{N_c^2(3 \log (z) +1)^2
   z^4}{9{c_+}^2}\right)
    \,+\,{\cal O}(z^9),\label{bbexpansionzz}\\
		b^{(r)} &=&-6 z^3 \log (z)
\,+\,{\cal O}(z^9)  \,,\\
		\Phi^{(r)} &=&\frac{1}{4} \log \left(\frac{e^{4 \Phi_o}}{18
   c_+^3}\right) +  
   \frac{ N_c^2 (12 \log (z)+1)
   z^4}{48 c_+^2}
   +\frac{N_c^4 (72 \log (z) (3 \log (z) (4 \log (z)+7)+13)+257)
   z^8}{3456 c_+^4}
    \,+\,{\cal O}(z^9) \,,\\
		x^{(r)} &= &\frac{1}{8} \left(\log
   \left(\frac{9 e^{4 \Phi_o} N_c^8}{536870912
   c_+^3}\right)+4 \log (-12 \log (z)-1)\right)
   \\ && 
   +\frac{N_c^2 (3 \log (z)+1) (6 \log (z) (6 \log
   (z)+7)+19) z^4}{9 c_+^2 (12 \log (z)+1)}
 \nonumber   \\ && 
   +\frac{\left(14400 c_+^3-37
   c_-+12 \log (z) \left(3 \log (z) \left(2304 \log (z)
   c_+^3+7872 c_+^3-c_-\right)-8 \left(c_--1392
   c_+^3\right)\right)\right) z^6}{1728 c_+^3 (12 \log
   (z)+1)} 
 \nonumber   \\ && +\frac{ z^8}{497664 c_+^4
   (12 \log (z) N_c+N_c)^2}
   \left\{\frac{}{}-1218038 N_c^6+27 \left(c_--576
   c_+^3\right) \left(c_--192 c_+^3\right)
   \right.  \nonumber \\ && \left. \frac{}{}+12 \log (z)
   \left[\frac{}{}-326360 N_c^6+27 \left(258048 c_+^6-1152 c_-
   c_+^3+c_-^2\right)
   +144 \log (z) \left(\frac{}{}5737 N_c^6-864
   c_+^3 \left(c_--480 c_+^3\right)
    \right.\right.\right.  \nonumber\\ && \left.\left.\left.\frac{}{}+8 \log (z)
   \left(124416 c_+^6+5143 N_c^6+3 N_c^6 \log (z) (72
   \log (z) (\log (z) (12 \log
   (z)+37)+58)+3667)\right)\right)\right]\right\}
     \,+\,{\cal O}(z^9)  \nonumber \,,\\     
		p^{(r)} &= &
		\frac{1}{24} \left(-\log \left(\frac{9 c_+^5 e^{4\Phi_o}}{8192}\right)-4 \log \left(-\frac{N_c^2 (12 \log (z)+1)}{24
   c_+^2}\right)\right)
   \\ &&
   -\frac{{N_c}^2 z^4 (9 \log (z) (3 \log (z) (4 \log
   (z)+7)+16)+58)}{108 {c_+}^2 (12 \log (z)+1)}
 \nonumber   \\ &&
 +  \frac{z^6 \left(-25344 {c_+}^3+65 {c_-}+12 \log (z) \left(-17088
   {c_+}^3+7 {c_-}+6 \log (z) \left(-2304 \log (z)
   {c_+}^3-4416 {c_+}^3+{c_-}\right)\right)\right)}{10368
   {c_+}^3 (12 \log (z)+1)}
  \nonumber  \\ &&
   -\frac{z^8}{1492992c_+^4 (12
   \log (z)N_c+{N_c})^2}
\left\{ \frac{}{}-1271030N_c^6+27 \left({c_-}-576
  c_+^3\right) \left({c_-}-192c_+^3\right)\right. \nonumber \\ && \left. \nonumber
  +12 \log (z)
   \left[\frac{}{}-457112N_c^6+27 \left(258048c_+^6-1152c_-
  c_+^3+{c_-}^2\right)+144 \log (z) \left(\frac{}{}-3131
  N_c^6-864c_+^3 \left({c_-}-480c_+^3\right)
  \right.\right.\right.\\  \nonumber && \left.\left.\left.
  +4
   \log (z) \left(248832c_+^6+2078N_c^6+3N_c^6 \log
   (z) (72 \log (z) (8 \log (z) (3 \log
   (z)+7)+65)+2861)\right)\frac{}{}\right)\frac{}{}\right]  
 \frac{}{}  \right\}     \,+\,{\cal O}(z^9) \nonumber \,, \\
		\tilde{g}^{(r)} &=&-\frac{z^2 N_c(1+3 \log(z))}{3c_+}+
\frac{z^6 \left(-648 {c_+}^3+35 {N_c}^3+9 {N_c}^3 \log (z)
   (24 \log (z) (\log (z)+1)+17)\right)}{324 {c_+}^3} \\ &&
+   \frac{{N_c} z^8 \left(1152 {c_+}^3-{c_-}+3 \log (z)
   \left(2304 \log (z) {c_+}^3+768
   {c_+}^3-{c_-}\right)\right)}{1728 {c_+}^4}
    \,+\,{\cal O}(z^9) \nonumber \,,\\
		\partial_{\r}h_1^{(r)} &=& 
\sqrt[4]{\frac{e^{4 \Phi_o}}{18c_+^3}} N_c
\left\{\frac{1}{2}+\frac{144 N_c^2 \left(\log ^2\left(z^{12}\right)+\log
   \left(z^{132}\right)+19\right) z^4}{20736 c_+^2}+4 (3 \log (z)+1)
   z^6
  \right. \\ && \left. \frac{}{}  -\frac{N_c^4 (24 \log (z) (36 \log (z) (2 \log (z) (12
   \log (z)+7)+7)+101)-239) z^8}{20736 c_+^4}\right\} \,+\,{\cal O}(z^9) \,,\nonumber \\
		h_2^{(r)} &=&\sqrt[4]{\frac{e^{4 \Phi_o}}{18c_+^3}}   \frac{N_c z^3}{2 }
   (3 \log (z)+1) \left(1+\frac{N_c^2}{c_+^2} z^4\left(\frac{1}{24}+\frac{1}{2} \log (z) \right)
   \right)
    \,+\,{\cal O}(z^9)  \,.
\label{bbexpansionfinal}\eeqs

It is useful to look more in detail at the two specific quantities
\beqs
x^{(r)}+3p^{(r)}&=&
\frac{1}{2}\log\frac{3}{2}\,
+\frac{{N_c}^2 z^4 (\log (z) (3 \log (z)+4)+2)}{4 {c_+}^2} 
+z^6 \left(-\frac{{c_-}}{384 {c_+}^3}+6 \log (z)+1\right)\nonumber\\ 
&&
+\frac{{N_c}^4 z^8 (3 \log (z)+1) (3 \log (z) (3 \log (z) (3 \log
   (z)+7)+20)+23)}{216 {c_+}^4}
  \,+\,{\cal O}(z^9)\,,\\
  e^{2\tilde{g}^{(r)}}+a^{(r)\,2}-1&=&
-\frac{2N_c z^2 (3 \log (z)+1)}{3c_+}
+\frac{2 z^4 N_c^2 (3 \log (z)+1)^2}{9c_+^2}
+\frac{{N_c}^3 z^6 (3 \log (z)+1)}{6c_+^3}\nonumber\\ &&
-\frac{{N_c} z^8}{2592
  c_+^4}  \left(\frac{}{}-3456c_+^3+352N_c^3
 +3
  c_-+3 \log (z) \left(-2304c_+^3+832N_c^3
  \right.\right.\nonumber \\ && \left.\frac{}{}\left(+3
  c_-+288 \log (z) \left(-24c_+^3+7N_c^3+2
  N_c^3 \log (z) (3 \log (z)+4)\right)\right)\right) \,+\,{\cal O}(z^9)\,,
\eeqs
both of which are unaffected by the rotation.

\section{Curvature invariants for KS deformations}\label{curvatureccc}

In Section~\ref{Sec:Curvature}, a few curvature invariants for the rotated solutions were studied in order to understand the nature of the singularity in the IR. For comparison, we perform here the same analysis for the singular deformations of Klebanov-Strassler obtained by taking the integration constant $f_0$ of Eq.~\eqref{eq:KSf} to be non-zero.

All functions in the background are determined analytically except $x$ which satisfies the equation of motion Eq.~\eqref{eq:KSeomx}, the IR expansion of which determines $x$ as
\SP{
	x = x_0+\frac{32 \rho ^3}{9 f_0}-\frac{8}{9} \left(e^{\Phi _{\infty }-2 x_0} N^2\right) \rho
   ^4+\frac{128 \rho ^5}{45 f_0}+\frac{64}{135} \left(e^{\Phi _{\infty }-2 x_0}
   N^2-\frac{40}{f_0^2}\right) \rho ^6 + \mathcal O \left(\rho^7\right),
}
where $x_0$ is an integration constant. This, in turn, implies that $R$, $R_{\mu\nu}R^{\mu\nu}$, and $R_{\mu\nu\tau\sigma}R^{\mu\nu\tau\sigma}$ have IR expansions given by
\SP{
	R = \frac{1}{2} e^{\Phi _{\infty }-3 x_0} N^2 f_0-\frac{4}{3} \left(e^{\Phi _{\infty }-3 x_0} N^2
   f_0\right) \rho ^2+\mathcal O\left(\rho ^4\right),
}
\SP{
	R_{\mu\nu}R^{\mu\nu} = \frac{5}{8} e^{2 \Phi _{\infty }-6 x_0} N^4 f_0^2-\frac{10}{3} \left(e^{2 \Phi _{\infty }-6
   x_0} N^4 f_0^2\right) \rho ^2+ \mathcal O\left(\rho ^4\right),
}
\SP{
	R_{\mu\nu\tau\sigma}R^{\mu\nu\tau\sigma} = \frac{45 e^{-2 x_0} f_0^2}{256} \rho^{-8} -\frac{3 \left(e^{-2 x_0} f_0^2\right)}{8}\rho^{-6}+\frac{5 e^{-2 x_0} f_0}{8}  \rho^{-5}+ \mathcal O\left(\rho^{-4}\right).
}
As with the rotated solutions, $R$ and $R_{\mu\nu}R^{\mu\nu}$ stay finite in the IR, while $R_{\mu\nu\tau\sigma}R^{\mu\nu\tau\sigma}$ diverges as $\rho^{-8}$.

For completeness, let us also give the IR expansions for the non-singular solution obtained by putting $f_0 = 0$, i.e. the original solution of Klebanov-Strassler. Now, we have that
\SP{
	e^x = \tilde x \rho +\left(\frac{4 \tilde x}{15}-\frac{16 e^{\Phi _{\infty }} N^2}{9 \tilde x}\right) \rho
   ^3+\left(-\frac{128 e^{2 \Phi _{\infty }} N^4}{81 \tilde x^3}+\frac{32 e^{\Phi _{\infty }}
   N^2}{45 \tilde x}+\frac{16 \tilde x}{525}\right) \rho ^5+ \mathcal O\left(\rho ^7\right),
}
where $\tilde x$ is an integration constant. (Note that this expansion is radically different from the one in the case of non-zero $f_0$.) This leads to
\SP{
	R = \frac{16 e^{\Phi _{\infty }} N^2}{3 \tilde x^3}-\frac{128 \left(e^{\Phi _{\infty }} N^2 \left(\tilde x^2-2
   e^{\Phi _{\infty }} N^2\right)\right)}{9 \tilde x^5} \rho^2 + \mathcal O\left(\rho ^4\right),
}
\SP{
	R_{\mu\nu}R^{\mu\nu} = \frac{640 e^{2 \Phi _{\infty }} N^4}{9 \tilde x^6}+\frac{2048 e^{2 \Phi _{\infty }} N^4 \left(122
   e^{\Phi _{\infty }} N^2-45 \tilde x^2\right)}{243 \tilde x^8}  \rho^2 + \mathcal O\left(\rho ^4\right),
}
\SP{
	R_{\mu\nu\tau\sigma}R^{\mu\nu\tau\sigma} =& \frac{32 \left(440 e^{2 \Phi _{\infty }} N^4+81 \tilde x^4\right)}{135 \tilde x^6}+ \\& \frac{512 \left(6000
   e^{3 \Phi _{\infty }} N^6-2600 e^{2 \Phi _{\infty }} \tilde x^2 N^4+ 270 e^{\Phi _{\infty }} \tilde x^4
   N^2-243 \tilde x^6\right)}{2025 \tilde x^8} \rho^2 + \mathcal O\left(\rho ^3\right),
}
which, as expected, all stay finite in the IR.


\end{document}